\documentclass{optica-article}

\journal{opticajournal} % for journals or Optica Open

\articletype{Research Article}

\usepackage{lineno}
\usepackage{amsmath}
\usepackage{mathtools}
\usepackage{amsfonts}
\usepackage{booktabs}
\usepackage{bbold}
\usepackage{soul}
\usepackage{lscape}
\usepackage[title]{appendix}
\usepackage{comment}

\begin{document}

\title{Four-Qubit Variational Algorithms in Silicon Photonics with Integrated Entangled Photon Sources}

\author{ Alessio Baldazzi,\authormark{1,*} 
Matteo Sanna,\authormark{1,2}
Massimo Borghi,\authormark{1,3}
Stefano Azzini,\authormark{1}
and Lorenzo Pavesi\authormark{1}}

\address{
\authormark{1} 
Department of Physics, University of Trento, Trento, Italy\\
\authormark{2} current affiliation:
Rotonium S.r.L., Padova, Italy\\
\authormark{3} current affiliation:
Department of Physics, University of Pavia, Pavia, Italy
}

\email{\authormark{*}alessio.baldazzi@unitn.it} %% email address is required; see note below about the corresponding author designation

% use {asbstract*} to suppress the copyright line. Copyright information will be added in production

\begin{abstract*} 
Variational quantum algorithms are hybrid quantum-classical approaches extensively studied for their potential to leverage near-term quantum hardware for computational advantages.
In this work, we successfully execute two variational quantum algorithms on a silicon photonic integrated circuit at room temperature: a variational quantum eigensolver for the Hydrogen molecule and a variational quantum factorization for semi-prime numbers. 
In our reconfigurable silicon photonic circuit, four identical spontaneous-four-wave-mixing-based integrated photon pair sources are used to prepare two path-entangled ququarts, whose correlation gives rise to the resource for generic trial states' preparation.
This marks a first demonstration of variational quantum algorithms on a photonic quantum simulator with integrated photon pair sources.
% Thus, we demonstrate how these algorithms can exploit the currently available photonic processors to accurately solve specific tasks. 

\end{abstract*}

%%%%%%%%%%%%%%%%%%%%%%%%%%  body  %%%%%%%%%%%%%%%%%%%%%%%%%%
\section{Introduction}
\label{sec:intro}

\begin{figure}[t]
    \centering
    \includegraphics[width=\textwidth]{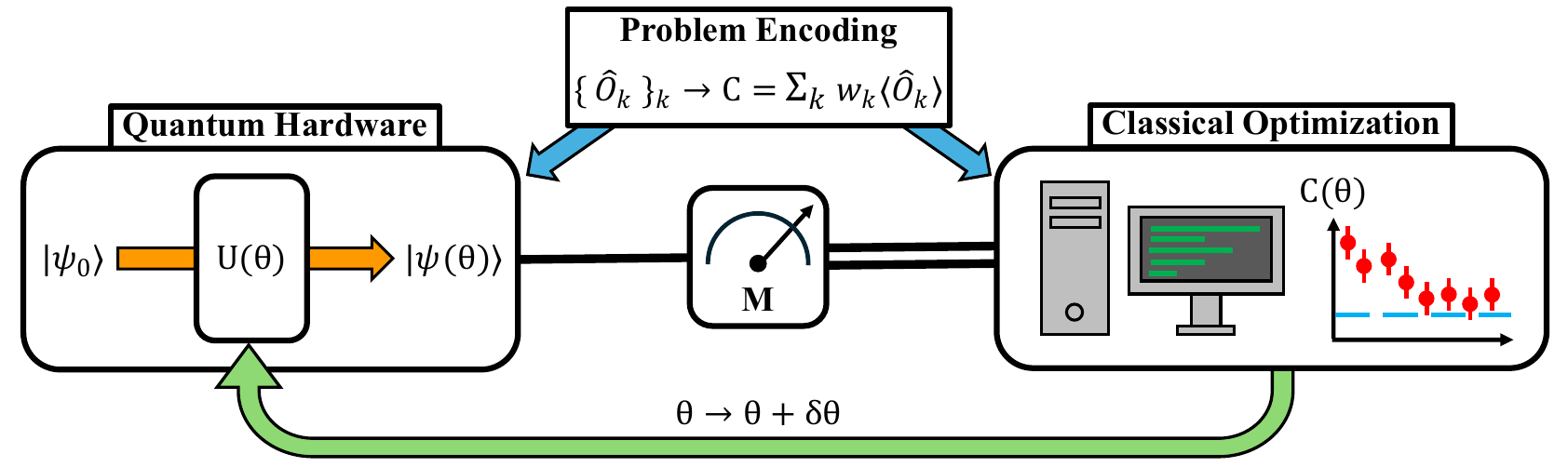}
    \caption{Schematic workflow of variational quantum algorithms. The procedure starts with the problem encoding: depending on the problem/task and the qubit register of the Quantum Hardware, a set of observables $\{\hat{O}_k\}_k$ is identified and the cost function ${\rm C}$ is constructed as a weighted sum of the observables expectation values, $\{\langle \hat{O}_k \rangle\}_k$.
    Then, the Quantum Hardware is configured to prepare the trial states $|\psi(\uptheta)\rangle$ by starting from an initial state $|\psi_0\rangle$ and applying the transformation ${\rm U}$, which depends on the variational parameters $\uptheta$. The measurements of the trial state give the observables' expectation values. By processing the collected data, the classical machine calculates the cost function for each trial state and, after that, it sets new parameters on the quantum part. This cycle is driven by an optimization routine made by the classical part and it is repeated until the cost function converges to an extremal point.}
    \label{fig:algorithm}
\end{figure}

Quantum computing (QC)~\cite{Feynman82,divincenzo_physical_2000,bennett2000quantum,nielsen_chuang_2010,johnson2014quantum} is based on a new computational paradigm, that promises an exponential speed-up over classical machines for certain classes of problems~\cite{lloyd1996universal,R_nnow_2014}. 
However, nowadays quantum processors are typically limited in terms of number of qubits, interconnection topology, fidelity of the entangling gates and losses. Because of these limitations, they are termed 'Noise Intermediate Scale Quantum' (NISQ) devices~\cite{Preskill_2018}. Variational quantum algorithms (VQAs)~\cite{cerezo} are investigated to achieve quantum utility~\cite{IBMQutility} through NISQ processors. In fact, they require a modest number of qubits and a limited circuit depth.
In a nutshell, the workflow of VQAs is sketched in Fig.~\ref{fig:algorithm}. 
First, the problem is encoded into a collection of measurable observables, $\{\hat{O}_k\}_k$, based on the qubit register of the Quantum Hardware (QH). A cost function ${\rm C}$ is defined as a weighted sum of expectation values as follows
\begin{equation}
    {\rm C} = \sum_k w_k \langle \hat{O}_k \rangle \,,
    \label{eq:costfunction}
\end{equation}
where $\{\langle \hat{O}_k \rangle\}_k$ are the expectation values of the corresponding observables evaluated on a quantum state $|\psi(\uptheta)\rangle$, parametrized by a list of parameters $\uptheta$, and the classically-computed coefficients $\{w_k\}_k$ depend on the specific problem to address.
Once the problem is encoded, the reconfigurable QH starts preparing trial states and measurements are performed. The collected raw data are sent to the classical machine for post-processing, where the cost function is evaluated. An optimization routine is executed by the classical part and the quantum part is updated with a new setting of parameters $\uptheta$, accordingly to the cost function result. This procedure is repeated until the cost function converges to an extremal value. 
In this way, the minimization of the cost function and the optimal setting of parameters for the trial state preparation is achieved.
For a generic set $\{ \hat{O}_k \}_k$, the observables don't commute, and the evaluation of their expectation values can be simultaneously done only within the same commuting group (CG)~\cite{Tilly_2022,josa1994fidelity,james2001measurement,darino2003quantum,altepeter20044qubit}.
With respect to Quantum Phase Estimation and Shor's algorithm~\cite{shor,kitaev1995,Cleve_1998,nielsen_chuang_2010}, VQAs are hybrid quantum-classical approaches and they don't require a set of universal gates or quantum error-correction codes. Such algorithms show robustness against error~\cite{McClean_2016,O_Malley_2016} and can run on shallow circuits~\cite{Kandala_2017}. What is needed is a reconfigurable QH able to prepare states that can be mapped to the Hilbert space configurations of the addressed problem. Indeed, the whole procedure shown in Fig.~\ref{fig:algorithm} must be tailored to a specific problem, since the coefficients $\{w_k\}_k$ and the observables $\{\hat{O}_k\}_k$ are determined by the addressed task and the qubit register.
The problems can be classified as classical, e.g. combinatorics problems, and quantum, e.g. energy levels and scattering amplitudes. In both cases, the advantage with respect to classical algorithms could come from the information encoding and the quantum parallelization, given by the exponential scaling of the Hilbert space dimension with the number of qubits, and the use of superposition and entanglement~\cite{nielsen_chuang_2010}.

The first experimental implementation of a variational quantum eigensolver (VQE) has been realized on a photonic integrated circuit (PIC) with an external photon pair source~\cite{Peruzzo_2014}. Then, the whole VQE theory has been formalized in~\cite{McClean_2016,Tilly_2022}.
VQE has been also demonstrated with superconducting qubits~\cite{O_Malley_2016,Kandala_2017,Colless_2018,qing2023usevqecalculateground} and trapped ions~\cite{Shen_2017,hempel_2018,Kokail_2019}, In these platforms, chemical accuracy of VQE has been reported~\cite{zhang2022,Kandala_2019,nam2020ground}. In the same family of variational algorithms, we can find the variational quantum factoring (VQF)~\cite{VQF,zhang2023variational}, a factorization algorithm tailored to the number to factorize. Also in this case, the algorithm has been successfully performed with superconducting qubits~\cite{selvarajan2021prime}, and, only very recently, on a PIC with an external photon pair source~\cite{agresti2024}. 
In general, despite several experimental implementations, it is still not clear if VQAs can lead to a useful quantum advantage~\cite{McClean_2016,Tilly_2022}, with major issues related to the minimization procedure of the cost function, such as the barren plateau phenomenon~\cite{cerezo}. 
Thus, even if NISQ processors satisfy the requirements of VQAs, the desired accuracy and the duration of the specific algorithm may prevent VQAs from surpassing classical techniques.
For these reasons, further theoretical and experimental investigations are needed to understand the full potential of this method for specific-purpose tasks.

Integrated quantum photonics~\cite{hamilton2017gaussian,Wang_2018,adcock2019programmable,wang2019boson,Llewellyn_2019,Bartlett2020Universal,wang2020integrated,Vigliar_2021,Adcock_2021,Lee_24} is a particularly promising platform, since photons are a natural choice to transfer information thanks to low decoherence and high data rates.
Moreover, the use of different degrees of freedom of photons~\cite{o2007optical,tan_resurgence_2019} makes it possible to encode the information with a wide variety of choices by using linear operations with high fidelity, robustness and flexibility. For example, path encoding~\cite{Kok_2007} relies on Mach-Zehnder interferometers (MZI) or MZI networks~\cite{reck_experimental_1994,clements_optimal_2016} to implement any transformations of a generic qudit, i.e. $d$-dimensional qubit. 
Knill et al.~\cite{knill_scheme_2001} showed that linear optical quantum computing (LOQC)~\cite{cerf_optical_1998,pittman2001probabilistic,knill_scheme_2001,Ralph_2001,pittman2002demonstration,postsel_CZ,post_sel_2,obrien_demonstration_2003,gasparoni2004realization,zhao2005experimental,browne2005resource,fiuravsek2006linear,bao2007optical,okamoto2011realization,li2011reconfigurable,tan_resurgence_2019,li2021heralded,lee2022controlled,liu2022universal,li2022quantum,li2022chip,liu2023linear,Kwon_24,baldazzi2024} can be realized through post-selection. 
% In particular, by choosing specific post-selection/heralding criteria, an effective non-linearity~\cite{Scheel_2003} emerges and allows to execute multi-photon entangling gates even if the success probability is lower than 100$\%$.
This paradigm for photon entangling gates enables the existing technology to achieve photonic QC~\cite{o2007optical,o2009photonic,ladd2010quantum,laing2010high,carolan2015universal,Bartlett2020Universal,zhong20,alexander2024manufacturableplatformphotonicquantum} with both approaches to quantum algorithms: measurement-based~\cite{briegel2001persistent,raussendorf2001one} and gate-based~\cite{nielsen_chuang_2010} QC.
Because of the well-established photonic technology, fault-tolerant large-scale quantum photonic architectures could be achieved in the next few years leveraging the latest achievements of integrated optics~\cite{harris_2016,qiang2018large,bogaerts2018silicon,bogaerts2020programmable,ba0_22,moody2022,luo2023recent}. In particular, thanks to high-quality single photon sources and high performant PIC, a significant improvement of photonic based VQEs has been recently demonstrated using gate-based computation~\cite{Maring24}: four times faster than~\cite{Peruzzo_2014} and a chemical accuracy on par with superconducting and trapped-ion qubits~\cite{zhang2022,Kandala_2019,nam2020ground}. However, this latest achievement is based on a quantum-dot single photon source working at 5 K and on integrated photonic probabilistic CNOT gates, two limiting aspects that could be improved.

Quantum silicon photonics allows generating high-quality entangled photons at room temperature via spontaneous four-wave mixing (SFWM)~\cite{Fukuda_05,Clemmen_09,Helt_10,Azzini2012,Engin2013,SchmittbergerMarlow_20}. These states provide an extremely valuable quantum resource without the need for cryogenic temperatures. In the low-gain regime, a pair of energy-time entangled photons is generated~\cite{Mancini_2002,energy_time1,energy_time2}. Such quantum correlation can be utilized as an entanglement resource for photonic processors. 

This work aims to significantly advance the capabilities of integrated photonic VQAs by presenting a room-temperature silicon photonic integrated circuit (Si-PIC) implementing a four-qubit variational quantum eigensolver able to solve a quantum chemistry problem and a factorization task with high accuracy. The Si-PIC uses four spiral waveguides as sources of correlated photon pairs to create multidimensional entangled states, which serve as trial states for the implemented VQAs. For the first time, to the best of our knowledge, this work addresses proof-of-principle instances of quantum chemistry and factorization problems by means of a small-scale quantum photonic processor with integrated quantum light sources. Moreover, an interesting feature of our implementation exploits the time-energy entanglement of the photon pairs to obtain path entanglement, thus showing the possibility to avoid, at least to some extent, the use of probabilistic CNOT gates.

The paper is organized as follows.
Section~\ref{sec:circuit} describes the Si-PIC used to implement the VQAs.
Sec.~\ref{sec:H2} and Sec.~\ref{sec:factorize} show the results of the VQE for the Hydrogen molecule and of VQF for the semiprime numbers, respectively.
In Sec.~\ref{sec:disc}, we comment on scaling and improvement of our scheme.
Finally, conclusions are given in Sec.~\ref{sec:conclu}, providing also some comparison with the results on other platforms. All the details are reported in the various appendices.

\section{Photonic integrated circuit for four-qubit variational algorithms}
\label{sec:circuit}

\begin{figure}[t]
    \centering
    \includegraphics[width=\textwidth]{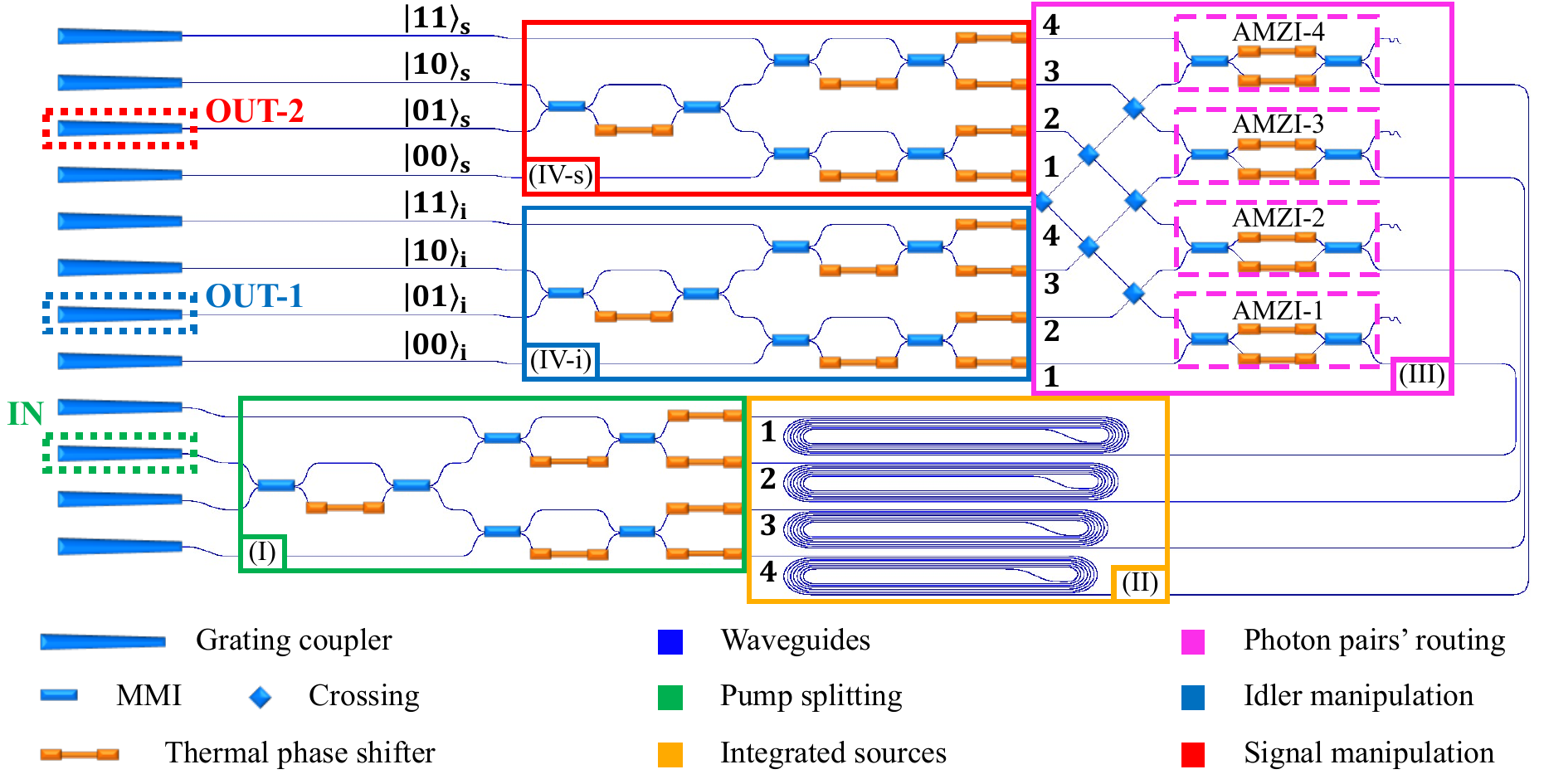}
    \caption{
    The layout of the silicon photonic integrated circuit utilized for four-qubit variational quantum algorithms.
    Blue lines are waveguides, solid blue components are passive integrated optical devices (grating couplers, MMIs and crossing waveguides) and solid orange components are thermal phase shifters.
    IN denotes the input where light is injected, and OUT-1 and OUT-2 stand for the outputs where light is collected. 
    The circuit is divided into five parts: (I) pump splitting in the green box, (II) integrated sources in the golden box, (III) photon pairs' routing (composed of AMZIs and crossing waveguides) in the purple box, and finally, (IV-i and IV-s) idler and signal manipulation in blue and red boxes, respectively.
    The numbers on the inputs of stage (II) and outputs of stage (III) denote the four spatial modes used to encode the two-ququart states, while the associated qubit states are reported on the left of stages (IV-i/s).
    }
    \label{fig:circuit}
\end{figure}

Fig.~\ref{fig:circuit} reports the scheme of the Si-PIC, whose footprint is 1.5$\times$5 mm$^2$.
The building blocks of this circuit are grating couplers~\cite{grating}, multimode-interference-based integrated beam splitters (MMIs)~\cite{Soldanoilgrande}, crossing waveguides~\cite{chen2006low} and thermal phase shifters (PSs)~\cite{Harris_14,Jacques_19}.
In particular, the composition of MMIs and PSs gives rise to integrated MZIs.
A detailed discussion of the Si-PIC design, the description of the setup and the characteristics of the utilized photonic components, both in terms of theoretical modeling and measured performances, are reported in the Methods and App.~\ref{app:preliminaries}-\ref{app:circuit}-\ref{app:setup}-\ref{app:lin_nonlin_char}.

The Si-PIC has four input and eight output ports. The ports are based on grating couplers and only one input and two output ports were used in this work (label IN, OUT-1 and OUT-2 in Fig.~\ref{fig:circuit}). The circuit is composed of five parts: (I) pump splitting (green box of Fig.~\ref{fig:circuit}), (II) integrated sources (golden box of Fig.~\ref{fig:circuit}), (III) photon pairs' routing (purple box of Fig.~\ref{fig:circuit}), (IV-i) and (IV-s) linear manipulation of the idler and signal photons (blue and red boxes of Fig.~\ref{fig:circuit}), respectively.

Stage (I) is used to coherently pump the 1.5 cm-long spiral-waveguide-based integrated sources, present in stage (II), with arbitrary amplitudes to generate correlated photon pairs through SFWM. 
We utilize fundamental-TE-intramodal non-degenerate SFWM in silicon spiral waveguides and, as usual, we denote the two generated photons as idler and signal, where the first is the photon with shorter wavelength and the latter the photon with longer wavelength.
We choose spiral waveguides as photon pair sources to obtain a high degree of spectral indistinguishability~\cite{indi_photo22,Lee_2023}, which is an important requirement for the VQAs, whose implementation is reported in Sec.~\ref{sec:H2} and Sec.~\ref{sec:factorize}.
In stage (III), asymmetric MZIs (AMZIs)~\cite{shamy_22} and crossing waveguides route idlers to stage (IV-i) and signals to stage (IV-s). 
In the low squeezing regime and through the coincidence basis of idlers and signals, we post-select the events where only one source emits a pair. 
Under this setting, the single-photon spatial modes of idler and signal can be utilized to encode ququart computational basis states, $\{ | m \rangle \}_{m=1\ldots4}$, and the output state of stage (III) reads as follows
\begin{equation}
    % | \mathbf{1}_m \rangle \to | m \rangle
    % \quad\implies\quad
    % | \Psi_{\rm III} \rangle \to 
    |\psi_{\rm III}^{(4)}\rangle \equiv \sum_{m=1}^4 \alpha_m  | m \rangle_i | m \rangle_s \,,
    \label{eq:stageiii_ququart}
\end{equation}
where the complex parameters $\{\alpha_m\}_m$ depend on the setting of stage (I) and the subscripts $i/s$ refer to idler and signal. 
Through the separation and the routing, the energy-time correlation of the twin photons~\cite{Mancini_2002,energy_time1,energy_time2} is converted into a spatial correlation. Neglecting losses, if an idler photon enters the $m$-th spatial mode of stage (IV-i), its twin signal photon arrives in the $m$-th spatial mode of stage (IV-s).
For a general setting of stage (I), we can achieve any separable two-ququart computational basis states or any $d$-dimensional entangled state of dimension $(2,3,4)$, where the first ququart is encoded in the idler photon and the second in the signal photon.
Then, by utilizing binary numbers, the two-ququart state can be mapped to the following four-qubit state
\begin{equation}
    |\psi^{(4)}_{\rm III} \rangle = 
    \alpha_{1} | 00 \rangle_i | 00 \rangle_s + 
    \alpha_{2} | 01 \rangle_i | 01 \rangle_s +
    \alpha_{3} | 10 \rangle_i | 10 \rangle_s +
    \alpha_{4} | 11 \rangle_i | 11 \rangle_s \,.
    \label{eq:stageiii_qubit}
\end{equation}
The labeling of the spatial modes and the two-qubit states is reported in  Fig.~\ref{fig:circuit} on the right and left of the blue and red boxes, respectively.
The first and second qubits are encoded in the idler photon and the third and fourth qubits in the signal photon.
For the rest of the manuscript whenever the subscripts $i/s$ are omitted, we are following the order in the previous equation for a state $|x_1 x_2 x_3 x_4\rangle$, with $\{x_i\}_{i=1\ldots4}=\{ 0,1 \}$, written in the qubit states.
Then, the final stages (IV-i) and (IV-s) perform linear transformations on each photon belonging to a given correlated photon pair. 
Thus, the final state written in the ququart computational basis becomes~\cite{Plesch_2011}
\begin{equation}
\begin{split}
   |\psi_{\rm IV}^{(4)}\rangle &= 
   \left( U^{(i)} \!\otimes\! U^{(s)} \right) \!|\psi_{\rm III}^{(4)} \rangle 
   =\sum_{m=1}^4 \alpha_m 
   \, | \xi_m \rangle_i | \xi_m \rangle_s \quad,\quad
   \mbox{where}\,\,\,
   | \xi_m \rangle_{i/s} \equiv U^{(i/s)} \,| m \rangle_{i/s} \,,
\end{split}
   \label{eq:stageiv_ququart}
\end{equation}
and $U^{(i/s)}$ represents the linear transformation on the idler/signal photon implemented in stages (IV-i/s). 
From the previous equation, we can note that the two ququarts are independently manipulated in stages (IV-i) and (IV-s). 
% Therefore, once the separated or entangled state of the two ququart is created through the first three stages, only local operations on each ququart are possible.
If $U^{(i/s)}$ is a generic SU(4)-transformation~\cite{Vidal_2004}, the state in the previous equation is exactly the well-known Schmidt decomposition of a bipartite system~\cite{Huber_2013}.
In this case, it is possible to prepare the generic trial state in the two-ququart Hilbert space.
The triangular MZI scheme~\cite{Wang_2018,Vigliar_2021} of stages (IV-i/s) cannot implement the generic unitary transformation of ququart states, since such action can be achieved by schemes like Reck~\cite{reck_experimental_1994} and Clements~\cite{clements_optimal_2016}. Instead, what the final stages can execute are generic projections of any ququart state encoded in the idler and signal photons on OUT-1 and OUT-2, respectively. As shown in Fig.~\ref{fig:circuit}, these outputs correspond to the qubit state $| 01 \rangle_{i/s}$, or equivalently the ququart state $| 2 \rangle_{i/s}$
% , and the qubit state $| 01 \rangle_{s}$, or equivalently the ququart state $| 2 \rangle_{s}$, respectively
~\cite{Wang_2018,Vigliar_2021}. 
Finally, the idler and signal photons are collected from OUT-1 and OUT-2 respectively and sent to two single-photon detectors (SPDs).
This operation can be ideally described by the Von Neumann projector on the ququart state $|2 \rangle_i|2 \rangle_s$, or equivalently the qubit state $| 01 \rangle_i| 01 \rangle_s$.
Within this configuration, the non-universality of ququart manipulation and the presence of one detector for each ququart is overcome by using multiple projective measurements.
This choice uses the minimal number of detectors and it is achieved by executing different projectors in the final stages of the circuit given the observable to be measured.
The number of linearly independent projectors is 16, i.e. a set of four projectors relative to four independent input states of stage (IV-i/s) for each ququart.
The choice of 16 projectors represents the measurement setting that we want to implement, and thus, it depends on the set of CGs of observables $\{\hat{O}_k\}_k$ in the cost function to be measured~\cite{josa1994fidelity,james2001measurement,darino2003quantum,altepeter20044qubit}.

\begin{figure}[t]
    \centering
    \includegraphics[width=\textwidth]{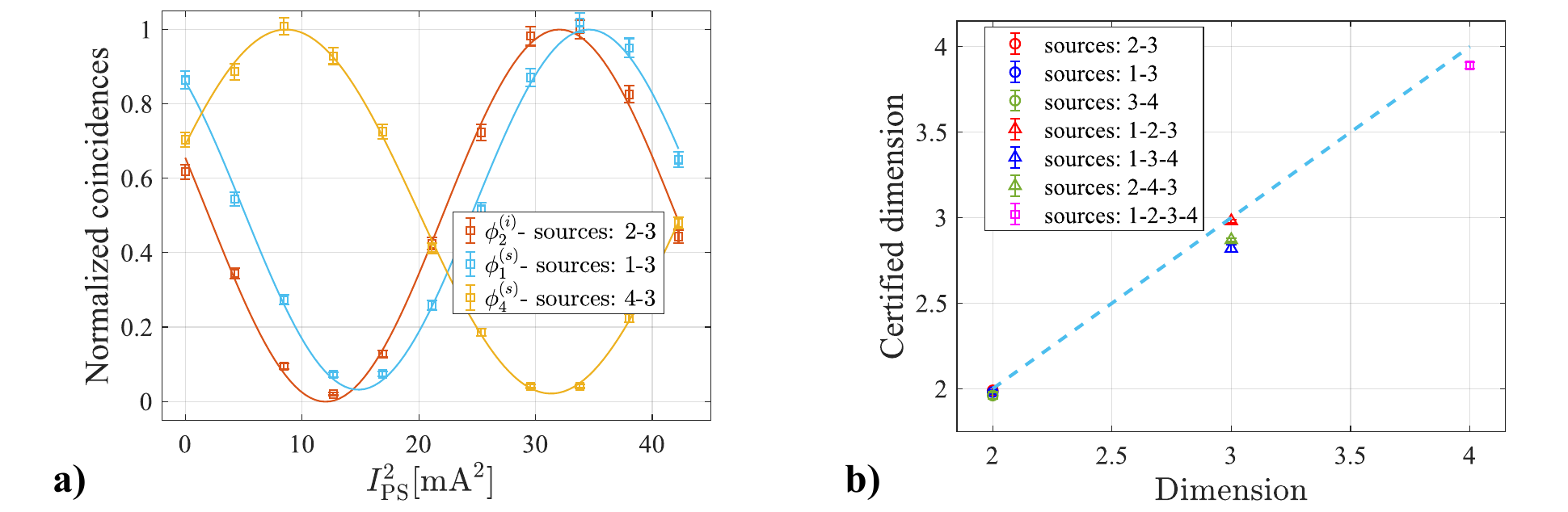}
    \caption{ 
    (a) Normalized coincidences of correlated photons measured at the outputs OUT-1 and OUT-2 in heralded single-photon interference experiment. 
    % Pairs are generated in a balanced superposition of two sources and manipulated with a relative phase $\phi$ followed by balanced beam-splitter action as a function of the current $I_{\rm PS}$ applied to the PS that introduces $\phi$.
    The colors are associated with three different pairs of sources and the sinusoidal fit is reported as a continuous line.
    The PSs used for the three interference patterns are: $\phi_2^{(i)}$ (the PS in the second spatial mode at the input of stage (IV-i)) for sources' pair (2-3), $\phi_{1/4}^{(s)}$ (PS in the first/fourth spatial mode at the input of stage (IV-s)) for sources' pair (1/4-3).
    (b) Measured certified dimension of generated entangled states. Different colors and symbols represent dimensions and the utilized sources, while the light blue line refers to ideal values.
    Error bars are smaller than the markers.
    }
    \label{fig:inter_dim}
\end{figure}

The quantum resource of our photonic processor is given by the multi-dimensional entangled states of photon pairs. At every run of the VQAs, the two single photons are independently manipulated and there is no multi-photon interference between independent photon pairs~\cite{Silverstone_2013,Faruque_2018}. Therefore, purity~\cite{Vernon_2017,signo_hera,Paesani_2020} is not of concern.
What matters is the indistinguishability of the sources present in our bipartite system, since it affects the single-photon interference visibility. 
Heralded single-photon interference experiments are described in App.~\ref{app:her_single_inter} and the results are shown in Fig.~\ref{fig:inter_dim}(a). Moreover, measurements to certify the effective multidimensional entanglement have also been carried out as explained in App.~\ref{app:cert_dim}, and the related results are plotted in Fig.~\ref{fig:inter_dim}(b). Overall, these results demonstrate a good degree of sources' indistinguishability ($99.3\pm 0.5$ visibility for sources 2-3) and high quality of bipartite entanglement.

Based on the Si-PIC, and given a set of binary observables $\hat{O}_k$, the cost function given in Eq.~\eqref{eq:costfunction} can be rewritten in terms of the ingredients of the Si-PIC itself as follows %(see App.~\ref{app:vqaimple} for details)
\begin{equation}
\begin{split}
    {\rm C}(\boldsymbol{\alpha}) 
    &=   
    \sum_{k} \sum_{{\rm I} \in {\rm CG} } w_k \!\!\!\sum_{m_1,m_2=1}^4 \!\!\!\boldsymbol{\pi}\!\left[k,{\rm I},m_1,m_2\right]\, 
    \Big| \langle 2 |_i \langle 2 |_s \! \left( U^{(i)}\!\left[{\rm I},m_1\right] \otimes U^{(s)}\!\left[{\rm I},m_2\right] \right) \!|\psi_{\rm III}^{(4)}\! (\boldsymbol{\alpha}) \rangle \Big|^2 ,
    \end{split}
    \label{eq:costfunc_pro_rotnew}
\end{equation}
where $\boldsymbol{\alpha}$ are the variational parameters of our QH, the index I indicates the measurement settings associated with different CGs, $U^{(i/s)}\!\left[{\rm I},m\right]$ is the projection of the $m$-th element among the four eigenstates of the $k$-th observable on the state $|2 \rangle_{i/s}$ within the same measurement setting and $\boldsymbol{\pi}\!\left[k,{\rm I},m_1,m_2\right]$ is the product of the eigenvalues of the corresponding eigenstates pairs.
The variational parameters depend on the pump splitting performed in stage (I), while the measurement settings are set in stages (IV-i) and (IV-s). In particular, as we measure idler-signal photons coincidence events at OUT-1 and OUT-2 of the Si-PIC, we use them to estimate the probabilities contained in Eq.~\eqref{eq:costfunc_pro_rotnew}, upon a normalization by the total coincidence events recorded with the same measurement setting. In this case, Eq.~\eqref{eq:costfunc_pro_rotnew} can be rewritten as follows 
\begin{equation}
\begin{split}
    & {\rm C}(\boldsymbol{\alpha}) =   
    \sum_{k} \sum_{{\rm I} \in {\rm CG} } w_k \!\sum_{m_1,m_2=1}^4 \boldsymbol{\pi}\!\left[k,{\rm I},m_1,m_2\right]\, 
    \frac{{\rm CC}[\boldsymbol{\alpha},k,{\rm I},m_1,m_2] }{{\rm CC}^{\rm tot}[\boldsymbol{\alpha},k,{\rm I}]} \,,\\
    & \mbox{where}\,\,\,{\rm CC}^{\rm tot}[\boldsymbol{\alpha},k,{\rm I}] \equiv \sum_{m_1,m_2=1}^4
    {\rm CC}[\boldsymbol{\alpha},k,{\rm I},m_1,m_2] \,.
    \end{split}
    \label{eq:cost_CC}
\end{equation}
Here ${\rm CC}$ refers to the coincidence counts measured with a specific setting of variational parameters $\boldsymbol{\alpha}$ and the $(m_1,m_2)$ projective measurement associated with the I-th measurement setting, and ${\rm CC}^{\rm tot}$ to the total coincidence counts of all the projective measurements within the I-th measurement setting.
The evaluation of the cost function is achieved by utilizing the coincidence counts provided by our photonic processor and by the weighted sum shown in Eq.~\eqref{eq:cost_CC} on the digital computer, i.e. a PC.

\section{Solving the Hydrogen molecule}
\label{sec:H2}

\begin{figure}[t]
    \centering
    \includegraphics[width=\textwidth]{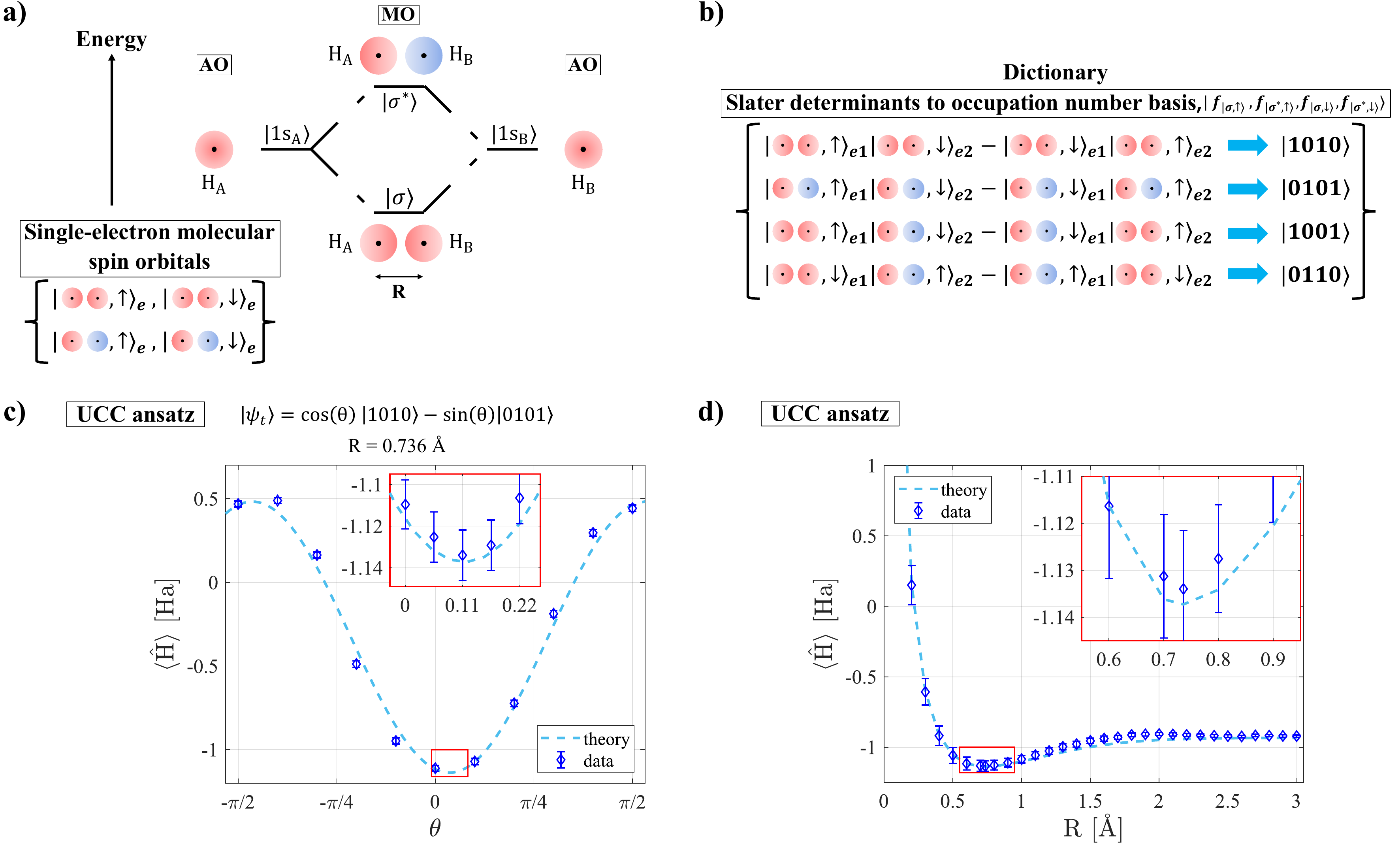}
    \caption{
    (a) Graphical representation of atomic orbitals (AOs) and molecular orbitals (MOs) for the Hydrogen molecule ${\rm H}_2$ together with the list of single-electron molecular spin orbitals. The two MOs states $|\sigma\rangle$ and $|\sigma^*\rangle$ are the bonding and anti-bonding orbital states.
    (b) Slater determinants associated with the four configurations of the two electrons in ${\rm H}_2$ mapped to the occupation number states. 
    (c) Experimental and theoretical values of the ${\rm H}_2$ ground state energy, measured in Hartree, as a function of the variational parameter $\uptheta$, which controls the amount of contribution of the anti-bonding orbital to the equilibrium configuration. The energy is calculated within the UCC ansatz, reported above the graph, and using the coefficients $\{ w_k \}_k$ associated with an atomic distance R equal to 0.736~\AA~\cite{psi4}. The red-border inset shows additional data points obtained during the optimization procedure assessing the minimum energy value.
    (d) Experimental and theoretical values of the ${\rm H}_2$ ground state energy, measured in Hartree, as a function of the atomic distance. Each point is the minimum energy for a fixed atomic distance within the UCC ansatz. The red-border inset shows a zoomed version of the data around the energy minimum.
    }
    \label{fig:H2_ucc}
\end{figure}

By utilizing the bipartite spatially-entangled system of two photons described in Sec.~\ref{sec:circuit}, we address a quantum chemistry problem. In this case, the algorithm is called VQE~\cite{Peruzzo_2014,McClean_2016} and the cost function is given by the expectation value of the electronic Hamiltonian of molecules. 
The foundation of the algorithm is the Rayleigh-Ritz variational principle~\cite{helgaker2014molecular}. In App.~\ref{app:H2}, we show more details on the encoding of quantum chemistry problems.

For our demonstration, we choose the simplest molecule, i.e. the Hydrogen molecule ${\rm H}_2$.
Using the minimal STO-3G basis set~\cite{STO3G}, composed of 1s orbitals of the individual H atoms with two spin values, we have a total number of atomic orbitals (AOs) equal to 4. Molecular orbitals (MOs) are given by linear combinations of AOs and represent the global and delocalized nature of the electrons in a molecule.
The AOs and the MOs of ${\rm H}_2$ are sketched in Fig.~\ref{fig:H2_ucc}(a-b) and the 4 MOs are
\begin{equation}
\begin{split}
    & \Big\{ 
    | \sigma , \uparrow \rangle_{[e_1}  | \sigma , \downarrow \rangle_{e_2]} \,\,\,,\,\,\,
    | \sigma^* , \uparrow \rangle_{[e_1}  | \sigma^* , \downarrow \rangle_{e_2]} \,\,\,,\,\,\,
    | \sigma , \uparrow \rangle_{[e_1}  | \sigma^* , \downarrow \rangle_{e_2]} \,\,\,,\,\,\,
    | \sigma , \downarrow \rangle_{[e_1}  | \sigma^* , \uparrow \rangle_{e_2]}
    \Big\}
    \\
    & \mbox{where} \qquad
    \begin{cases}
    | \sigma , \uparrow/\downarrow \rangle_e \equiv 
    \frac{1}{\sqrt{N_+}} \left( | {\rm A} , \uparrow/\downarrow \rangle +| {\rm B} , \uparrow/\downarrow \rangle \right) \,,\,
    % | \sigma , \uparrow/\downarrow \rangle_e \equiv 
    % \frac{1}{\sqrt{N_+}} \left( | {\rm A} , \downarrow \rangle +| {\rm B} , \downarrow \rangle \right)
    \\
    | \sigma^* , \uparrow/\downarrow \rangle_e \equiv 
    \frac{1}{\sqrt{N_+}} \left( | {\rm A} , \uparrow/\downarrow \rangle -| {\rm B} , \uparrow/\downarrow \rangle \right) \,,
    \end{cases}
\end{split}
\end{equation}
the square parenthesis denote anti-symmetrization, $N_\pm$ are normalization factors, $| {\rm A/B} , \uparrow/\downarrow \rangle$ denotes the AO 1s associated with the atom A/B and the spin state $\uparrow/\downarrow$. The state $|\sigma\rangle$ is called bonding molecular orbital (BMO), while the state $|\sigma^*\rangle$ anti-bonding molecular orbital (AMO).
The system of two electrons in ${\rm H}_2$ has 4 possible configurations made of combinations of the bonding and anti-bonding orbitals, reported in the left part of Fig.~\ref{fig:H2_ucc}(b). 
We utilize the Jordan-Wigner mapping~\cite{McArdle_2020} to translate the occupation number states $|f\rangle$ of each MO involved in the ground state of ${\rm H}_2$ into qubit states. The qubit encoding dictionary is given in Fig.~\ref{fig:H2_ucc}(b), where each qubit represents the occupation number of MOs with the following order: $\{ f_{|\sigma, \uparrow \rangle}, f_{|\sigma^*, \uparrow \rangle}, f_{|\sigma, \downarrow\rangle}, f_{|\sigma^*, \downarrow  \rangle} \}$. Moreover, the electronic Hamiltonian is mapped to the equivalent bosonic Hamiltonian given by a linear combination of binary operators made of the identity and Pauli operators, $\{ \mathbf{1}, \hat{X}, \hat{Y}, \hat{Z} \}$.
Then, these operators can be organized into two CGs, which implies two different sets of basis states and their corresponding measurement settings for stages (IV-i) and (IV-s) of our Si-PIC. The observables and parameters~\cite{psi4} composing the ${\rm H}_2$ Hamiltonian are given in App.~\ref{app:H2}.

\begin{figure}[t]
    \centering
    \includegraphics[width=\textwidth]{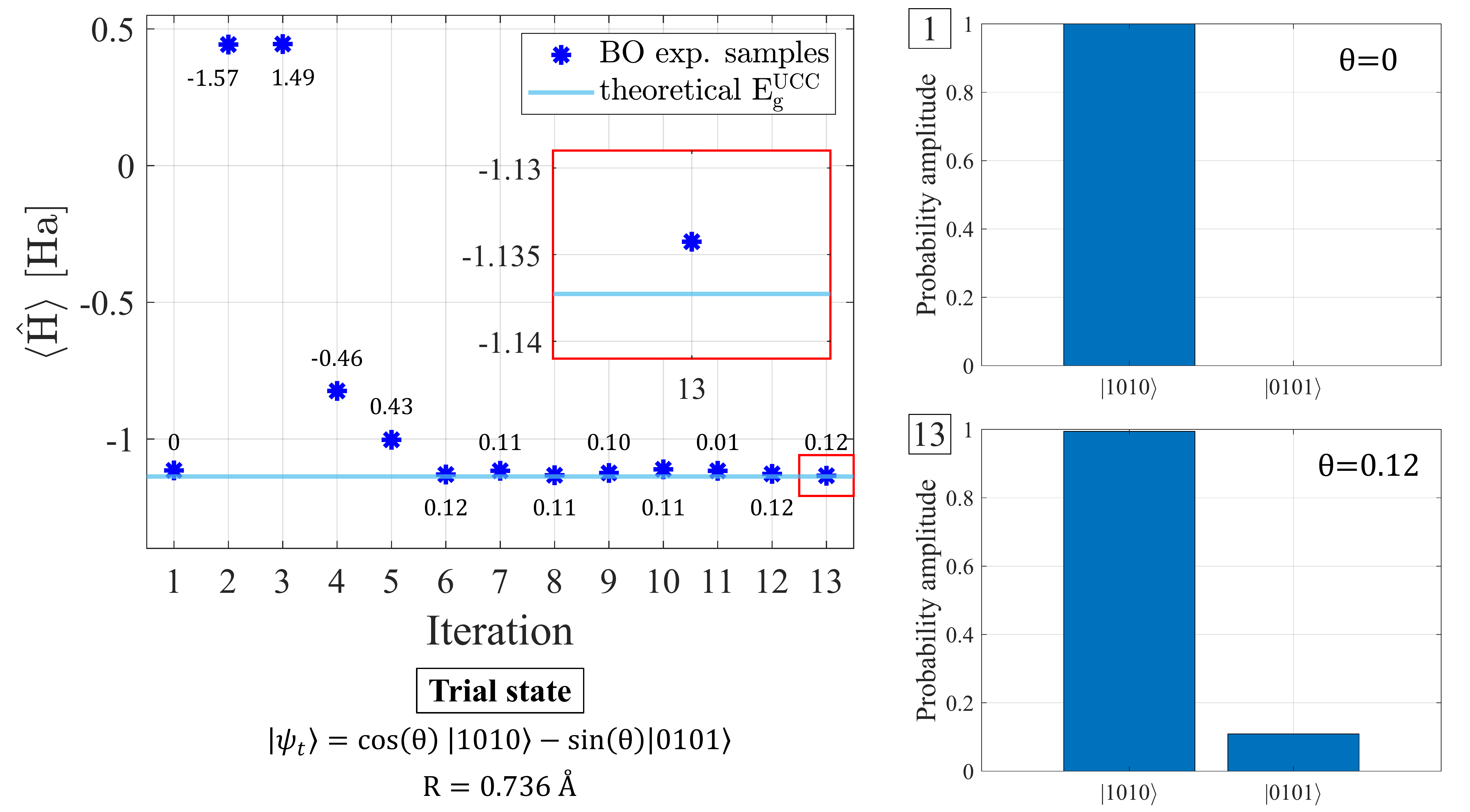}
    \caption{On the left, the evolution of the minimum Bayesian search for the Hydrogen molecule within the minimal STO-3G basis set and the UCC ansatz and using the coefficients associated with an atomic distance R equal to 0.736~\AA~\cite{psi4}, as reported below the graph. 
    Blue stars are the experimental samples of the expectation value of the Hamiltonian, measured in Hartree, evaluated during the Bayesian optimization, and the light blue line indicates the theoretical value of the ground state energy of the Hydrogen molecule within the minimal STO-3G basis set and the UCC ansatz. 
    Close to every data point, the associated value of the set variational parameter is reported.
    The red-border inset shows a zoom of the last datum.
    On the right, the probability amplitudes associated with the first (Hartree-Fock state) and the last steps of the algorithm.
    }
    \label{fig:BO}
\end{figure}

Under the Unitary Coupled Cluster (UCC) ansatz~\cite{hoffmann1988unitary,bartlett1989alternative,romero2018strategies}, the trial state for ${\rm H}_2$ is given by a linear combination of BMO-BMO and AMO-AMO states. In the qubit register, the UCC trial state is parameterized as 
\begin{equation}
    | \psi (\uptheta) \rangle
    =
    \cos\uptheta \,|1010\rangle
    - \sin\uptheta \,|0101\rangle 
    \quad
    \leftrightarrow \quad
    | \psi (\uptheta) \rangle=\cos\uptheta \,|3\rangle_i|3\rangle_s
    - \sin\uptheta \,|2\rangle_i|2\rangle_s \,,
    \label{eq:trialstateH2}
\end{equation}
where the variational parameter $\uptheta$ controls the amount of contribution of AMO-AMO states to the equilibrium configuration and we present the state written in terms of qubits as well as ququarts.
Using the path encoding shown in Sec.~\ref{sec:circuit}, 
% where the first two qubits are assigned to the idler photon and the last two qubits to the signal photon, 
we realize the state in Eq.~\eqref{eq:trialstateH2} by exciting two sources through stage (I) of the Si-PIC. In particular, as it is shown in Eq.~\eqref{eq:stageiii_qubit}, we can prepare the desired state by setting to zero two $\alpha$ parameters, corresponding to routing all the optical input power to only two output paths of stage (I).
% Thus, analogously to the preparation of two-dimensional maximally entangled states, Eq.~\eqref{eq:twodimentstate}, 
We chose sources (2,3), because of the high indistinguishability, as shown in Fig.~\ref{fig:inter_dim}(a), and we utilized the leftmost MZI of stage (I) to control the optical power ratio between sources (2,3). Thus, the relative phase $\uptheta$ between the two arms of the MZI represents the variational parameter $\uptheta$ of the UCC ansatz, Eq.~\eqref{eq:trialstateH2}.
Regarding the measurement settings, we have to calculate the expectation values of two CGs of observables. The first CG is composed of projectors associated with the computational basis, while for the second group the rotated projectors are constructed starting from the eigenvalues of operators $\hat{X}\otimes \hat{Y}$ and $\hat{Y}\otimes \hat{X}$. App.~\ref{app:H2} contains the details of the two measurement settings.

After setting the problem, we explore the configurations of BMO-BMO and AMO-AMO states in Eq.~\eqref{eq:trialstateH2} by varying the variational parameter $\uptheta$. In other words, we are looking at all the Hilbert space of ${\rm H}_2$ within the minimal STO-3G basis set and the UCC ansatz.
The experimental results together with the theoretical ones are reported in Fig.~\ref{fig:H2_ucc}(c-d).
In Fig.~\ref{fig:H2_ucc}(c) we show the estimated and theoretical values of the ${\rm H}_2$ energy as a function of the variational parameter $\uptheta$, calculated using the coefficients, $\{ w_k \}_k$ in Eq.~\eqref{eq:costfunction} ~\cite{psi4}, associated with an atomic distance R equal to 0.736~\AA.
For each distance, i.e. for different weights $\{ w_k \}_k$ in the evaluation of the ground state energy, we fit the data and identify a minimum energy value.
Fig.~\ref{fig:H2_ucc}(d) summarizes the collection of estimated ${\rm H}_2$ ground state energies as a function of the atomic distance together with the theoretical values.
The analysis of all the Hilbert space allows to calculate the ground state energy for all the distances by the classical post-processing on the raw coincidences' data.
From this analysis, we found the experimental equilibrium ground energy expectation value of $-1.1340\pm0.0124$ Ha for $\uptheta=0.11\pm0.01$ and R=0.736~\AA\,, which is compatible with the theoretical result of $-1.1373$ Ha~\cite{graves2023}.

Once the photonic processor has been tested in preparing all the trial states described by Eq.~\eqref{eq:trialstateH2} and performing the two measurement settings associated with the two CGs, the VQA needs a classical optimization routine.
We implemented two methods, one based on gradient descent~\cite{boyd2004convex} and another one based on Bayesian optimization~\cite{bayopt,Garnett_2023}.
In both cases, we start from the trial state with $\uptheta=0$, i.e. the BMO-BMO configuration, and we choose the coefficients $\{ w_k \}_k$ associated with an atomic distance equal to 0.736~\AA. This state is exactly the Hartree-Fock (HF) state of ${\rm H}_2$ and in our encoding is represented by the separable two-ququart state $|3\rangle_i|3\rangle_s$ or equivalently $|1010\rangle$ in the qubit basis.
We observed that the convergence of the gradient-based method is strongly limited by the statistical error and the long measurement time in the gradient evaluation.
This limit is well known in the literature, and it is fundamentally due to the quantum projection noise~\cite{shot_noise}, manifested through the large number of measurement shots required by the hybrid quantum-classical nature of these algorithms~\cite{Tilly_2022,Scriva_2024}. This aspect is a subject of very active research in the field of variational quantum circuits, which recently brought about a new solution known as parameter shift rule~\cite{Schuld_PSR,shiftrule}. Very recently, this technique has been reported also for VQAs implemented on PICs ~\cite{hoch2024,pappalardo2024}.

\begin{table}[!h]
\centering
\scriptsize
\begin{tabular}{|c||cc||cc|}
\hline
 &  \multicolumn{2}{c||}{$\uptheta$}  & \multicolumn{2}{c|}{$\langle \hat{\rm H} \rangle$ } \\ 
\hline
& UCC theory & Experiment & UCC theory & Experiment\\ 
\hline
Hartree-Fock ansatz & $0$ & $0\pm0.01$ & $-1.1168$ Ha & $-1.1149\pm0.0055$ Ha    \\ 
\hline
Minimum & $0.11$ & $0.12\pm0.01$ & $-1.1373$ Ha & $-1.1343\pm0.0062$ Ha \\ 
\hline
\end{tabular}
\caption{
Theoretical and experimental values of energy expectation value and the variational parameter for the initial ansatz, i.e. Hartree-Fock state, and last trial state of the Bayesian search, shown in Fig.~\ref{fig:BO}.
}
\label{tab:H2results}
\end{table}

Given the known limits of classical gradient-based methods that we also observed, we decided to opt for Bayesian optimization~\cite{Iannelli_2022rZ}, a gradient-free method where the acquisition function is given by the lower confidence bound equal to $2.5\%$. The details on the Gaussian Process~\cite{mockus2005bayesian,bayopt,Garnett_2023} are given in App.~\ref{app:H2}.
The robustness of this method relies on its gradient-free nature: a minor number of evaluations for the cost function and, consequently, a lower machine time. Moreover, the outcome of this algorithm gives global information about the behavior of the cost function and generally needs fewer trials for the variational parameter to converge.
Fig.~\ref{fig:BO} summarizes the outcome achieved with the Bayesian optimization. With a good degree of reproducibility and 6-to-13 numbers of iterations, we obtained a very good agreement between the theory and the experiments, with an average value featuring an accuracy of 0.003 Ha with respect to the UCC outcome. These results are presented in Table~\ref{tab:H2results}, where the first and last steps associated with the Bayesian search shown in Fig.~\ref{fig:BO} are summarized.
We point out that the choice of the HF ansatz does not affect the number of iterations in the case of the Bayesian optimization contrary to the gradient-descent case.

\section{Factorizing semiprime numbers}
\label{sec:factorize}
% semiprime: number made of two prime numbers (also squares)
% sphenic: number made of three prime numbers (also cubes)

In this section, we address the factorization problem using the VQF algorithm~\cite{VQF,zhang2023variational}. Given a semiprime number $N$, the goal consists of finding the two prime numbers $(p,q)$ such that $N= p\,q$. 
As before, the algorithm starts with the definition of a cost function, whose construction is based on the following Hamiltonian~\cite{selvarajan2021prime}
\begin{equation}
\begin{split}
        {\rm H}_{\rm fact} &= (N - p q)^2 \,,
        \qquad 
        \mbox{where}\quad
        \begin{cases}
            p = 1+\sum_{i=1}^{m_x} x_i 2^i \,,\\
            q = 1+\sum_{i=1}^{m_y} y_i 2^i \,.
        \end{cases}
\end{split}
\label{eq:hfact_ansatz}
\end{equation}
Here, $m_{x/y}$ is a number between 1 and $\lceil\log_2 N\rceil-1$ and $\{ x_i , y_i \}_i$ are either 0 or 1.
Note that $(p,q)$ are assumed to be odd numbers different from one and they are expressed in binary numbers.
${\rm H}_{\rm fact}$ can be expanded and rewritten as a weighted sum of monomials of the variables $\{ x_i \}_i$ and it is exactly zero for the factorization solution. 
After inserting the ansatz for $(p,q)$ in ${\rm H}_{\rm fact}$, we encode possible solutions in a string of qubits, i.e. $|x_1 \ldots x_{\lfloor\log_2 N\rfloor}\rangle$, and promote the classical variables to quantum binary operators as follows: $x_i \to (\mathbb{1}-\hat{Z}_i)/2$, where $\hat{Z}_i$ is the Z-Pauli applied to the $i$-th qubit. Finally, we obtain the weighted sum of operators containing only the identity and Z-Pauli operators, 
\begin{equation}
\begin{split}
        \hat{\rm H}_{\rm fact} &= \sum_k w_k \hat{O}_k \,,
    \quad \mbox{where}\,\,
    \hat{O}_k \equiv \bigotimes_{ \hat{\sigma}_j \in \{ \mathbf{1}, \hat{Z} \}} \hat{\sigma}_j \,.
\end{split}
\label{eq:hfacto}
\end{equation}
The coefficients $\{ w_k \}_k$ depend on $N$ and on the ansatz in Eq.~\eqref{eq:hfact_ansatz} for $p$ and $q$. 
Note that we need only the measurement setting associated with the computational basis. % , which are the eigenvectors of the Z-Pauli operators.
% The overall time to calculate the cost function of one trial state is given by the time to acquire enough statistics times 16, i.e. the number of projectors.

\begin{figure}[t]
    \centering
    \includegraphics[width=\textwidth]{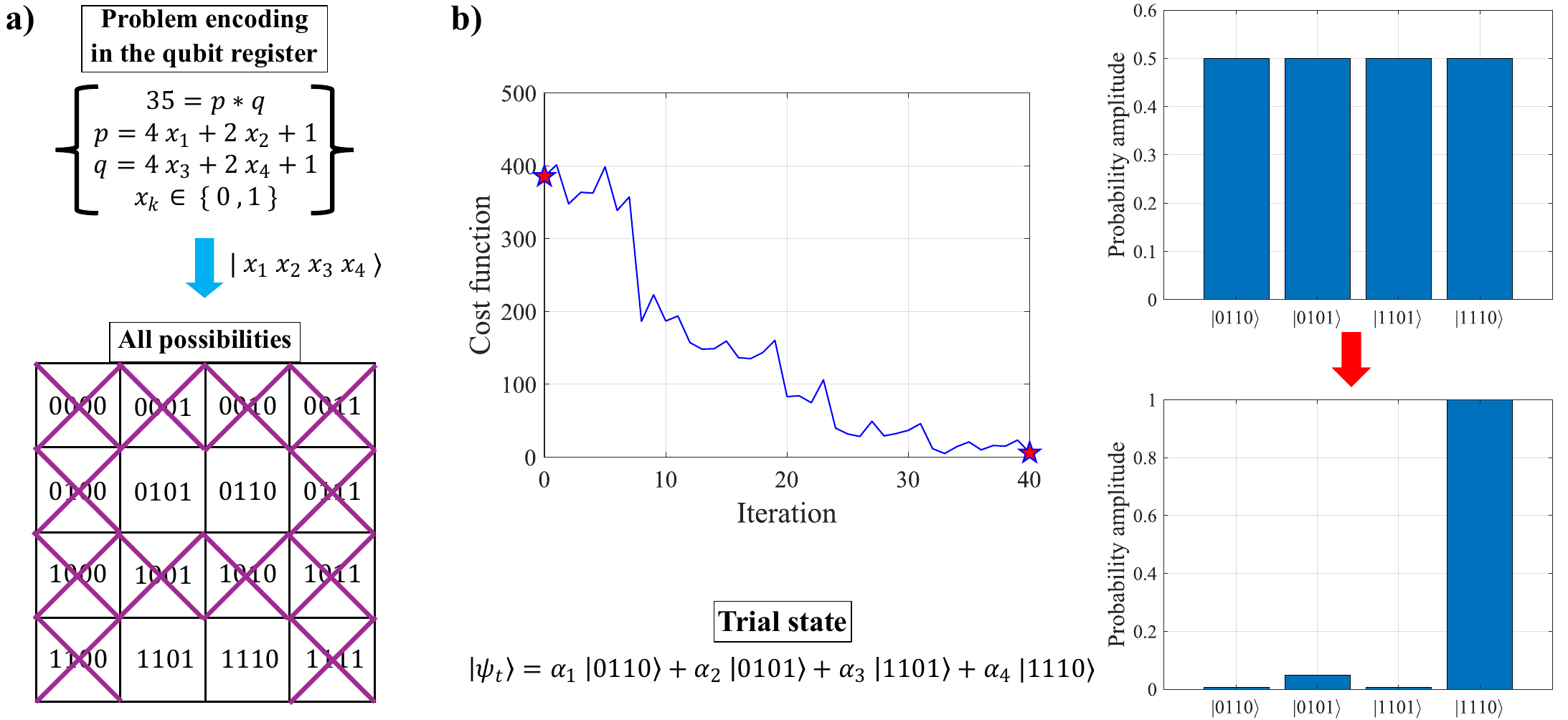}
    \caption{(a) The problem encoding in our qubit register for the factorization of 35 followed by the 16 possible solutions where we cross out the impossible trivial choices.
    (b) On the left, the evolution of the minimum gradient descent search for the 35 factorization using the trial state reported below the graph. 
    On the right, the probability amplitudes associated with the "democratic" initial guess and the configuration with the lowest result for the cost function, which corresponds to $(7,5)$.
    }
    \label{fig:facto35}
\end{figure}

We chose $N=35$ and the ansatz $p = 4x_1+2x_2+1$ and $q = 4x_3+2x_4+1$. Among the 16 possibilities for the variables $\{ x_i \}_i$, we exclude the trivial choices with one factor equal to one or $p=q$ (square numbers) and the options that are symmetric in the exchange of $p\leftrightarrow q$.
After this fast cleaning, we maintain only one exception among the squares and we end up with four possibilities: $\{ 0110\to(3,5),0101\to(3,3),1101\to(7,3),1110\to(7,5)\}$. In Fig.~\ref{fig:facto35}(a), we present the problem encoding for this factorization and all the possible solutions, where we cross out the excluded ones.
Then, the trial state has the following form
\begin{equation}
    | \psi (\boldsymbol{\alpha}) \rangle
    =
    \alpha_1 \,|0110\rangle
    + \alpha_2 \,|0101\rangle
    + \alpha_3 \,|1101\rangle
    + \alpha_4 \,|1110\rangle\,,
    \label{eq:trialstatefacto}
\end{equation}
where $\sum_{i=1}^4 |\alpha_i|^2 =1$ and the parameters $\{\alpha_i\}_{i\in1\ldots4}$ depend on three phases corresponding to the PSs of stage (I).
Modulo a linear transformation, the previous state has the same form as the state in Eq.~\eqref{eq:stageiii_qubit}.
For the minimization of the expectation value of the Hamiltonian in Eq.~\eqref{eq:hfacto}, we have implemented only the gradient descent algorithm.
The list of projectors, parameters $\{w_k\}_k$ and the details on the optimization method are given in App.~\ref{app:35}. 
The initial guess is given by a "democratic" choice of the $\alpha$ parameters: $\alpha_1=\alpha_2=\alpha_3=\alpha_4=1/2$, so we are assigning $25\%$ probability to the four selected options. 
In the central and right parts of Fig.~\ref{fig:facto35}, we report the evolution of the cost function during the variational algorithm and the initial and last values of the probability amplitudes of the four selected choices. 
At the end of the procedure, all the optical power is driven to one source, the fourth one, and the final separable trial state is $|1110\rangle$, which is given by setting $\alpha_4=1$ and $\alpha_1=\alpha_2=\alpha_3=0$ in Eq.~\eqref{eq:trialstatefacto} and corresponds to the correct solution $(7,5)$.
% Contrary to the previous case, the classical nature of this problem is found in the separability of the solution.
The convergence reported in Fig.~\ref{fig:facto35} demonstrates the successful factorization of 35 by utilizing our Si-PIC. The same machinery can be used to factorize semiprime numbers up to 49 by suitable choices of the starting ansatz and weights $\{w_k\}_k$ in Eq.~\eqref{eq:hfacto}. 
Contrary to the previous case, the gradient descent algorithm shows more robustness in the result convergence, and we didn't implement the Bayesian optimization. The main reason can be found in the smaller evaluation time needed to evaluate the cost function for one trial, since there is only one CG.

\section{Discussion on the PIC scaling}
\label{sec:disc}

In Sections \ref{sec:H2} and \ref{sec:factorize}, we have presented the implementation of two different proof-of-principle VQAs relying on the use of path-entangled ququarts. These have been executed in silicon photonics exploiting on-chip photon pair sources. In quantum silicon photonics, the integration of up to 32 on-chip photon pair sources based on spiral waveguides has already been demonstrated~\cite{Bao23}. Therefore, silicon photonics VQA-based processors able to address large molecules such as water or large semiprime numbers are already within reach of current technology. 
However, preparing $d$-dimensional entangled states in bipartite structures, like the one presented in this work, requires $(d-1)$ MZIs in stage (I), $d$ photon pair sources in stage (II), $d$ AMZIs in stage (III), $(d-1)$ MZIs and an array of $d$ PSs in both stages (IV-i) and (IV-s) and two SPDs connected to outputs $| \lceil\frac{d}{2}\rceil\rangle_{i/s}$. The simplicity of having two SPDs is paid with $d^2$ measurement setting to retrieve all the various output correlations. In this case, the time required for the implemented VQAs scales quadratically with the dimension $d$~\cite{Wang_2018,Vigliar_2021}. If the projections and the multiple projective measurements are substituted with a generic SU$(d)$ transformation, the resources in the final stages are $d(d-1)$ in terms of MZIs~\cite{reck_experimental_1994,clements_optimal_2016}, $2d$ in terms of SPDs and the computational time does not depend on $d$.
In both cases, if we use the fact that for $d=2^n$ we can encode $n$ qubits, the scaling of the spatial and/or time resources grows manifestly as an exponential of the qubit register dimension. 
Nevertheless, to further increase the number of available qubits,
more Si-PICs with the same structure could be combined to prepare and distribute entanglement. Thus, each Si-PIC would become a building block to initialize entangled bipartite states, which can be routed and collected to solve bigger classical and quantum problems~\cite{pirker2018modular}. To further improve the scalability, one can consider the fusion-based LOQC~\cite{bartolucci_fusion-based_2023} where two-qubit gates, called fusion gates, have 50$\%$ of success probability in the manipulation stage.
Therefore, such an improved scheme would involve the generation of different photon pairs followed by photonic networks designed to create cluster states~\cite{briegel2001persistent,raussendorf2001one,bartolucci_fusion-based_2023}, which can be used to explore bigger Hilbert spaces for the desired VQA.

\section{Conclusion}
\label{sec:conclu}

In this work, we have reported on the successful demonstration of solving a quantum chemistry problem (H$_2$ molecule) as well as a factorization problem ($N=35$) by means of the same silicon photonics variational quantum circuit using two ququarts, or equivalently four qubits. For the first time to the best of our knowledge, the VQAs implementation is based on two ququarts that are path-entangled thanks to the use of on-chip photon pair sources. For both types of problems, all obtained outcomes are in very good agreement with the predicted results. In particular, in the solution of the H$_2$ molecule ground state an accuracy of 0.003 Ha is reported. Thus, we demonstrate how these algorithms can exploit the currently available photonic processors to accurately solve specific tasks.

The simulation of the Hydrogen molecule is described on Qiskit library for python~\cite{anaya2022simulatingmoleculesusingvqe}, and Refs.~\cite{qing2023usevqecalculateground,Bentellis23} show the VQE calculation of the ground energy for Hydrogen molecule on IBM Quantum (superconducting circuits) and AQT (trapped ions) QH, obtaining results compatible with ours. Bigger molecules, like water~\cite{nam2020ground}, have been simulated with trapped ions and qubit registers of order 10.
The desired chemical accuracy is approximately $1.59\,10^{-3}$ Ha per particle, compatible with our results.
When the dimension and the degree of correlations of the molecule increase, quantum error mitigation~\cite{Lee_22}, entanglement measurements~\cite{Lee_2024} and suitable choices for the classical optimization are needed to reach such an accuracy.

Integrated photonics represents a promising platform for VQAs, since generating and detecting entangled photons, a potentially relevant resource, can be achieved using room temperature technology, contrary to e.g. superconducting and trapped-ion qubits which require ultra-low temperatures and vacuum conditions. Our experiments, showing VQAs based on on-chip generation and manipulation of single photons at room temperature, mark a significant step along the path to realise a fully-integrated photonic quantum processor. Indeed, due to e.g. the well-established silicon photonics technology and the high fidelity of the operations, larger-scale integrated photonic circuits such as the one reported in \cite{Bao23} can already be utilized to tackle larger problems on a room-temperature and compact device, thus witnessing both relevance and impact of our results.

% Integrated photonics represents a fruitful platform for VQAs, since generating and detecting entangled photons, relevant resource for VQAs, can be achieved using room temperature technology, contrary to e.g. superconducting and trapped-ion qubits which require ultra-low temperatures and vacuum conditions. 
% % Quandela VQE result shows that integrated quantum photonics is on par with the other platforms, and here, with a different sort of integrated photonic hardware we show that (next paragraph)
% Due to the well-established technology and the high fidelity of the operations, large-scale integrated photonic circuits can be utilized to tackle larger problems on room-temperature and compact devices.
% Along the path to reach a fully integrated photonic processor, we showed VQAs based on on-chip generation and manipulation of single photons at room temperature. 
% %, and thus the detection is the last ingredient to be integrated.
% To the best of our knowledge, this is the first photonic implementation of VQAs on a Si-PIC with integrated sources of photon pairs. 
% We successfully showed how to solve a quantum chemistry problem and a factorization problem by implementing suitable VQAs. 
% In both cases, all the obtained outcomes are in good agreement with the predicted results. 

\section{Methods}
\label{sec:method}

The pump light comes from a TUNICS-BT NetTest Wavelength Tunable CW Laser Diode Source followed by Thorlabs' EDFA100x core-pumped erbium-doped fiber amplifier.
The Dense Wavelength Division Multiplexing modules (OpNeti and Precision Microptics) have 200 GHz bandwidth and 100 GHz FWHM: among the sixteen channels from 21 to 51, we used channels 35, 27 and 41, whose spectral responses are reported in App.~\ref{app:setup}.

The SOI photonic chip was fabricated using e-beam lithography based on nano-fabrication by SiPhotonic Technologies ApS via a commercial MPW service. The photonic circuit for the VQAs fits within a size of 1.5x5 mm$^2$. The silicon waveguide core is $220$ nm-thick and $500$ nm-wide. We used the foundry's PDKs for grating couplers, MMIs and crossing waveguides. The thermal phase shifters are made of titanium, 100 $\mu$m-long and have a tuning efficiency of around 0.14 rad/mA$^2$. The experiments are performed by using fundamental electric-transverse mode, and, in order to set the correct polarization, the transmission is maximized through manual fiber polarization controllers based on the different insertion losses of the waveguide modes.
In order to improve the visibility of the camera and have an easier coupling, we used a lidless fiber array (Meisu Optics). 
The fiber array is put on Thorlabs' 6-Axis NanoMax Stage, whose (X,Y,Z) are connected to Thorlabs' 150 V USB Closed-Loop 3-channel piezo-controller. 

The detection of the residual pump is done with Thorlabs' PM100USB and Thorlabs' PDA20CS-EC (InGaAs Amplified detector), respectively PM-1 and PM-2 in Fig.~\ref{fig:setup}.
The single photons are detected through two ID-Quantique single photon detector modules, respectively id210 for idler photons and id201 for signal photons. Both single-photon detectors have 15$\%$ efficiency, 20 ns gate width and 80 $\mu$s deadtime. Id201 is set in external gating mode and it is triggered at 500 kHz by id210 which is set in internal gating mode.
The output counts of the detectors are collected by time-tagging electronics (Swabian Instruments) connected to a PC.

The photonic chip is glued to a printed board circuit, designed by the electronic workshop of the University of Trento.
Current modules (National Instruments) are attached to a power supply (E3631A 80W Triple Output Power Supply) and provide the currents for the thermal phase shifters of the PIC through the printed board circuit. 
We used MatLab on our PC to run the self-alignment routine for the piezo-controller, to acquire the data coming from the time-tagging electronics, the power meters and to set the current at the thermal phase shifters.
In particular, in order to maintain the temperature change locally confined, the counts associated with each phase shifter' setting are collected in two/four intervals of 30 s until the total number of counts is approximately 2000/4000 twofold events, which directly determines the statistical error.
The overall time to calculate the energy expectation value associated with one trial state is given by the time to acquire enough statistics times the number of commuting groups associated with the desired cost function, times 16, i.e. the number of projectors.

We used an on-chip power equal to 2 mW per excited source, which means 4 mW on-chip power for the VQE and 8 mW on-chip power for the VQF. This choice is based on the non-linear characterization results.
The overall run time of the analysis summarized in Fig.~\ref{fig:H2_ucc} is around 17 hours, and the overall time of the Bayesian optimization shown in Fig.~\ref{fig:BO} is around 12 hours.
Finally, the overall time of the optimization summarized in Fig.~\ref{fig:facto35} is around 21 hours.

\newpage

% \begin{comment}

\renewcommand{\thefigure}{A\arabic{figure}}
\renewcommand{\theequation}{A\arabic{equation}}
\renewcommand{\thetable}{A\arabic{table}}

\setcounter{figure}{0}
\setcounter{equation}{0}
\setcounter{table}{0}

\appendix
\begin{appendices}

\section{Integrated linear optical devices}
\label{app:preliminaries}

In this section, we show how to describe the behavior of linear integrated optical devices present in the photonic circuits shown in Fig.~\ref{fig:circuit} and described in Sec.~\ref{sec:circuit}.

The fundamental building block is the MZI, whose ingredients are balanced beam splitters (BSs) and phase shifters (PSs).
Balanced ideal BS can be represented with the following matrix
\begin{equation}
    U_{\rm BS} = \frac{1}{\sqrt{2}}
    \begin{pmatrix}
        1 & \pm{\rm i} \\
        \pm{\rm i} & 1
    \end{pmatrix} \,,
    \label{eq:beam_ideal}
\end{equation}
modulo a global phase. The plus sign corresponds to the MMIs case, while the minus sign to the directional coupler case. In our circuit, we have MMIs. 
Then, considering a set of $n$ waveguides, PSs on each path can introduce different additional phases. This configuration involves $n$ inputs and $n$ outputs and it can be described by the following matrix
\begin{equation}
    \left[U_{\rm PS}^{(n)}(\uptheta)\right]_{jk} = 
        {\rm e}^{{\rm i}\uptheta_j} \mathbb{1}_{jk}^{(n)} 
        \,,
    \label{eq:phase_shift}
\end{equation}
where $\uptheta = (\uptheta_1,\uptheta_2,\ldots,\uptheta_n)$ and $\mathbb{1}^{(n)}$ is the identity.
Each waveguide goes straight without any mixing and only the relative phases can be measured, if we look just at such a set of $n$ waveguides. 
Finally, since our MZIs are the result of cascading one MMI, one PS in the upper/lower internal arm and another MMI, the unitary matrix associated with MZI is 
\begin{equation}
    \begin{split}
    U_{\rm MZI}( \uptheta, \pm )  &\equiv U_{\rm BS} \cdot U_{\rm PS}^{(2)}(\pm \uptheta,0) \cdot U_{\rm BS} \\
    &= {\rm i}\, {\rm e}^{\pm {\rm i}\uptheta/2} 
    \begin{pmatrix}
        \pm \sin(\uptheta/2) & \cos(\uptheta/2) \\
        \cos(\uptheta/2) & \mp \sin(\uptheta/2)
    \end{pmatrix} \,,
    \end{split}
\label{eq:MZI_matrixapp}
\end{equation}
where the plus or minus sign depends on the PS's position between the internal arms of the MZI.

In order to describe the MZI network, we define a generic embedded $2\times2$ matrix $U$ in a scheme with $m$ spatial modes. The transformation of the $k$-th and $(k+1)$-th inputs reads 
\begin{equation}
U^{(k,k+1)} \equiv
    \begin{pmatrix}
    1 & 0 &  \ldots & \ldots & \ldots & 0 \\
    0 & \ddots & & & &\vdots \\
    \vdots & & \left(U_{\rm MZI}\right)_{11} & \left(U_{\rm MZI}\right)_{12} & & \vdots \\
    \vdots & & \left(U_{\rm MZI}\right)_{21} & \left(U_{\rm MZI}\right)_{22} & & \vdots \\
    \vdots & & & & \ddots & 0 \\
    0 & \ldots & \ldots & \ldots & 0 & 1 \\
    \end{pmatrix} \,.
    \label{Ukemb}
\end{equation}
This transformation is leaving unaffected all inputs different from $k$ and $(k+1)$ ones, which are evolving accordingly to $U$.
Thus, it is possible to construct any unitary matrix $m\times m$ from $2\times2$ sub-matrices. Concretely, this is equivalent to creating a generic linear $m\times m$ device assembling linear $2\times2$ devices, like MZIs.
Indeed, the Reck and Clements~\cite{reck_experimental_1994,clements_optimal_2016} schemes give exactly two prescriptions to obtain a generic unitary operation by assembling MZIs.
Finally, to describe the evolution of the generic photon state, we can use the Fock basis and the ladder operators. In this way, the action on the creation operator of the $k$-mode of the network $a^\dagger_k$ reads as follows~\cite{Kok_2007}
\begin{equation}
    a^\dagger_k  \to \sum_{j=1}^m U^{-1}_{kj} a^\dagger_j \,,
    \label{eq:mxmgen}
\end{equation}
where $U^{-1}$ is the inverse of $U$, the transformation associated with the MZI network.

\begin{figure}[t]
    \centering
    \includegraphics[width=\textwidth]{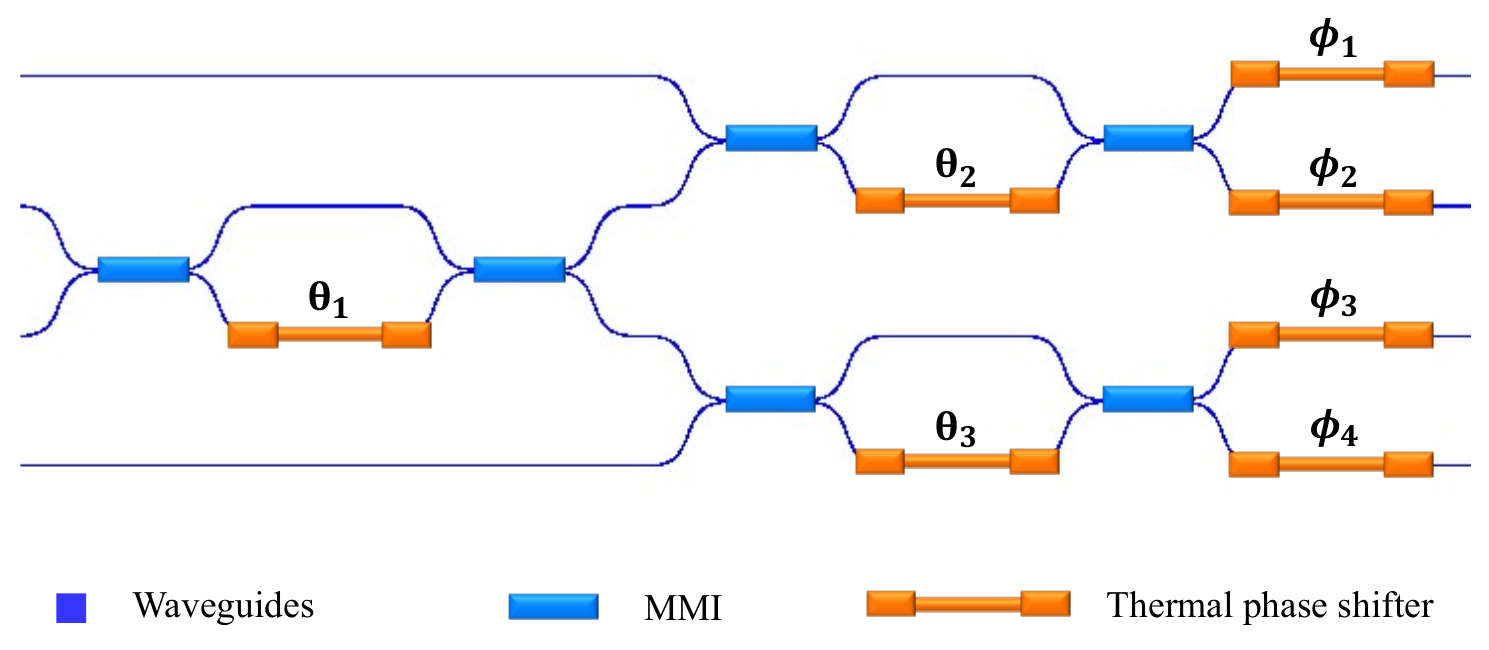}
    \caption{
    Graphical representation of the triangular network of MZIs, contained in the silicon photonic integrated circuit utilized for variational quantum algorithms.
    Blue lines are waveguides, solid blue components are multimode-interference-based integrated beam splitters (MMIs) and orange components are thermal phase shifters.
    Stages (I), (IV-i) and (IV-s) have the following triangular scheme, Fig.~\ref{fig:circuit}, for pump splitting and projection.
    }
    \label{fig:MZI_triangle}
\end{figure}

Looking at Fig.~\ref{fig:circuit}, stages (I), (IV-i) and (IV-s) have the triangular scheme shown in Fig.~\ref{fig:MZI_triangle}, but in the case of pump splitting and projection the inputs and the outputs are inverted.
Therefore, the matrix associated with stage (I) reads as follows 
\begin{equation}
\begin{split}
&U_{\rm TS}^{(p)}( \uptheta_1^{p}, \uptheta_2^{p}, \uptheta_3^{p}, \upphi_1^{p}, \upphi_2^{p}, \upphi_3^{p}, \upphi_4^{p} )  \\
&\equiv
U_{\rm PS}^{(4)}( \upphi_1^{p}, \upphi_2^{p}, \upphi_3^{p}, \upphi_4^{p} ) \cdot U_{\rm MZI}^{(3,4)}( \uptheta_3^{p},- )\cdot U_{\rm MZI}^{(1,2)}( \uptheta_2^{p},- ) \cdot U_{\rm MZI}^{(2,3)}( \uptheta_1^{p},- )  
\\
&=
\left(
\begin{array}{cccc}
 {\rm e}^{{\rm i} \upphi_1^{p}} \sin (\uptheta_2^{p}) & {\rm e}^{{\rm i} \upphi_1^{p}} \sin (\uptheta_1^{p}) \cos (\uptheta_2^{p}) & {\rm e}^{{\rm i} \upphi_1^{p}} \cos (\uptheta_1^{p}) \cos
   (\uptheta_2^{p}) & 0 \\
 {\rm e}^{{\rm i} \upphi_2^{p}} \cos (\uptheta_2^{p}) & -{\rm e}^{{\rm i} \upphi_2^{p}} \sin (\uptheta_1^{p}) \sin (\uptheta_2^{p}) & -{\rm e}^{{\rm i} \upphi_2^{p}} \cos (\uptheta_1^{p}) \sin
   (\uptheta_2^{p}) & 0 \\
 0 & {\rm e}^{{\rm i} \upphi_3^{p}} \cos (\uptheta_1^{p}) \sin (\uptheta_3^{p}) & -{\rm e}^{{\rm i} \upphi_3^{p}} \sin (\uptheta_1^{p}) \sin (\uptheta_3^{p}) & {\rm e}^{{\rm i} \upphi_3^{p}} \cos
   (\uptheta_3^{p}) \\
 0 & {\rm e}^{{\rm i} \upphi_4^{p}} \cos (\uptheta_1^{p}) \cos (\uptheta_3^{p}) & -{\rm e}^{{\rm i} \upphi_4^{p}} \sin (\uptheta_1^{p}) \cos (\uptheta_3^{p}) & -{\rm e}^{{\rm i} \upphi_4^{p}} \sin
   (\uptheta_3^{p}) \\
\end{array}
\right) \,,
\end{split}
\label{eq:UTSp}
\end{equation}
while the one associated with stages (IV-i) and (IV-s) is 
\begin{equation}
\begin{split}
&U_{\rm TS}^{(i/s)}( \uptheta_1^{i/s}, \uptheta_2^{i/s}, \uptheta_3^{i/s}, \upphi_1^{i/s}, \upphi_2^{i/s}, \upphi_3^{i/s}, \upphi_4^{i/s} )  \\
& \equiv
U_{\rm MZI}^{(2,3)}( \uptheta_1^{i/s},+ ) \cdot U_{\rm MZI}^{(1,2)}( \uptheta_2^{i/s},+ ) \cdot U_{\rm MZI}^{(3,4)}( \uptheta_3^{i/s},+ ) \cdot  U_{\rm PS}^{(4)}( \upphi_1^{i/s}, \upphi_2^{i/s}, \upphi_3^{i/s}, \upphi_4^{i/s} )
\\
&=
\left(
\begin{array}{cccc}
 {\rm e}^{{\rm i} \upphi_1^{i/s}} \!\! \sin (\uptheta_2^{i/s}) & {\rm e}^{{\rm i} \upphi_2^{i/s}} \!\! \cos (\uptheta_2^{i/s}) & 0 & 0 \\
 {\rm e}^{{\rm i} \upphi_1^{i/s}} \!\! \sin (\uptheta_1^{i/s}) \cos (\uptheta_2^{i/s}) & -{\rm e}^{{\rm i} \upphi_2^{i/s}} \!\! \sin (\uptheta_1^{i/s}) \sin (\uptheta_2^{i/s}) & {\rm e}^{{\rm i} \upphi_3^{i/s}} \!\! \cos
   (\uptheta_1^{i/s}) \sin (\uptheta_3^{i/s}) & {\rm e}^{{\rm i} \upphi_4^{i/s}} \!\! \cos (\uptheta_1^{i/s}) \cos (\uptheta_3^{i/s}) \\
 {\rm e}^{{\rm i} \upphi_1^{i/s}} \!\! \cos (\uptheta_1^{i/s}) \cos (\uptheta_2^{i/s}) & -{\rm e}^{{\rm i} \upphi_2^{i/s}} \!\! \cos (\uptheta_1^{i/s}) \sin (\uptheta_2^{i/s}) & -{\rm e}^{{\rm i} \upphi_3^{i/s}} \!\! \sin
   (\uptheta_1^{i/s}) \sin (\uptheta_3^{i/s}) & -{\rm e}^{{\rm i} \upphi_4^{i/s}} \!\! \sin (\uptheta_1^{i/s}) \cos (\uptheta_3^{i/s}) \\
 0 & 0 & {\rm e}^{{\rm i} \upphi_3^{i/s}} \!\! \cos (\uptheta_3^{i/s}) & -{\rm e}^{{\rm i} \upphi_4^{i/s}} \!\! \sin (\uptheta_3^{i/s}) \\
\end{array}
\right) \,,
\end{split}
\label{eq:UTSis}
\end{equation}
where the labeling of the phases is reported in Fig.~\ref{fig:MZI_triangle} and the apexes $(p,i,s)$ refer to stages (I), (IV-i) and (IV-s), respectively.

We conclude this section by writing the relations between the variational parameter $\alpha$s, present in Eq.~\eqref{eq:stageiii_ququart}, and the phases $\uptheta$ of stage (I).
Using Eq.~\eqref{eq:UTSp} and the fact that the input IN in Fig.~\ref{fig:circuit} represents the second spatial mode, it is possible to show that the following relations hold:
\begin{equation}
\begin{split}
    & \alpha_1 = {\rm e}^{{\rm i} \upphi_1^{p}} \sin (\uptheta_1^{p}) \cos (\uptheta_2^{p})
    \,,\qquad \alpha_2 = -{\rm e}^{{\rm i} \upphi_2^{p}} \sin (\uptheta_1^{p}) \sin (\uptheta_2^{p})
    \,,\\
    & \alpha_3 = {\rm e}^{{\rm i} \upphi_3^{p}} \cos (\uptheta_1^{p}) \sin (\uptheta_3^{p})
    \,,\qquad \alpha_4 = {\rm e}^{{\rm i} \upphi_4^{p}} \cos (\uptheta_1^{p}) \cos (\uptheta_3^{p})
    \,,
\end{split}
\label{eq:thetatoalpha}
\end{equation}
where it is easy to verify the probability conservation relation, i.e. $\sum_{m=1}^4 |\alpha_m|^2 = 1$.

\section{Description of the four-qubit photonic integrated circuit}
\label{app:circuit}

Fig.~\ref{fig:circuit} reports the scheme of the Si-PIC, which is composed of five parts: (I) pump splitting (green box of Fig.~\ref{fig:circuit}), (II) integrated sources (golden box of Fig.~\ref{fig:circuit}), (III) photon pairs' routing (purple box of Fig.~\ref{fig:circuit}), (IV-i) and (IV-s) linear manipulation of the idler and signal photons (blue and red boxes of Fig.~\ref{fig:circuit}), respectively.
% Using the same notation of the FA's corresponding coupled fibers, 
The bright pump laser is injected in Si-PIC input IN. 
The light entering the Si-PIC can be described by a coherent state, i.e.
\begin{equation}
    | \Psi_{\rm IN} \rangle \propto \exp\left[\int \!\!\!{\rm d}\omega
   f(\omega)  \hat A_{\rm IN}^\dagger(\omega) \right] | {\rm vac} \rangle \,,
   \label{eq:input}
\end{equation}
where $f$ is the continuous wave (CW) pump amplitude function centered at 1549.3 nm,  $A_{\rm IN}^\dagger$ denotes the single-photon creation operator and $| {\rm vac} \rangle$ the vacuum state.

Stage (I) is used to coherently split the pump light with arbitrary amplitudes on its four output modes by tuning the phases of the MZI triangular network, which composes this initial part.
The implemented transformation in stage (I) is the following:
\begin{equation}
    \hat A^\dagger_{\rm IN}
    \to \sum_{m=1}^4 \alpha_m \hat A^\dagger_{m} \,,
    \label{eq:stagei}
\end{equation}
where $\hat A^\dagger_{ m}$ denotes the single-photon creation operator in the $m$-th output spatial mode. The labeling of output modes relative to this initial stage is reported in the golden box of Fig.~\ref{fig:circuit} and the same labeling is used for the sources, which compose the next stage.
The parameters $\{ \alpha_m \}_m$ depend on the phases of the pump splitting stage and they ideally satisfy the probability conservation relation, i.e. $\sum_{m=1}^4 |\alpha_m|^2 = 1$. 
Eq.~\eqref{eq:thetatoalpha} show the explicit relations between the phases of stage (I) and the parameters $\{ \alpha_m \}_m$.

In stage (II), 1.5cm-long spiral-waveguide-based integrated sources are coherently pumped and pairs of correlated photons are generated through SFWM. 
SFWM~\cite{Fukuda_05,Clemmen_09,Helt_10,Azzini2012,Engin2013,SchmittbergerMarlow_20} is a quantum parametric non-linear optical process able to convert photons with the same wavelength into photons with different wavelengths or vice versa. 
Depending on the input configuration, it is possible to create non-degenerate (two-mode) or degenerate (single-mode) squeezed vacuum states~\cite{Lvovsky_14,Bagchi_2020,Quesada_2022}.
In the low-gain regime, a pair of photons is generated and multi-photon contamination can be neglected. 
The generated twin photons are correlated in energy and time~\cite{Mancini_2002,energy_time1,energy_time2}, and such property can be utilized to achieve sources of single photons through the heralding procedure~\cite{Bonneau_2015,signo_hera}. 
Equivalently, it is also possible to manipulate both generated photons and then consider only coincidences. This logic is analogous to the concept of coincidence basis, where post-selection measurements are used as the criteria to manage the raw data coming from the manipulation of identical single photons~\cite{post_sel_2,bartolucci_fusion-based_2023}.
% We utilize TE0-intramodal non-degenerate SFWM on silicon spiral waveguides and, 
As usual, the two generated photons are denoted as idler and signal, where the first is the photon with shorter wavelength and the latter the photon with longer wavelength.
% We choose the spiral waveguides as photon pair sources to obtain a high degree of indistinguishability~\cite{Lee_2023}, which is an important requirement for the VQAs implemented in Sec.~\ref{sec:H2} and Sec.~\ref{sec:factorize}.
Since we are operating in the low-gain regime, the corresponding state reads as follows
\begin{equation}
    \! | \Psi_{\rm II} \rangle \propto \exp\!\left[\sum_{m=1}^4 \!\!\alpha_m\!\!\int \!\!\!{\rm d}\omega 
   f(\omega) \hat{A}^\dagger_m(\omega) \right]\!
   \exp\!\left[ \!\frac{1}{2}\!\sum_{m=1}^4 \!\!\alpha_m \xi_m \!\!\int \!\!\!{\rm d}\omega_1 {\rm d}\omega_2
    J_m(\omega_1,\omega_2) \hat a_m^\dagger(\omega_1) \hat a_m^\dagger(\omega_2) \!\right] \!\!
   | {\rm vac} \rangle ,
   \label{eq:stageii}
\end{equation}
where $\xi_m$ and $J_m$ are respectively the squeezing parameter and the Joint Spectral Amplitude (JSA)~\cite{zielnicki2018joint} of the $m$-th source and $\hat a^\dagger_m$ denotes the single-photon creation operator in the $m$-th source's spatial mode. 
In the previous equation, the first and second factors represent respectively the residual pump light and the generated twin photons.

In stage (III), asymmetric MZIs (AMZIs)~\cite{shamy_22} and crossing waveguides are used to separate the generated non-degenerate photon pairs and route idlers to stage (IV-i) and signals to stage (IV-s). 
Through the separation and the routing, the energy-time correlation of the twin photons~\cite{Mancini_2002,energy_time1,energy_time2} is converted into spatial correlation, and the state takes the following form
\begin{equation}
    | \Psi_{\rm III} \rangle \propto %\left[\mbox{res. pump} \right]
   \exp\!\left[ \frac{1}{2} \!\sum_{m=1}^4 \!\alpha_m \,\xi_m \!\!\int \!\!\!{\rm d}\omega_1 {\rm d}\omega_2\,
   \tilde{J}_m(\omega_1,\omega_2) \, \hat a_{m,i}^\dagger(\omega_1) \hat a_{m,s}^\dagger(\omega_2) \right] \!
   | {\rm vac} \rangle \,,
   \label{eq:stageiii}
\end{equation}
where we neglect the residual pump, $\tilde{J}_m$ is the Joint Spectral Amplitude (JSA) of the $m$-th source modified by the AMZI-$m$ spectral response and $\hat a^\dagger_{m, i/s}$ denotes the single-photon creation operator in the corresponding source's spatial mode of the stage (IV-i/s).
The labeling for the spatial modes of stages (IV-i) and (IV-s) is reported in  Fig.~\ref{fig:circuit} on the right of the blue and red boxes.
Since the spiral-waveguide-based sources have been designed with identical parameters and the AMZIs have identical output responses, these sources and their produced photons can be considered identical or indistinguishable~\cite{Lee_2023}. This implies that all the squeezing parameters and the absolute value of the JSAs at the output of stage (III) are ideally equal for all the sources.
% The experimental verification of the sources' indistinguishability has been performed and it is reported in the next section.
Therefore, neglecting the residual pump and expanding in powers of the squeezing parameter, in the low-pump power approximation where a single pair is generated, we obtain the following state
\begin{equation}
    \!| \Psi_{\rm III} \rangle \!\propto \!| {\rm vac} \rangle + 
   \frac{\xi}{2} \!\int \!\!\!{\rm d}\omega_1 {\rm d}\omega_2\,
    |\tilde{J}(\omega_1,\omega_2)|\!
   \sum_{m=1}^4 \!\!\alpha_m {\rm e}^{{\rm i} \delta_m }
   | \mathbf{1}_{m,i}(\omega_1), \!\mathbf{1}_{m,s}(\omega_2) \rangle
   + \mathcal{O}\!\left( \xi^2\right) \!,
   \label{eq:stageiii_exp}
\end{equation}
where $| \mathbf{1}_{m,i/s} \rangle \equiv \hat a_{m,i/s}^\dagger | {\rm vac} \rangle$ is the idler/signal single-photon state in the $m$-th spatial mode of stage (IV-i/s), the terms at order $\mathcal{O}\!\left( \xi^2\right)$ contain multi-photon states and $\delta_m$ are phases that depend on the pairs of input spatial modes of stages (IV-i) and (IV-s). 
In the previous equation, the phases $\delta_m$ are assumed to be independent of the frequency, since the pump is CW and the photon pair's sources are spiral waveguides.
Such phase terms are caused by the non-linear generation
process (encoded in the JSA phases), by the AMZI response, by different number of crossed crossing waveguides and by any path length difference in stage (III).
The spurious phases $\delta_m$ can be compensated by using the arrays of PSs present at the entries of stages (IV-i) and (IV-s).
Since these phases have contributions from the non-linear generation process, the calibration of these PSs is performed by detecting the coincidences of idler and signal photons.

By working with the coincidence basis of idlers and signals in the low squeezing regime, we can neglect the contribution from the residual pump and multi-photon noise and utilize path encoding for the generated single-photon states.
Note that we are looking at events where only one source emits a pair (low squeezing regime), which is post-selected by performing a coincidence measurement between any idler and signal photons.
In particular, we can map the single-photon spatial mode of idler and signal to the corresponding ququart computational basis state, $\{ | m \rangle \}_{m=1\ldots4}$, as follows
\begin{equation}
    | \mathbf{1}_m \rangle \to | m \rangle
    \quad\implies\quad
    | \Psi_{\rm III} \rangle \to |\psi_{\rm III}^{(4)}\rangle \equiv \sum_{m=1}^4 \alpha_m {\rm e}^{{\rm i} \delta_m } | m \rangle_i | m \rangle_s \,,
    \label{eq:stageiii_ququart_app}
\end{equation}
where the mapped state has the form of four-dimensional entangled states if all $\alpha$s are different from zero. For a general setting of pump splitting, we can achieve any separable two-ququart states or any $d$-dimensional entangled state of dimension $(2,3,4)$, where the first ququart is encoded in the idler photon and the second one in the signal photon. In the previous equation, we are tracing out the frequency degree of freedom, because we assume total indistinguishability of the sources. In the general case, the photon pair state is described by a density matrix,
\begin{equation}
    \rho_{III} = \epsilon\,| \Psi_{\rm III} \rangle \langle \Psi_{\rm III} | + (1-\epsilon)\,\rho_{\rm mixed} \,,
\end{equation}
where $\epsilon$ quantify the indistinguishability of the sources and $\rho_{\rm mixed}$ is the density matrix of a mixed state~\cite{borghi2023reconfigurable}.
The experimental results shown in App.~\ref{app:her_single_inter}-\ref{app:cert_dim} prove that the use of pure state is a good approximation for the system evolution in our case.\\
Then, the labeling for ququart states of stages (IV-i) and (IV-s) can be mapped to two-qubit states by utilizing binary numbers starting from zero as follows 
\begin{equation}
    | 00  \rangle \iff | 1 \rangle 
    \quad, \quad
    | 01  \rangle \iff | 2 \rangle 
    \quad, \quad
    | 10  \rangle \iff | 3 \rangle 
    \quad, \quad
    | 11  \rangle \iff | 4 \rangle \,.
    \label{eq:map_4to2}
\end{equation}
Thus, since two ququarts can be encoded in stages (IV-i) and (IV-s), the circuit can encode states of four qubits, and the input state for the final stages reads as follows
\begin{equation}
    |\psi^{(4)}_{\rm III} \rangle = 
    \alpha_{1} {\rm e}^{{\rm i} \delta_1 } | 00 \rangle_i | 00 \rangle_s + 
    \alpha_{2} {\rm e}^{{\rm i} \delta_2 } | 01 \rangle_i | 01 \rangle_s +
    \alpha_{3} {\rm e}^{{\rm i} \delta_3 } | 10 \rangle_i | 10 \rangle_s +
    \alpha_{4} {\rm e}^{{\rm i} \delta_4 } | 11 \rangle_i | 11 \rangle_s \,.
    \label{eq:stageiii_qubit_app}
\end{equation}
The labeling of the two-qubit states used in stages (IV-i) and (IV-s) is based on the mapping presented in Eq.~\eqref{eq:map_4to2} and is reported in  Fig.~\ref{fig:circuit} on the left of the blue and red boxes.
The idler photons encode the first and second qubits and the signal photons the third and fourth qubits.
As said in the manuscript whenever the subscripts $i/s$ are omitted, we follow this order for a state $|x_1 x_2 x_3 x_4\rangle$, with $\{x_i\}_{i=1\ldots4}=\{ 0,1 \}$, written in the qubit states.

The final stages (IV-i) and (IV-s) are composed of two identical MZI triangular schemes, calibrated at the wavelength of idler and signal photons, respectively. Each network performs the linear manipulation of each photon belonging to a given correlated photon pair. 
% In the low squeezing regime and coincidence basis, we neglect the residual pump and multi-photon states, 
Eq.~\eqref{eq:stageiii_exp} and Eq.~\eqref{eq:stageiii_ququart_app} are modified by the following transformation as follows
\begin{equation}
\begin{split}
    & | \mathbf{1}_{m,i/s} \rangle
    \to 
    U^{(i/s)} \,| \mathbf{1}_{m,i/s} \rangle
    \quad\iff\quad
    | m \rangle_{i/s}
    \to
    U^{(i/s)} \,| m \rangle_{i/s}
   %  &| \Psi_{\rm IV} \rangle \propto \left[\mbox{res. pump} \right]
   % \exp\!\left[ \frac{\xi}{2} \!\int \!\!\!{\rm d}\omega_1 {\rm d}\omega_2\,
   %  |\tilde{J}(\omega_1,\omega_2)|\!
   % \sum_{m=1}^4 \!\alpha_m \! \left( U_i \, \hat a_{m,i}^\dagger(\omega_1) \right) \!\left(  U_s \,\hat a_{m,s}^\dagger(\omega_2) \right) \right] \!
   % | {\rm vac} \rangle \\
   % &\propto \left[\mbox{res. pump} \right]
   % \left\{ |{\rm vac} \rangle + 
   % \frac{\xi}{2} \!\int \!\!\!{\rm d}\omega_1 {\rm d}\omega_2\,
   %  |\tilde{J}(\omega_1,\omega_2)|\!
   % \sum_{m=1}^4 \!\alpha_m  \!
   % \left( U_i \, \hat a_{m,i}^\dagger(\omega_1) \right) \!\left(  U_s \,\hat a_{m,s}^\dagger(\omega_2) \right) \! | {\rm vac} \rangle
   % + \mathcal{O}\!\left( \xi^2\right)  \right\} 
   \,,
   \label{eq:stageiv_tra_app}
\end{split}
\end{equation}
where $U^{(i/s)}$ is the linear transformation of the idler/signal photon implemented in stages (IV-i/s). 
% In the previous equations, we are assuming the indistinguishability of the sources and in the second line we expand in powers of the squeezing parameter, as in Eq.~\eqref{eq:stageiii_exp}.
Thus, the final state written in the ququart computational basis reads as follows
\begin{equation}
\begin{split}
   |\psi_{\rm IV}^{(4)}\rangle &= \sum_{m=1}^4 \alpha_m 
   \, | \xi_m \rangle_i | \xi_m \rangle_s \quad,\quad
   % \sum_{j1,j2=1}^4 \! u_{m,j_1}^{(i)} u_{m,j_2}^{(s)} \, | j_1 \rangle_i | j_2 \rangle_s \,,\\
   \mbox{where}\,\,\,
   | \xi_m \rangle_{i/s} \equiv \sum_{j=1}^4 \! u_{m,j}^{(i/s)} \, | j \rangle_{i/s}
\end{split}
   \label{eq:stageiv_ququart_app}
\end{equation}
and $u$ are the element of the corresponding $U^{(i/s)}$ matrix representation and the spurious phases present in Eq.~\eqref{eq:stageiii_exp} and Eq.~\eqref{eq:stageiii_ququart_app} are absorbed into $U^{(i/s)}$.
% From the previous equation, we can note that the two ququarts are independently manipulated in stages (IV-i) and (IV-s). Therefore, once the separated or entangled state of the two ququart is created through the first three stages, only local operations on each ququart are possible.
The final stages can execute the projections of any ququart state on OUT-1 and OUT-2, which correspond to the ququart state $| 2 \rangle_{i}$ and $| 2 \rangle_{s}$, respectively~\cite{Wang_2018,Vigliar_2021}. This means that a generic input state to stages (IV-i) and (IV-s) is transformed as follows
\begin{equation}
    % | 2 \rangle_{i/s} = U^{(i/s)} 
    \sum_{m=1}^4 \beta_{m,i/s}\,| m \rangle_{i/s} 
    \xrightarrow[]{U^{(i/s)}}
    | 2 \rangle_{i/s}
    \iff
    U^{(i/s)} \equiv   
    \sum_{m=1}^4 \bar{\beta}_{m,i/s}\,| 2 \rangle_{i/s} \,\langle m |_{i/s} 
    \,,
    \label{eq:proj_stageiv}
\end{equation} 
where $\beta_{m,i/s}$ are probability amplitudes of the input state of stage (IV-i/s) and the bar denotes the complex conjugate. 
Note that these projectors are not Von Neumann projectors~\cite{nielsen_chuang_2010}: in order to distinguish them, we will use a different notation.
An example of projectors is given by the ones associated with the ququart computational basis:
\begin{equation}
    \mathbb{P}_m^{(i/s)} =
    | 2 \rangle_{i/s} \langle m |_{i/s}
    \iff
    | 2 \rangle_{i/s} = 
    \mathbb{P}_m^{(i/s)} | m \rangle_{i/s}\,\,\,,
    \quad \mbox{for}\,m \in (1\ldots4) \,.
    \label{eq:proj_comp_basis}
\end{equation}
Generic projections can be seen as a generic ququart transformation $\mathbb{U}$ followed by a combination of the previous projections, i.e. $U^{(i/s)} = \mathbb{P}^{(i/s)}\cdot \mathbb{U}^{(i/s)} $.

The idler and signal photons together with the residual pump are collected from outputs OUT-1 and OUT-2 respectively.
% : these outputs corresponds to spatial modes $(2,i)$ and $(2,s)$. 
Then, the photons of these two channels pass through their corresponding frequency filters, which select their generation wavelength range and arrive at the detectors. 
% \st{Note that we work in the coincidence basis of generated idler and signal photons: this means that each photon of the correlated pairs is manipulated and their coincidences reveal the correct events.}
The effect of collecting the coincidence events of idler and signal photons from OUT-1 and OUT-2 is described by the following Von Neumann projector 
\begin{equation}
\begin{split}
    \hat{\mathbf{P}}_{\rm det}^{\rm tot} &\equiv \hat{\mathbf{ P}}_{\rm det}^{(i)}\otimes \hat{\mathbf{P}}_{\rm det}^{(s)}  \\
    \hat{\mathbf{P}}_{\rm det}^{(i/s)} &\equiv \! \int \!\!\!{\rm d}\omega\,
    F_{i/s}(\omega) \!\!\sum_{n\ne 0} \frac{1}{n!} \left(\hat a_{2,i/s}(\omega)^\dagger\right)^{n}\! | {\rm vac} \rangle
    \langle {\rm vac}| \left(\hat a_{2,i/s}(\omega)\right)^{n} \,,
    \end{split}
    \label{eq:det_op_app}
\end{equation}
where the functions $F_{i/s}$ represent the spectral responses of the filters before the detection and we assume ideal threshold detectors. For real detectors, the detection efficiency must be taken into account~\cite{genovese2006measuring}.
Note that the implemented operators take into account the two actions of the filters: removing the residual pump and modifying the JSAs. 
Indeed, the intramodal process generates the pairs of correlated photons on a continuous band centered at the wavelength of the input photons and the filtering selects the generation windows of interest~\cite{Helt_12}.
In particular, the filtering increases the sources' indistinguishability, since the information about possible differences is erased and the sources become less sensitive to deviations from nominal values~\cite{quantum_eraser,Lee_2023}.
In our path encoding, the previous projection can be mapped to the associated projector on the ququart state $|2 \rangle_i|2 \rangle_s$, or equivalently
\begin{equation}
\begin{split}
    \hat{\mathbf{P}}_{\rm det}^{\rm tot} & \to  |2 \rangle_i|2 \rangle_s
    \langle 2 |_i \langle 2 |_s 
    \quad \iff \quad
    | 01 \rangle_i| 01 \rangle_s
    \langle 01 |_i \langle 01 |_s 
    = | 0101 \rangle \langle 0101 |
    \,.
    \end{split}
    \label{eq:det_op_ququart}
\end{equation}
To sum up, the triangular schemes in the final stages of the circuit, Eq.~\eqref{eq:proj_stageiv}, and the Von Neumann projection executed with the filtering and the detection, Eq.~\eqref{eq:det_op_ququart}, imply the lack of universal ququart manipulation and the presence of one detector for each ququart set of computational states.
However, our choice can perform the generic transformation and the complete sampling on the outputs by using multiple projective measurements.
This is achieved by executing all possible combinations of projectors in the final stages of the circuit given the observable to be measured.
Moreover, in this way we reduce the resources in terms of components: universal MZI schemes, like Reck and Clements, have more PSs and the detection in such configurations requires one detector for each output. 
Generally, the number of linearly independent projectors is 16, four for each ququart. 
For example, if we want to sample over the computational basis of the two ququart, we need to perform all the following 16 combinations of projectors
\begin{equation}
    \{ \mathbb{P}^{(i)}_1, \mathbb{P}^{(i)}_2, \mathbb{P}^{(i)}_3, \mathbb{P}^{(i)}_4 \} \otimes \{ \mathbb{P}^{(s)}_1, \mathbb{P}^{(s)}_2, \mathbb{P}^{(s)}_3, \mathbb{P}^{(s)}_4 \} \,,
    \label{eq:vec_pro_comp_basis}
\end{equation}
where $\{\mathbb{P}^{i/s}_m\}_{m=1\ldots4}$ are defined in Eq.~\eqref{eq:proj_comp_basis}. In the general case, we need a set of four projectors relative to four independent input states for each ququart.
The choice of 16 projectors represents the measurement setting, and thus, it is defined by the different commuting groups of observables to be measured for the desired cost function~\cite{josa1994fidelity,james2001measurement,darino2003quantum,altepeter20044qubit}.

\section{Experimental setup}
\label{app:setup}

\begin{figure}[t]
    \centering
    \includegraphics[width=\textwidth]{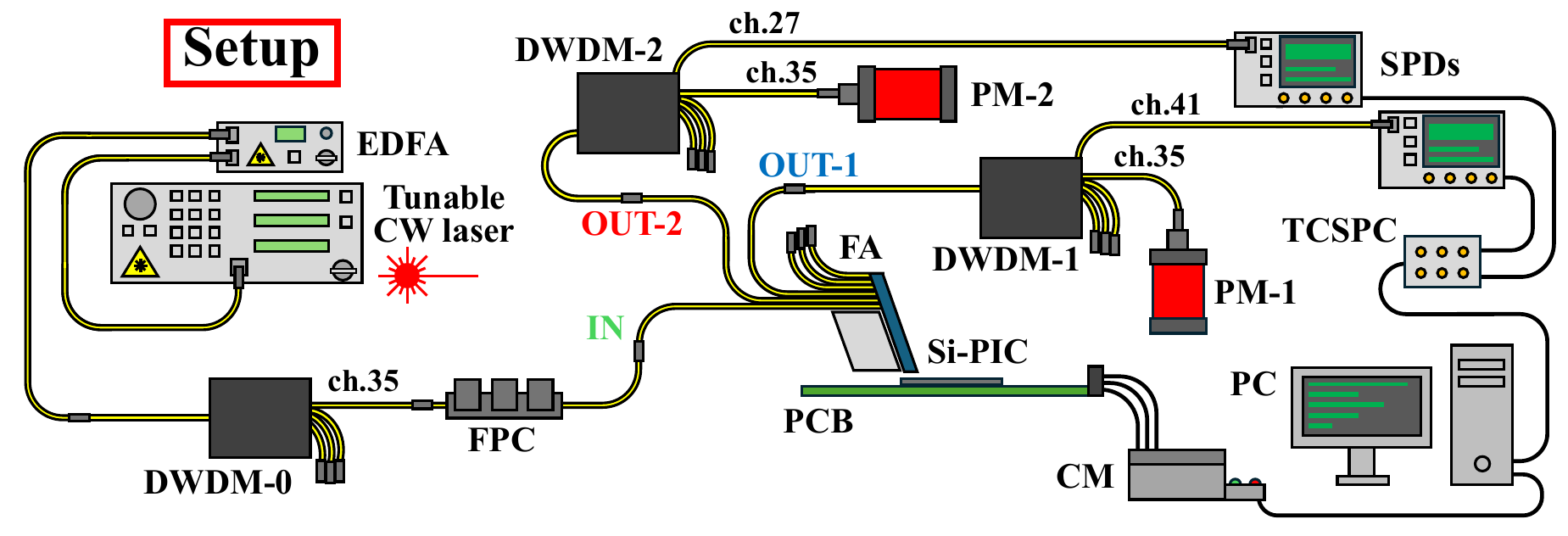}
    \caption{Graphical representation of the experimental setup used for the characterization and the validation of the four-qubit variational quantum photonic circuit.
    A tunable CW laser together with an EDFA is used as the light source. The chosen working wavelength is 1549.3 nm, which is the central wavelength of channel 35 in the subsequent DWDM module denoted as DWDM-0. 
    The polarization is set by utilizing a fiber polarization controller (FPC) before the fiber array (FA). 
    The input IN and the two outputs OUT-1 and OUT-2 channels of FA are coupled to the silicon photonic integrated circuit (Si-PIC) through grating couplers.
    The output fibers OUT-1 and OUT-2 are attached to two different DWDM modules, DWDM-1 and DWDM-2, respectively. In both cases, channels 35 are used to monitor the residual pump through power meters PM-1 and PM-2.
    Channel 41 (centered at 1544 nm) of DWDM-1 and channel 27 (centered at 1555.7 nm) of DWDM-2 are connected to two different Single-Photon Detectors (SPDs) to measure the pairs of correlated single photons.
    The SPDs are connected to a Time-Correlated Single-Photon Counting (TCSPC) module, which processes the arrival times of the photons and transfers them to a standard PC. 
    Finally, the current module (CM) is managed by the PC and provides the desired setting of currents to the printed-board circuit (PCB), where the Si-PIC is glued and wire-bonded.}
    \label{fig:setup}
\end{figure}

\begin{figure}[t]
    \centering
    \includegraphics[width=\textwidth]{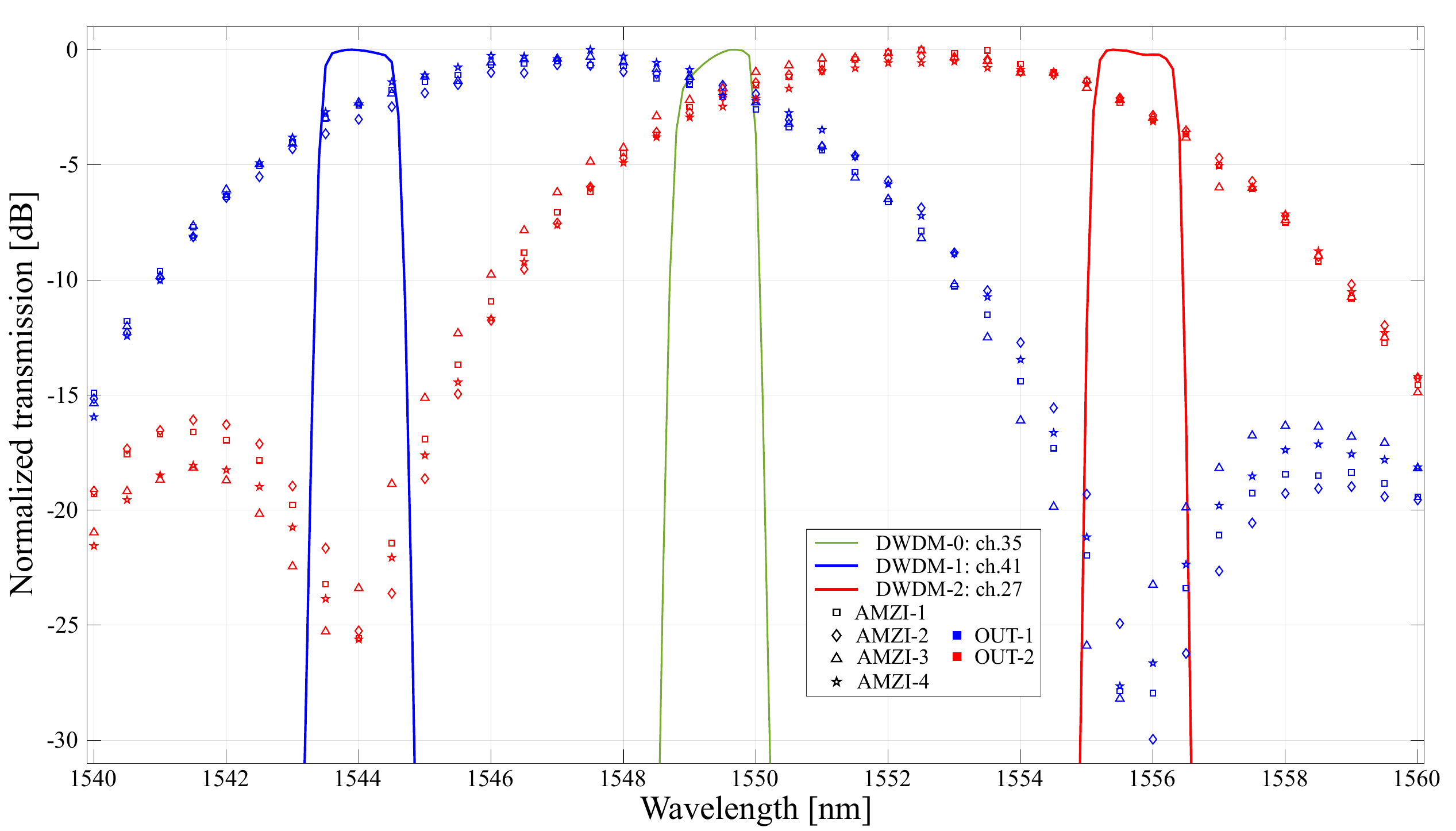}
    \caption{ 
    The measured response function of DWDM modules' channels used in the setup, shown in Fig.~\ref{fig:setup}, and the two outputs of each asymmetric MZI (AMZI), represented in Fig.~\ref{fig:circuit}, in the spectral region between 1540-1560 nm.
    The continuous green, blue and red lines correspond to channel 35 of DWDM-0, channel 41 of DWDM-1 and channel 27 of DWDM-2, while the blue and red symbols (square, diamond, triangle, pentagram) to the different AMZIs at the outputs OUT-1 and OUT-2.
    The DWDM modules are characterized by 100 GHz FWHM.
    The transmission of the DWDM modules and the AMZIs responses is normalized with respect to their maximum value: the insertion loss of the DWDMs is around 1.5 dB, while for the AMZIs is less than 1 dB.
    The AMZIs are tuned in such a way as to align the minimum/maximum transmission with the DWDMs on the outputs. This configuration allows the spatial separation of photons corresponding to the frequency domains of channels 27 and 41 of DWDMs.
    }
    \label{fig:dwdm_amzi}
\end{figure}

The setup is presented in Fig.~\ref{fig:setup}. 
A tunable CW laser is connected to an Erbium-doped Optical Fiber Amplifier (EDFA), whose output is inserted in a Dense Wavelength Division Multiplexing (DWDM) module, denoted as DWDM-0, to filter out the laser background emission. The laser wavelength is set to 1549.3 nm. 
The spectral response of channel 35 of DWDM-0 is shown with the continuous green line in Fig.~\ref{fig:dwdm_amzi}.
A fiber polarization controller (FPC) is used to set the polarization to TE at the input of the chip. Light is coupled in and out from the device using grating couplers and a fiber array (FA).
Three fibers of the FA are utilized: in Fig.~\ref{fig:setup}, IN stands for the input used to inject the pump light on the Si-PIC and OUT-1 and OUT-2 stand for the two outputs used to collect the out-coming light.
OUT-1 is connected to DWDM-1, whose channel 35 (centered at 1549.3 nm and 100 GHz full width at half maximum (FWHM)) is coupled to power meter PM-1 and channel 41 (centered at 1544 nm and 100 GHz FWHM) to an InGaAs single-photon detector (SPD). OUT-2 is connected to DWDM-2, whose channel 35 is coupled to power meter PM-2 and channel 27 (centered at 1555.7 nm and 100 GHz FWHM) to another InGaAs SPD. 
DWDM-1 and DWDM-2 are used to filter out the residual pump light and to select the two wavelength ranges for the generated twin photons.
The spectral responses of channel 41 of DWDM-1 and channel 27 of DWDM-2 are shown with the continuous red and blue lines in Fig.~\ref{fig:dwdm_amzi}.
The counts from the SPDs are collected and processed by a Time-Correlated Single-Photon Counting (TCSPC) module, connected to a standard PC.
The PC is utilized to manage the coincidences' raw data and to drive the current module (CM), which provides the desired setting of currents to the thermal phase shifters present in the Si-PIC through a printed-board circuit (PCB).

\section{Linear and non-linear characterizations}
\label{app:lin_nonlin_char}

In this section, we report the results of the linear characterization of the optical devices and the non-linear characterization of the integrated photon pair's sources.

\begin{figure}[t]
    \centering
    \includegraphics[width=\textwidth]{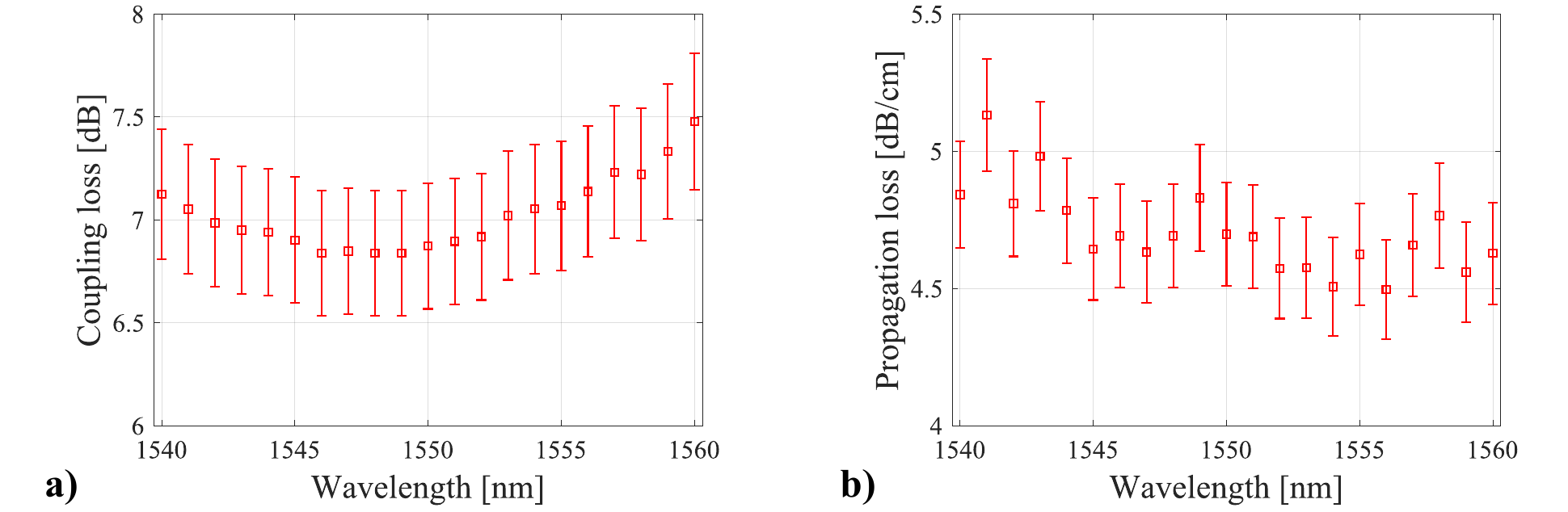}
    \caption{
    Insertion loss in the spectral region $1540-1560$ nm for fundamental TE mode. The single-mode silicon waveguide cross section is $220\times500$ nm$^2$. (a) Coupling loss per grating coupler (PDK of SiPhotonic Technologies ApS), and (b) propagation loss. 
    }
    \label{fig:IL}
\end{figure}

By injecting light in a set of spiral waveguides with different lengths, we quantified the insertion loss of the chip. In particular, Fig.~\ref{fig:IL}(a-b) show the coupling and the propagation losses in the spectral region $1540-1560$ nm for TE polarization, respectively. Both values are not at the state of the art for the SOI platform~\cite{shekhar2024roadmapping}. These losses are responsible for a lower photon pair generation rate, and thus a larger computational time is needed for VQAs.

\begin{figure}[t]
    \centering
    \includegraphics[width=\textwidth]{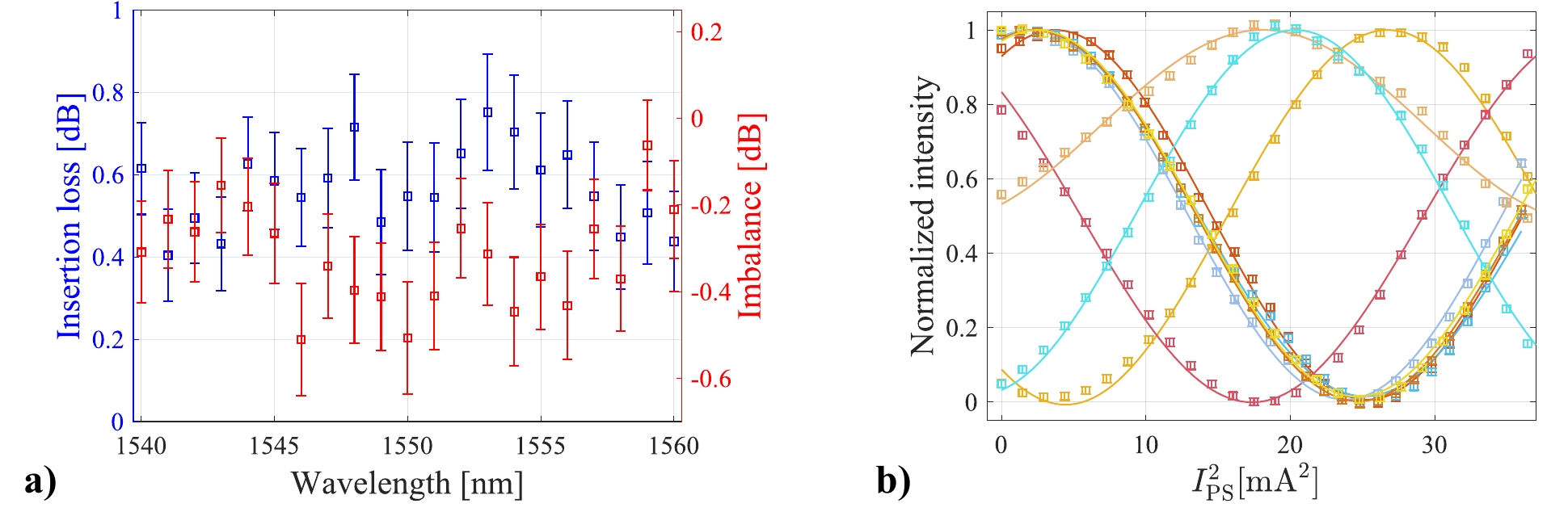}
    \caption{
    (a) Insertion loss (blue lines) and output imbalance (red lines) of MMIs (PDK of SiPhotonic Technologies ApS) in the spectral region $1540-1560$ nm for TE polarization. 
    % These are estimated by measuring the transmitted intensities from the output ports ($out_1$ and $out_2$). Then insertion losses are calculated by $10\times log(out_1+out_2)$ while the unbalance by $10\times log(out_2/out_1)$. $out$s are normalized with respect to a reference value. Lines represent experimental results, dashed lines represent simulations.
    (b) Normalized output optical intensity as a function of the squared current $I_{\rm PS}^2$ applied to the phase shifters belonging to all the MZIs used in stages (I), (IV-i) and (IV-s). In particular, the used wavelength for stage (I) is 1549.3 nm (pump wavelength), for stage (IV-i) 1544 nm (idler wavelength) and for stage (IV-s) is 1555.7 nm (signal wavelength).
    These curves have been used to calibrate all the phase shifters and obtain the relation between the applied current and the induced phase shift, reported as a continuous line. 
    The colors are associated with the three $\uptheta$ phases (red for $\uptheta_1$, blue for $\uptheta_2$ and yellow for $\uptheta_3$), while the shade with the three stages (normal for stage (I), low for stage (IV-i) and high for stage (IV-s)).
    }
    \label{fig:mmi_mzi}
\end{figure}

The imbalance and insertion loss of MMIs were quantified by measuring the output of an isolated component and they are reported in Fig.~\ref{fig:mmi_mzi}(a). The value of imbalance is related to the visibility of the interference of classical light in the MZI, and thus such a low value, ca 0.5 dB around 1550 nm, is important in order to achieve good fidelities in the manipulation, like pump splitting and projection. The insertion loss of MMIs is acceptable since our circuit is shallow and, like any loss, it affects the time to complete the sampling.

Fig.~\ref{fig:mmi_mzi}(b) shows the normalized intensity output (${\rm I}_{o}$) as a function of the PS squared current ($I_{\rm PS}^2$) of the nine MZIs present in stages (I), (IV-i) and (IV-s) by using the laser light set at the pump, idler and signal wavelength, respectively. Each curve is fitted and used to calibrate the PSs of the MZIs. This operation allows to find the parameter of the relation between the set phase $\uptheta_{\rm PS}$ and the applied current, assuming the following dependence
\begin{equation}
    {\rm I}_{o} = \frac{1}{2} \left[ 1 \pm \cos \left(2\uptheta_{\rm PS}\right)\right]\,\,,
    \qquad \mbox{and} \quad
    \uptheta_{\rm PS} = w_{\rm PS}\,I_{\rm PS}^2 + \uptheta_{\rm PS}^{(0)} \,,
    \label{eq:fit_phases}
\end{equation}
% which follows the Joule effect on the thermal PS and 
where $(w_{\rm PS},\uptheta_{\rm PS}^{(0)})$ are the parameters associated with a specific PS. Note that the plus /minus sign in the previous equation depends on the testing input-output configuration of the MZI, while $\uptheta_{\rm PS}^{(0)}$ is the spurious phase difference between the internal MZI arms.
Once we have linearly characterized all the PSs in the MZIs of stage (I) at the wavelength $1549.3$ nm, we are able to route all the optical power to only one source and linearly characterize the AMZIs present in stage (III) at the pump wavelength. Then, the operative setting of the AMZIs is given by splitting the pump equally at the AMZIs' outputs and routing idler and signal photons to stages (IV-i) and (IV-s), respectively. 
The spectral response of each AMZI with active control is reported in Fig.~\ref{fig:dwdm_amzi}.  
Note the alignment of the AMZIs' minimum transmission wavelength individually measured at OUT-1 with channel 27 of DWDM-2 and the ones at OUT-2 with channel 41 of DWDM-1. This implies that photons around 1544 nm are suppressed on OUT-2 and photons around 1555.7 nm are suppressed on OUT-1, or equivalently, all idler and signal photons are spatially separated and sent to stages (IV-i) and (IV-s), respectively.
Finally, the linear characterization of the final stage at the wavelengths of idler and signal photons allows to route the generated photons to OUT-1 and OUT-2 of our circuit, Fig.~\ref{fig:circuit}.
Therefore, to summarize, PSs in stage (I) are calibrated to work at 1549.3 nm (the bright pump wavelength), those in stage (IV-i) at 1544 nm (center of the detection window for the idler photons) and those in stage (IV-s) at 1555.7 nm (center of the detection window for the signal photons).

Table~\ref{tab:singlesource_setting} presents the phase setting of the MZIs belonging to stages (I), (IV-i) and (IV-s) in order to excite only one source and to project all the generated photons to OUT-1 and OUT-2.
Inserting these values in Eq.~\eqref{eq:UTSp} and Eq.~\eqref{eq:UTSis}, we obtain the matrices associated with the MZI networks in stages (I), (IV-i) and (IV-s). Note that within this pump splitting' setting the arrays of PSs at the entrance of the final stages do not matter, since they introduce a global phase.

\begin{table}[!h]
\centering
\scriptsize
\begin{tabular}{|c|ccc|ccc|ccc|}
\hline
Source & $\uptheta_1^p$ & $\uptheta_2^p$ & $\uptheta_3^p$ & $\uptheta_1^i$ & $\uptheta_2^i$ & $\uptheta_3^i$ & $\uptheta_1^s$ & $\uptheta_2^s$ & $\uptheta_3^s$\\ 
\hline
1 & $\frac{\pi}{2}$ & $0$ & $\frac{\pi}{2}$ & $\frac{\pi}{2}$ & $0$ & $\frac{\pi}{2}$ & $\frac{\pi}{2}$ & $0$ & $\frac{\pi}{2}$ \\ 
2 & $\frac{\pi}{2}$ & $\frac{\pi}{2}$ & $\frac{\pi}{2}$ & $\frac{\pi}{2}$ & $\frac{\pi}{2}$ & $\frac{\pi}{2}$ & $\frac{\pi}{2}$ & $\frac{\pi}{2}$ & $\frac{\pi}{2}$ \\ 
3 & $0$ & $\frac{\pi}{2}$ & $\frac{\pi}{2}$ & $0$ & $\frac{\pi}{2}$ & $\frac{\pi}{2}$ & $0$ & $\frac{\pi}{2}$ & $\frac{\pi}{2}$ \\ 
4 & $0$ & $\frac{\pi}{2}$ & $0$ & $0$ & $\frac{\pi}{2}$ & $0$ & $0$ & $\frac{\pi}{2}$ & $0$ \\ 
\hline
\end{tabular}
\caption{Phase shifters' setting of the MZIs in stages (I), (IV-i) and (IV-s) in order to send all the optical pump power to one source and then to route all the photons generated by that source to OUT-1 and OUT-2 of the circuit in Fig.~\ref{fig:circuit}.
The apexes $(p,i,s)$ refer to stages (I), (IV-i) and (IV-s), respectively.
The utilized $\{\upphi^{i/s}_k\}_k$ (see Fig.~\ref{fig:MZI_triangle}) are set to zero, even if they enters as global phases.
}
\label{tab:singlesource_setting}
\end{table}

\begin{figure}[t]
    \centering
    \includegraphics[width=\textwidth]{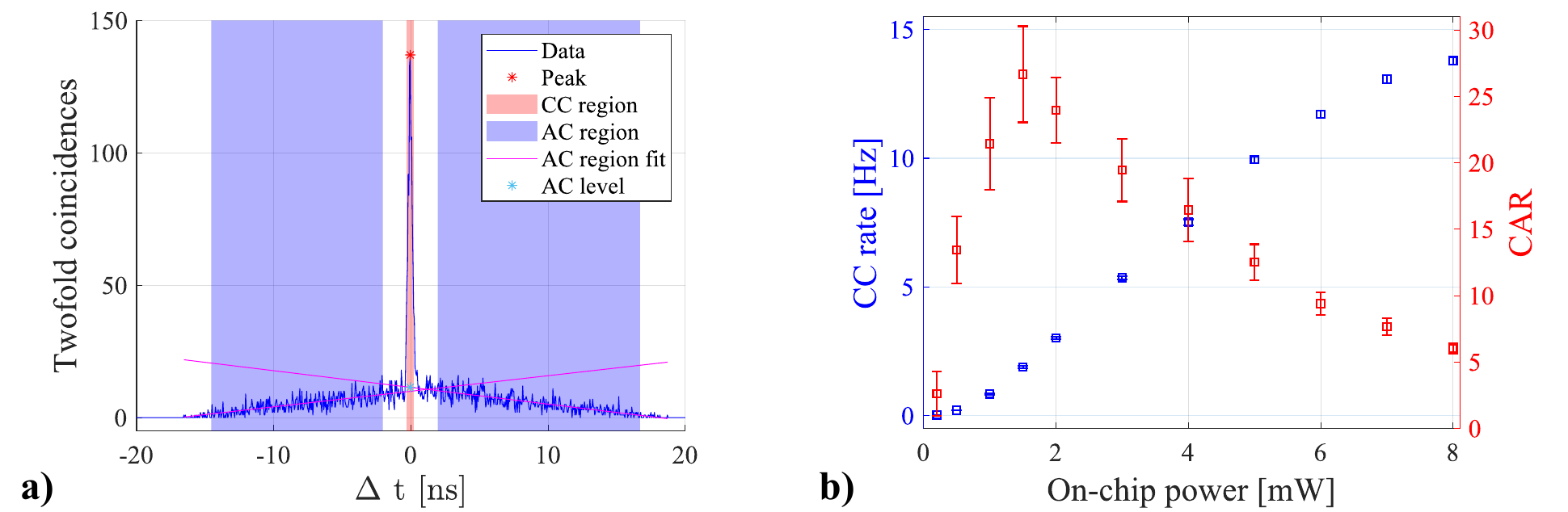}
    \caption{
    (a) Histogram displaying twofold coincidences as a function of the delay $\Delta t$ between idler (start) and signal (stop) detections with 8 mW on-chip input power in 60 s, driving all the power to the third spiral waveguide of stage (II), Fig.~\ref{fig:circuit}. The histogram features bins of coincidences of 50 ps, depicted in dark blue. The region containing accidental coincidences (AC region) is shaded in blue, while the utilized coincidence band (CC region) of 550 ps is shaded in red.
    For $\Delta t=0$ there are net coincidences (CC) and accidental coincidences (AC), and for $\Delta t\ne 0$ only AC convoluted with the gate width (20 ns). In order to estimate AC, we performed a linear fit (purple line) in the AC region to obtain the AC level (light blue dot) on the CC region.
    (b) Figures of merit for the non-linearly generated twin photons: the pump wavelength is 1549.3 nm, the idler wavelength is 1544 nm and the signal wavelength is 1555.7 nm. CAR (red dots) and CC rate (blue dots) are plotted as a function of on-chip pump power.
    }
    \label{fig:CC_car_rate}
\end{figure}

Given the previous PSs' settings, we performed a non-linear characterization of the individual sources. 
% We used the stage (I) to pump only one source of the circuit and the stage (III) to separate and drive the twin photons to the associated final stages. Finally, the generated idler and signal photons were routed to OUT-1 and OUT-2 through stages (IV-i) and (IV-s), respectively.
Fig.~\ref{fig:CC_car_rate}(a) shows the twofold coincidences' histogram, which is used to estimate the coincidence counts (CC) and the coincident to accidental ratio (CAR).
Fig.~\ref{fig:CC_car_rate}(b) summarizes the result of the non-linear characterization performed on the third source. The two figures of merit are the CC rate at the detectors and the CAR. Both values are limited by the insertion loss of the Si-PIC and the performances of the detectors, which have low efficiency and high deadtime value in the gating mode (see Methods for the details).
The other sources have compatible figures of merit.
Given these results, we decided to work close to their CAR maximum, which is achieved for ca 2 mW on-chip power.
This amount of optical power is chosen to be the one per excited source, in such a way as to have the same mean excitation condition per source. Thus, we coupled on-chip powers of $(4,6,8)$ mW to excite $(2,3,4)$ sources, respectively.

The $m$-th row of Table~\ref{tab:singlesource_setting} for $\uptheta^i$s and $\uptheta^s$s gives the phase shifters' setting for the projectors $\mathbb{P}_{m}^{(i)}$ and $\mathbb{P}_{m}^{(s)}$, respectively, defined in Eq.~\eqref{eq:proj_comp_basis}.
Through the detection of correlated photons, we can quantify the quality of the computational ququart basis projectors $\{ \mathbb{P}_{m_1}^{(i)} \otimes \mathbb{P}_{m_2}^{(s)} \}_{m_1,m_2=1\ldots4}$. 
To achieve this, we pumped only the $m$-th source and we collected all the coincidences in the measurement setting associated with the computational ququart basis. Then, we organized the result in a 4x4 matrix $M$ with (IV-i) setting on the rows and (IV-s) one on the columns. Such a matrix has ideally all the components equal to zero except for the diagonal $m$-th term.
We estimated the four fidelities as follows~\cite{clements_optimal_2016}
\begin{equation}
    {\rm F} = \left| \frac{{\rm Tr} \left[ M_{\rm ideal}^\dagger M_{\rm exp} \right]}{\sqrt{ {\rm Tr} \left[ M_{\rm exp}^\dagger M_{\rm exp}\right]}} \right|^2 \,,
    \label{eq:fidelity}
\end{equation}
where $M_{\rm ideal/exp}$ is the ideal/experimental realization of the $M$ matrix.
We obtained (99.8 $\pm$ 0.5, 99.2 $\pm$ 0.5, 99.9 $\pm$ 0.5, 99.5 $\pm$ 0.5) for the four pumped sources.

\section{Heralded single-photon interference}
\label{app:her_single_inter}

This section contains the details of the experiments performed with twin photons on heralded single-photon interference. The results allow to calibrate the PSs associated with the phases $\upphi$s in the final stages (IV-i) and (IV-s).

The arrays of PSs associated with the phases $\upphi$s (see Fig.~\ref{fig:MZI_triangle}) at the inputs of the final stages (IV-i/s) give a measurable effect only when we prepare a spatially-entangled state after stage (III). Moreover, as explained in App.~\ref{app:circuit}, we have spurious phases that depend on the integrated source and the different phase contributions from stage (III). 
Because of the spatial correlation possessed by the twin photons, only three PSs among the eight ones at the input of the final stages are needed and we chose the third spatial mode as the reference mode. Then, the calibration procedure to achieve the relation between the applied current and the phase for the PS has been done for the second spatial mode of (IV-i) and the first and fourth modes of (IV-s), denoted as $\{ \phi^{(i)}_2 , \phi^{(s)}_1 , \phi^{(s)}_4 \}$ respectively. Given the third mode as the reference one, such PS choice maximizes the distance among the working PSs at the final stages' inputs in order to reduce their thermal crosstalk.
The calibration procedure consists of the preparation of two-dimensional maximally entangled states $|\Phi^{(2)}_m \rangle$ by equally pumping pairs of sources $(m,3)$, for $m\in(1,2,4)$. Then, the final stages execute a phase shift on the $m$-th path, through the PS associated with $\phi^{(i/s)}_m$, and the projection of the balanced state between the spatial modes $(m,3)$ on the second spatial mode. 
% \st{a transformation which is the inverse of the one performed in stage (I).}
Table~\ref{tab:interf_setting} shows the phase setting of the MZIs belonging to stages (I), (IV-i) and (IV-s) in order to equally excite only two sources and perform the final transformation. 
Inserting the reported phase values in Eq.~\eqref{eq:UTSp} and Eq.~\eqref{eq:UTSis}, we get the matrices associated with the MZI networks in stages (I), (IV-i) and (IV-s).
The manipulation of the final stages is exactly the inverse of the transformation used in the preparation stage, modulo global phases.
\begin{table}[!h]
\centering
\scriptsize
\begin{tabular}{|c|ccc|ccc|ccc||c|}
\hline
Sources & $\uptheta_1^p$ & $\uptheta_2^p$ & $\uptheta_3^p$ & $\uptheta_1^i$ & $\uptheta_2^i$ & $\uptheta_3^i$ & $\uptheta_1^s$ & $\uptheta_2^s$ & $\uptheta_3^s$ & Visibility\\ 
\hline
2-3 & $-\frac{\pi}{4}$ & $\frac{\pi}{2}$ & $\frac{\pi}{2}$ & $-\frac{\pi}{4}$ & $\frac{\pi}{2}$ & $\frac{\pi}{2}$ & $-\frac{\pi}{4}$ & $\frac{\pi}{2}$ & $\frac{\pi}{2}$ & $99.3\pm0.5$\\ 
1-3 & $\frac{\pi}{4}$ & $0$ & $\frac{\pi}{2}$ & $\frac{\pi}{4}$ & $0$ & $\frac{\pi}{2}$ & $\frac{\pi}{4}$ & $0$ & $\frac{\pi}{2}$ & $97.1\pm0.5$\\ 
3-4 & $0$ & $\frac{\pi}{2}$ & $\frac{\pi}{4}$ & $0$ & $\frac{\pi}{2}$ & $\frac{\pi}{4}$ & $0$ & $\frac{\pi}{2}$ & $\frac{\pi}{4}$ & $97.9 \pm0.5$\\ 
\hline
\end{tabular}
\caption{Phase shifters' setting of the MZIs in stages (I), (IV-i) and (IV-s) in order to prepare and manipulate the two-dimensional maximally entangled states.
The last column contains the visibilities, $V=(\rm{max}-\rm{min})/(\rm{max}+\rm{min})$, of the heralded single-photon interference fringes, reported in Fig.~\ref{fig:inter_dim}, measured by varying the phases at the input of stages (IV-i/s).
These measurements are used to calibrate the thermal phase shifters $\{ \phi^{(i)}_2 , \phi^{(s)}_1 , \phi^{(s)}_4 \}$.
The apexes $(p,i,s)$ refer to stages (I), (IV-i) and (IV-s), respectively.}
\label{tab:interf_setting}
\end{table}

In coincidence basis and low squeezing regime, this decomposition of the ideal evolution performed in stages (IV-i) and (IV-s) reads as follows in the ququart computational basis 
\begin{equation}
\begin{split}
    &|\Phi^{(2)}_m \rangle = 
    \frac{1}{\sqrt{2}} \left( | 3 \rangle_i | 3 \rangle_s + {\rm e}^{{\rm i} \delta_m} | m \rangle_i | m \rangle_s \right)
    \to 
    \frac{1}{\sqrt{2}} \left( | 3 \rangle_i | 3 \rangle_s +
    {\rm e}^{{\rm i} \left( \delta_m+\phi_m\!(I_{\rm PS})\right)}| m \rangle_i | m \rangle_s \right)
    \\
    & \to
    \begin{cases}
        % \frac{{\rm e}^{{\rm i} \left(\delta_m+\phi_m\!(I_{\rm PS})\right)/2}}{\sqrt{2}} \left\{ 
    \cos \!\left(\frac{\delta_2+\phi_2^{(i)}\!(I_{\rm PS})}{2} \right)
    \left( | 2 \rangle_i |2 \rangle_s  + | 3 \rangle_i | 3 \rangle_s \right)  +
    {\rm i} \,\sin \!\left(\frac{\delta_2+\phi_2^{(i)}\!(I_{\rm PS})}{2}\right) 
    \left( | 2 \rangle_i | 3 \rangle_s + | 3 \rangle_i | 2 \rangle_s \right)\,,
    % \right\} 
    & \mbox{for}\quad m = 2 \,, \\
    \cos \!\left(\frac{\delta_1+\phi_1^{(s)}\!(I_{\rm PS})}{2} \right)
    \left( | 2 \rangle_i |2 \rangle_s  + | 3 \rangle_i | 3 \rangle_s \right)  +
    {\rm i} \,\sin \!\left(\frac{\delta_1+\phi_1^{(s)}\!(I_{\rm PS})}{2}\right) 
    \left( | 2 \rangle_i | 3 \rangle_s + | 3 \rangle_i | 2 \rangle_s \right)\,,
    % \right\} 
    & \mbox{for}\quad m = 1 \,,\\
    \cos \!\left(\frac{\delta_4+\phi_4^{(s)}\!(I_{\rm PS})}{2} \right)
    \left( | 2 \rangle_i |2 \rangle_s  + | 4 \rangle_i | 4 \rangle_s \right)  +
    {\rm i} \,\sin \!\left(\frac{\delta_4+\phi_4^{(s)}\!(I_{\rm PS})}{2}\right) 
    \left( | 2 \rangle_i | 4 \rangle_s + | 4 \rangle_i | 2 \rangle_s \right)\,,
    % \right\} 
    & \mbox{for}\quad m = 4 \,, \\
    \end{cases}
\end{split}
\end{equation}
where we neglected the global phases, $I_{\rm PS}$ is the current applied to the PS on the input mode $m$ and $\phi_m$ is the associated phase shift.
As in Eq.~\eqref{eq:fit_phases}, we used a linear fit between the phase $\phi_k$ and the squared current $I_{\rm PS}^2$.
The above evolution is analogous to two spatially-entangled single photons, each one individually in a superposition of both inputs of two different balanced beam-splitters.
As a result, we achieved the calibration of the three PSs chosen among the PS arrays at the inputs of the final stages. This allows the compensation of the spurious phases present at the outputs of stage (III).
Note that the amplitude squared of the component $| 2 \rangle_i | 2 \rangle_s$ is directly connected to the normalized coincidence counts measured at OUT-1 and OUT-2.
Using 4 mW on-chip power, the experimental results are shown in Fig.~\ref{fig:inter_dim}(a) and the observed interference fringes visibilities are reported in Table~\ref{tab:interf_setting}.
% characterized by visibilities equal to $\{99.3\pm0.5 , 97.1\pm0.5, 97.9 \pm0.5\}$ for the sources' pair $\{ (2,3), (1,3), (3,4) \}$, respectively. 
The power is twice the one used to excite one source, in such a way as to have the same power per source. The visibility is affected by the different modulus of the JSAs, by calibration errors for the PSs and by thermal crosstalk~\cite{Silverstone_2013}. 
% \st{These effects modify the previous equation by unbalancing the probability amplitudes associated with the prepared states and by operating not the correct manipulation and routing to the outputs.} 
The measured values give a good lower bound for the indistinguishability of the sources present in our Si-PIC, which is an important requirement for the implemented VQAs.

\section{Certified dimension}
\label{app:cert_dim}

Following the method explained in~\cite{Sikora_2016} and performed in~\cite{Wang_2018}, the Certified Dimension has been verified for $d$-dimensional entangled states of dimension $(2,3,4)$ with different choices of sources.
The certification involves two parties and an entangled state.
The two parties receive one element of the entangled state and perform local operations and measurements on such a bipartite quantum system. In particular, one user, Alice, has one measurement setting and the other, Bob, two measurement settings associated with two different bases.
In our case, Alice is represented by stage (IV-i), while Bob by stage (IV-s).

The maximally entangled states of dimension $(2,3,4)$ with different choices of sources have the following forms
\begin{equation}
\begin{split}
    |\Phi^{(2)}_m \rangle &= 
    \frac{1}{\sqrt{2}} \left( | m \rangle_i | m \rangle_s +
     | 3 \rangle_i | 3 \rangle_s \right)
    \,\,, \quad 
    \mbox{for}\,\,m \in \{1,2,4\}\,,
    \\
    |\Phi^{(3)}_{m_1,m_2} \rangle &= 
    \frac{1}{\sqrt{3}} \left( | m_1 \rangle_i | m_1 \rangle_s + | m_2 \rangle_i | m_2 \rangle_s
    + | 3 \rangle_i | 3 \rangle_s \right)
    \,\,, \quad
    \mbox{for}\,\,\begin{cases}
        (m_1,m_2) \in \{1,2,4\} \,,\\
     m_1 \ne m_2 \,,
    \end{cases} 
    \\
    |\Phi^{(4)} \rangle &= 
    \frac{1}{2} \sum_{m=1}^4 | m \rangle_i | m \rangle_s \,.
\end{split}
\end{equation}
Table~\ref{tab:dim_cert_prep} exhibits the phase setting of the MZIs belonging to stage (I) in order to equally excite two, three and all sources of stage (II) and prepare the maximally entangled states of different dimensions reported in the previous equation.
\begin{table}[!h]
\centering
\scriptsize
\begin{tabular}{|c|ccc|}
\hline
Sources & $\uptheta_1^p$ & $\uptheta_2^p$ & $\uptheta_3^p$  \\ 
\hline
\hline
2-3 & $-\frac{\pi}{4}$ & $\frac{\pi}{2}$ & $\frac{\pi}{2}$   \\ 
1-3 & $\frac{\pi}{4}$ & $0$ & $\frac{\pi}{2}$   \\ 
3-4 & $0$ & $\frac{\pi}{2}$ & $\frac{\pi}{4}$  \\ 
\hline
\hline
1-2-3 & $\mbox{atan}(\sqrt{2})$ & $-\frac{\pi}{4}$ & $\frac{\pi}{2}$   \\ 
1-3-4 & $\mbox{atan}(1/\sqrt{2})$ & $0$ & $\frac{\pi}{4}$  \\ 
2-3-4 & $-\mbox{atan}(1/\sqrt{2})$ & $\frac{\pi}{2}$ & $\frac{\pi}{4}$  \\ 
\hline
\hline
all & $\frac{\pi}{4}$ & $-\frac{\pi}{4}$ & $\frac{\pi}{4}$  \\  
\hline
\end{tabular}
\caption{Phase shifters' setting of the MZIs in stage (I) in order to prepare maximally entangled states of dimension $(2,3,4)$ with different choices of sources.
% The apexes $(p)$ refer to stages (I).
}
\label{tab:dim_cert_prep}
\end{table}

The projectors associated with the different measurement settings are restricted to the analysis of the spatial mode involved in the generation process.
For all the maximally entangled states, Alice uses the measurement setting of the computational basis of the ququart, while Bob the computational basis and the following basis
\begin{equation}
\begin{split}
    |e^{(2)}_m \rangle_s & \in 
    \{ | m \rangle_s + | 3 \rangle_s ,
    | m \rangle_s - | 3 \rangle_s\}
    \,\,, \quad
    \mbox{for}\,\,m \in \{1,2,4\} \,,
    \\
    |e^{(3)}_{m_1,m_2} \rangle_s & \in 
    \{  | m_1 \rangle_s , | m_2 \rangle_s + | 3 \rangle_s , 
    | m_2 \rangle_s - | 3 \rangle_s \}
    \,\,, \quad
    \mbox{for}\,\,\begin{cases}
        (m_1,m_2) \in \{1,2,4\} \,,\\
     m_1 \ne m_2 \,,
    \end{cases} 
    \\
    |e^{(4)} \rangle_s & \in 
    \{ | 1 \rangle_s,  | 2 \rangle_s + | 3 \rangle_s , | 2 \rangle_s - | 3 \rangle_s , | 4 \rangle_s \} \,,
    \end{split}
\end{equation}
depending on the dimension of the entangled state. The projections executed in the final stages of our circuit are associated with these measurement settings and the phases to achieve these operations are presented in Table~\ref{tab:dim_cert_meas}.
\begin{table}[!h]
\centering
\scriptsize
\begin{tabular}{|c|ccc|ccc|}
\hline
Sources & $\uptheta_1^i(x)$ & $\uptheta_2^i(x)$ & $\uptheta_3^i(x)$ & $\uptheta_1^s(y_1; y_2)$ & $\uptheta_2^s(y_1; y_2)$ & $\uptheta_3^s(y_1; y_2)$ \\ 
\hline
\hline
2-3 & $[ \frac{\pi}{2}, 0 ]$ & $[\frac{\pi}{2}, \frac{\pi}{2}]$ & $[\frac{\pi}{2}, \frac{\pi}{2}]$ & 
$\begin{bmatrix} \frac{\pi}{2} & 0\\ -\frac{\pi}{4} & \frac{\pi}{4} \end{bmatrix}$ & 
$\begin{bmatrix} \frac{\pi}{2} & \frac{\pi}{2}\\ \frac{\pi}{2} & \frac{\pi}{2} \end{bmatrix}$ & 
$\begin{bmatrix} \frac{\pi}{2} & \frac{\pi}{2}\\ \frac{\pi}{2} & \frac{\pi}{2} \end{bmatrix}$  \\ 
1-3 & $[ \frac{\pi}{2}, 0 ]$ & $[0, 0]$ & $[\frac{\pi}{2}, \frac{\pi}{2}]$ & 
$\begin{bmatrix} \frac{\pi}{2} & 0\\ -\frac{\pi}{4} & \frac{\pi}{4} \end{bmatrix}$ & 
$\begin{bmatrix} 0 & 0\\ 0 & 0 \end{bmatrix}$ & 
$\begin{bmatrix} \frac{\pi}{2} & \frac{\pi}{2}\\ \frac{\pi}{2} & \frac{\pi}{2} \end{bmatrix}$  \\ 
3-4 & $[ 0, 0]$ & $[\frac{\pi}{2}, \frac{\pi}{2}]$ & $[\frac{\pi}{2}, 0]$ & 
$\begin{bmatrix} 0 , 0\\ 0 , 0 \end{bmatrix}$ & 
$\begin{bmatrix} \frac{\pi}{2} , \frac{\pi}{2}\\ \frac{\pi}{2} , \frac{\pi}{2} \end{bmatrix}$ & 
$\begin{bmatrix} \frac{\pi}{2} , 0\\ -\frac{\pi}{4} , \frac{\pi}{4} \end{bmatrix}$\\ 
\hline
\hline
1-2-3 & $[ \frac{\pi}{2}, \frac{\pi}{2}, 0 ]$ & $[ 0, \frac{\pi}{2}, \frac{\pi}{2} ]$ & $[ \frac{\pi}{2}, \frac{\pi}{2}, \frac{\pi}{2} ]$ & 
$\begin{bmatrix} \frac{\pi}{2} , \frac{\pi}{2} , 0 \\ \frac{\pi}{2} , -\frac{\pi}{4}, \frac{\pi}{4}  \end{bmatrix}$  & 
$\begin{bmatrix} 0, \frac{\pi}{2}, \frac{\pi}{2} \\ 0, \frac{\pi}{2}, \frac{\pi}{2}  \end{bmatrix}$ & 
$\begin{bmatrix} \frac{\pi}{2}, \frac{\pi}{2}, \frac{\pi}{2} \\ \frac{\pi}{2}, \frac{\pi}{2}, \frac{\pi}{2} \end{bmatrix}$  \\ 
1-3-4 & $[\frac{\pi}{2}, 0, 0]$ & $[ 0, 0, 0 ]$ & $[ \frac{\pi}{2}, \frac{\pi}{2}, 0 ]$ & 
$\begin{bmatrix} \frac{\pi}{2}, 0, 0 \\ \frac{\pi}{2}, 0, 0 \end{bmatrix}$ & 
$\begin{bmatrix} 0, 0, 0 \\ 0, 0, 0 \end{bmatrix}$ & 
$\begin{bmatrix} \frac{\pi}{2}, \frac{\pi}{2}, 0 \\ 0, -\frac{\pi}{4}, \frac{\pi}{4} \end{bmatrix}$ \\ 
2-3-4 & $[ \frac{\pi}{2}, 0, 0 ]$ & $[ \frac{\pi}{2}, \frac{\pi}{2}, \frac{\pi}{2} ]$ & $[ \frac{\pi}{2}, \frac{\pi}{2}, 0 ]$ & 
$\begin{bmatrix} \frac{\pi}{2}, 0, 0 \\ \frac{\pi}{2}, 0, 0 \end{bmatrix}$ & 
$\begin{bmatrix} \frac{\pi}{2}, \frac{\pi}{2}, \frac{\pi}{2} \\ \frac{\pi}{2}, \frac{\pi}{2}, \frac{\pi}{2} \end{bmatrix}$ & 
$\begin{bmatrix} \frac{\pi}{2}, \frac{\pi}{2}, 0 \\ 0, -\frac{\pi}{4}, \frac{\pi}{4} \end{bmatrix}$  \\ 
\hline
\hline
all & $[ \frac{\pi}{2}, \frac{\pi}{2}, 0, 0]$ & $[ 0, \frac{\pi}{2}, \frac{\pi}{2}, \frac{\pi}{2} ]$ & $[ \frac{\pi}{2}, \frac{\pi}{2}, \frac{\pi}{2}, 0 ]$ & 
$\begin{bmatrix} \frac{\pi}{2} , \frac{\pi}{2} , 0 , 0\\ \frac{\pi}{2} , -\frac{\pi}{4} , \frac{\pi}{4} , 0 \end{bmatrix}$ & 
$\begin{bmatrix} 0, \frac{\pi}{2}, \frac{\pi}{2}, \frac{\pi}{2} \\ 0, \frac{\pi}{2}, \frac{\pi}{2}, \frac{\pi}{2} \end{bmatrix}$ & 
$\begin{bmatrix} \frac{\pi}{2}, \frac{\pi}{2}, \frac{\pi}{2}, 0 \\ \frac{\pi}{2}, \frac{\pi}{2}, \frac{\pi}{2}, 0 \end{bmatrix}$ \\  
\hline
\end{tabular}
\caption{Phase shifters' setting of the MZIs in stages (IV-i) and (IV-s) in order to execute the chosen measurement settings for the dimensional certification analysis.
Each element in the arrays and the matrices is associated with the implemented projectors. The elements in the same row represent the phases for a specific measurement setting. Alice has one measurement setting (one row), and Bob two measurement settings (two rows).
The apexes $(i,s)$ refer to stages (IV-i) and (IV-s), respectively.
The utilized $\{\upphi^{i/s}_k\}_k$ (see Fig.~\ref{fig:MZI_triangle}) are set to zero.
}
\label{tab:dim_cert_meas}
\end{table}

For all the cases, we construct the correlation matrices $\mathcal{P}^{(d)}_{1/2}$ between the outcomes of Alice and the ones of Bob by looking at the coincidence counts with the chosen measurement settings of the final stages. The correlation matrices are defined and estimated as follows
\begin{equation}
\begin{split}
    & \mathcal{P}^{(d)}_{1/2}[a,b] \equiv \mbox{Tr} \left[\! |\Phi^{(d)}\rangle \langle \Phi^{(d)}|  U^{(i)}[a] \otimes U^{(s)}_{1/2}[b] \!\right]
    \, \to \,
    \mathcal{P}^{(d)}_{1/2}[a,b] = \frac{{\rm CC}_{1/2}[d,U^{(i)}[a], U^{(s)}_{1/2}[b]] }{{\rm CC}^{\rm tot}_{1/2}[d]} ,\\
    & \mbox{where}\,\,\,{\rm CC}^{\rm tot}_{1/2}[d] \equiv \sum_{a,b=1}^d
    {\rm CC}_{1/2}[d,U^{(i)}[a], U^{(s)}_{1/2}[b]] \,,
\end{split}
\end{equation}
where $\{U^{(i)}[a]\}_{a=1\ldots d}$ denote the projectors in Alice's measurement setting, $\{U^{(s)}_{1/2}[a]\}_{a=1\ldots d}$ are the two sets of projectors in the two Bob's measurement setting and ${\rm CC}_{1/2}[d,U^{(i)}[a], U^{(s)}_{1/2}[b]]$ represents the coincidence counts collected on OUT-1 and OUT-2 for the $d$-dimensional entangled state measured by Alice with $\{U^{(i)}[a]\}_{a=1\ldots d}$ and by Bob with $\{U^{(s)}_{1/2}[a]\}_{a=1\ldots d}$. The Certified Dimension is derived from the two correlation matrices as follows
\begin{equation}
    \mathcal{D}^{-1}\equiv
\sum_{b_1,b_2=1}^d  \left( \sum_{a=1}^d  \sqrt{\mathcal{P}_1[a,b_1]\, \mathcal{P}_2[a,b_2] } \right)^2
    \,,
\end{equation}
and it quantifies the quality of the prepared entangled state by looking at the two sets of correlation.
Setting the on-chip power to $(4,6,8)$ mW for $d=(2,3,4)$ respectively, the experimental certified dimensions are shown in Fig.~\ref{fig:inter_dim}(b) for different dimensions and sources' choice. As in the previous case, deviation from the ideal behavior is due to MZI calibration errors and partial distinguishability of the sources.

\section{Variational quantum algorithms with a bipartite photonic system}
\label{app:vqaimple}

In this section, we make a pair of simple examples to give general ideas about the measurement setting choice and how to evaluate the cost function from the raw data in the coincidence basis.

Using the qubit basis, Eq~\eqref{eq:map_4to2}, and our ordering for the four qubits, we consider the expectation value of the operator $\hat Z_1\otimes\mathbf{1}_2$, which is the $Z$-Pauli applied to the first qubit and the identity to the second qubit. In this case, there is no need to rotate the computational basis, and therefore, the measurement setting to be used is given in Eq.~\eqref{eq:vec_pro_comp_basis}. The same conclusion holds for any operators composed of identity and $Z$-Pauli.
Instead, if we consider the expectation value of the operator $\hat X_1\otimes\mathbf{1}_2$, which is the $X$-Pauli applied to the first qubit, the measurement setting is chosen based on the eigenvectors of $\hat X_1$, i.e.
\begin{equation}
    \{ |+0\rangle_{i}, |-0\rangle_{i}, |+1\rangle_{i}, |-1\rangle_{i} \} \otimes \{ |00\rangle_{s}, |01\rangle_{s}, |10\rangle_{s}, |11\rangle_{s} \} \,,
\end{equation}
where $|\pm\rangle = ( |0\rangle \pm |1\rangle )/\sqrt{2}$ and such states take the following form in the ququart state basis 
\begin{equation}
    \Big\{ \frac{\left( |1\rangle_{i} + |3\rangle_{i} \right)}{\sqrt{2}} , \frac{\left( |1\rangle_{i} - |3\rangle_{i} \right)}{\sqrt{2}} , \frac{\left( |2\rangle_{i} + |4\rangle_{i} \right)}{\sqrt{2}} , \frac{\left( |2\rangle_{i} - |4\rangle_{i} \right)}{\sqrt{2}}  \} \otimes \{ |1\rangle_{s}, |2\rangle_{s}, |3\rangle_{s}, |4\rangle_{s} \} \,,
\end{equation}
where we used the mapping of Eq.~\eqref{eq:map_4to2}.
Finally, for a generic operator composed of Pauli operators, the strategy consists of looking at the eigenvectors of the operator and finding the corresponding projectors on the output $| 2 \rangle_i| 2 \rangle_s$, which corresponds to $|0101\rangle$. 
% Equivalently, one can find the transformation that diagonalizes such an operator and combine it with the 16 projectors in Eq.~\eqref{eq:vec_pro_comp_basis}.

We conclude this section by explaining how to calculate a generic cost function with our architecture. This step allows us to connect the raw data coming from our QH with its classical counterpart and implement the desired VQA.
First of all, given the set of observables $\{\hat{O}_k\}_k$ needed to construct the weighted sum with the parameters $\{w_k\}_k$, we have to identify the commuting group (CG) of observables~\cite{McClean_2016,Tilly_2022}. This clustering is required since it fixes the measurement setting for each CG.
In particular, without loss of generality in the final form of the cost function, we can consider binary operators constructed as a generic combination of identity and Pauli operators $\{ \mathbf{1}, \hat{X}, \hat{Y}, \hat{Z} \}$~\cite{jordanwigner},
\begin{equation}
    \hat{O}_k^{({\rm I})}
    =  
     \bigotimes_{ \hat{\sigma}_j^{(k,{\rm I})} \in \{ \mathbf{1}, \hat{X}, \hat{Y}, \hat{Z} \}} \hat{\sigma}_j^{(k,{\rm I})} \,,
     \label{eq:genbinoperator}
\end{equation}
where the index I indicates the belonging CG.
Since $\{ \mathbf{1}, \hat{X}, \hat{Y}, \hat{Z} \}$ are two dimensional binary operators, the dimension of $\hat{O}_k^{({\rm I})}$ is $2^n$ for $n$ total Pauli operators involved in the construction of $\hat{O}_k^{({\rm I})}$. This means that the quantum register must have the same dimension, i.e. $2^n$ quantum states, which is exactly the dimension of $n$ qubits. For simplicity, in this part we will denote the computational state basis as $\{ | m \rangle \}_{m=1\ldots2^n}$, which can be converted to qubit notation by using binary numbers starting from zero.
Note that the computational basis is composed of eigenvectors of any operators composed of $\{ \mathbf{1}, \hat{Z} \}$.
Then, by using the spectral theorem~\cite{Moretti_2013cma}, each previous binary observable can be written as follows
\begin{equation}
\begin{split}
    & \hat{O}_k^{({\rm I})}
    =
    \sum_{{\mathbf{e}}} \boldsymbol{\pi}\!\left[k,{\rm I},{\mathbf{e}}\right]\, 
    \hat{\mathbf{P}}_{\mathbf{e}} 
    =
    \sum_m \boldsymbol{\pi}\!\left[k,{\rm I},m\right]\, 
    U_{\rm R}[{\rm I}]^\dagger \, \hat{\mathbf{P}}_m \,  U_{\rm R}[{\rm I}] \\
    & \mbox{where}\quad
        \hat{\mathbf{P}}_{\mathbf{e}}  = | \mathbf{e} \rangle \langle \mathbf{e} | \quad,\quad
        \hat{\mathbf{P}}_m = | m \rangle \langle m | \quad,\quad
        | m \rangle = U_{\rm R}[{\rm I}]\,| \mathbf{e} \rangle
        \quad,
\end{split}
\end{equation}
and $\{|\mathbf{e}\rangle\}$ are the eigenvectors of $\hat{O}_k^{({\rm I})}$, $\boldsymbol{\pi}\!\left[k,{\rm I},{\mathbf{e}}\right]$ is the eigenvalue associated with the eigenvector $|\mathbf{e}\rangle$, $\hat{\mathbf{P}}_{\mathbf{e}}$ and $\hat{\mathbf{P}}_m$ are the Von Neumann projectors on the states $| \mathbf{e} \rangle$ and $| m \rangle$ and $U_{\rm R}[{\rm I}]$ is the unitary transformation for the mapping of the basis $\{|\mathbf{e}\rangle\}$ to the computational basis $\{|m\rangle\}$.
In this way, it is possible to express the cost function, Eq.~\eqref{eq:costfunction}, as follows
\begin{equation}
\begin{split}
    {\rm C}(\boldsymbol{\alpha}) 
    &= \sum_k w_k
    \langle \psi (\boldsymbol{\alpha}) | \, \hat{O}_k \, | \psi (\boldsymbol{\alpha}) \rangle \\
    &=
    \sum_{k} \sum_{{\rm I} \in {\rm CG} } \! w_k \! \sum_m \boldsymbol{\pi}\!\left[k,{\rm I},m\right]\, \Big\langle U_{\rm R}[{\rm I}]\psi (\boldsymbol{\alpha}) \Big| \, \hat{\mathbf{P}}_m \, \Big| U_{\rm R}[{\rm I}]\psi (\boldsymbol{\alpha}) \Big\rangle   \\
    &=   
    \sum_{k} \sum_{{\rm I} \in {\rm CG} } \! w_k \! \sum_m \boldsymbol{\pi}\!\left[k,{\rm I},m\right]\, 
    \Big| \langle m  | U_{\rm R}[{\rm I}] | \psi (\boldsymbol{\alpha}) \rangle \Big|^2\,.
    \end{split}
\end{equation}
From the previous equation, we can note that by rotating the trial state $| \psi (\boldsymbol{\alpha}) \rangle $, the sampling on the computation basis is enough to find the desired expectation values. Thus, once the CGs are chosen and the rotation $U_{\rm R}[{\rm I}]$ is defined for each CG, we need to apply such transformation to the trial state before the measurement on the computational state basis.

We now need to specialize the previous treatment to our implementation.
In our case, the computational basis can be labeled with the two ququart states, i.e. $\{ | m_1\rangle_i |m_2 \rangle_s \}_{m_1,m_2 = 1\ldots4}$, and the projectors implemented by stages (IV-i/s) satisfy the following relations
\begin{equation}
    \langle m_1 |_i \langle m_2 |_s = 
    \langle 2 |_i \langle 2 |_s \, \mathbb{P}_{m_1}^{(i)} \otimes \mathbb{P}_{m_2}^{(s)} \,,
\end{equation}
which are equivalent to Eq.~\eqref{eq:proj_comp_basis}.
Therefore, the cost function becomes
\begin{equation}
\begin{split}
    {\rm C}(\boldsymbol{\alpha}) 
    &=   
    \sum_{k} \sum_{{\rm I} \in {\rm CG} } w_k \!\!\sum_{m_1,m_2=1}^4 \!\!\boldsymbol{\pi}\!\left[k,{\rm I},m_1,m_2\right]\! 
    \Big| \langle 2 |_i \langle 2 |_s \! \left(\mathbb{P}_{m_1}^{(i)}\cdot U_{\rm R}^{(i)}[{\rm I}] \right) \otimes \left(\mathbb{P}_{m_2}^{(s)}  \cdot U_{\rm R}^{(s)}[{\rm I}] \right) \! |\psi (\boldsymbol{\alpha}) \rangle \Big|^2 \,.
    \end{split}
    \label{eq:costfunc_pro_rot}
\end{equation}
The cost function is calculated by sampling the rotated trial state over all the combinations of computational basis projectors. The terms in squared modulus inside the previous equation are the probabilities that the trial state is measured in state $| 2 \rangle_i | 2 \rangle_s$ after being rotated with $U_{\rm R}^{(i)}[{\rm I}]\otimes U_{\rm R}^{(s)}[{\rm I}]$ and projected with $\{ \mathbb{P}_{m_1}^{(i)} \otimes \mathbb{P}_{m_2}^{(s)} \}_{m_1,m_2=1\ldots4}$.
In our Si-PIC, the first three stages prepare the state in Eq.~\eqref{eq:stageiii_ququart} and Eq.~\eqref{eq:stageiii_qubit} and variational parameters $\boldsymbol{\alpha}$ depend on the setting of stage (I). The prepared state is equal to $|\psi (\boldsymbol{\alpha}) \rangle$ modulo a linear transformation. Then, the stages (IV-i) and (IV-s) execute the projections $\{ \mathbb{P}_{m_1}^{(i)} \otimes \mathbb{P}_{m_2}^{(s)} \}_{m_1,m_2=1\ldots4}$ rotated by $U_{\rm R}^{(i)}[{\rm I}]\otimes U_{\rm R}^{(s)}[{\rm I}]$.
Since the generic trial state $|\psi (\boldsymbol{\alpha}) \rangle$ cannot be prepared after stage (III), the preparation of the trial state is generally finished together with the rotated projection in stages (IV-i) and (IV-s). 
Finally, since we work in the coincidence basis by collecting pairs of photons from OUT-1 and OUT-2, we can estimate the probabilities contained in Eq.~\eqref{eq:costfunc_pro_rot} by counting the coincidence events of idler and signal photons for each projection $\{ \mathbb{P}_{m_1}^{(i)} \otimes \mathbb{P}_{m_2}^{(s)} \}_{m_1,m_2=1\ldots4}$ and normalizing that value for the overall coincidences within the same measurement setting.
This concept can be expressed by modifying Eq.~\eqref{eq:costfunc_pro_rot} and inserting the raw data of our QH as follows
\begin{equation}
\begin{split}
    & {\rm C}(\boldsymbol{\alpha}) =   
    \sum_{k} \sum_{{\rm I} \in {\rm CG} } w_k \!\sum_{m_1,m_2=1}^4 \boldsymbol{\pi}\!\left[k,{\rm I},m_1,m_2\right]\, 
    \frac{{\rm CC}[\boldsymbol{\alpha},k,{\rm I},m_1,m_2] }{{\rm CC}^{\rm tot}[\boldsymbol{\alpha},k,{\rm I}]} \,,\\
    & \mbox{where}\,\,\,{\rm CC}^{\rm tot}[\boldsymbol{\alpha},k,{\rm I}] \equiv \sum_{m_1,m_2=1}^4
    {\rm CC}[\boldsymbol{\alpha},k,{\rm I},m_1,m_2] \,,
    \end{split}
\end{equation}
and ${\rm CC}[\boldsymbol{\alpha},k,{\rm I},m_1,m_2]$ represents the coincidences collected on OUT-1 and OUT-2 for the prepared state $|\psi (\boldsymbol{\alpha}) \rangle$ rotated with $U_{\rm R}^{(i)}[{\rm I}]\otimes U_{\rm R}^{(s)}[{\rm I}]$ and projected with $ \mathbb{P}_{m_1}^{(i)} \otimes \mathbb{P}_{m_2}^{(s)} $.
Therefore, by utilizing the coincidence counts provided by our QH and by making specific weighted sums on the PC, the evaluation of the cost function is achieved. 

\section{Hydrogen molecule and Variational Quantum Eigensolver}
\label{app:H2}

For the general eigenvalue problem, 
\begin{equation}
    \hat{\rm H} \,| \psi_n \rangle = E_n \,| \psi_n \rangle
\end{equation}
the Hamiltonian of a generic molecule can be written in first quantization as follows
\begin{equation}
    \hat{\rm H}_{\rm f}^{(1)} = -\sum_i \frac{\hbar^2}{2M_i}\nabla_{R_i}^2 -\sum_i \frac{\hbar^2}{2m_e}\nabla_{r_i}^2
    -\sum_{i,j} \frac{Z_i\,e^2}{|R_i-r_j|}
    +\sum_{i,j<i} \frac{Z_i Z_j \,e^2}{|R_i-R_j|}
    +\sum_{i,j<i} \frac{e^2}{|r_i-r_j|} \,,
\end{equation}
where the parameters $(R,M,Z)$ are the nuclei positions, masses and atomic numbers, while the parameters $(r,m_e,e)$ are the electrons' positions and the electron mass and charge.

Because of the big difference between the nuclei masses and electron mass, the nuclei are treated as fixed classical point charges with the Born-Oppeneimer approximation.
Then, a basis $\phi$ is chosen to represent the electronic wave function
molecular orbitals as linear combinations of atomic orbitals (LCAO). The atomic orbitals are derived from variations of Hydrogen-like atomic orbitals, and typically the basis functions are expressed as the sum of Gaussian functions rather than the original Slater-type orbitals~\cite{froyen1979elementary,ching2012electronic}.

In the second quantization, the previous Hamiltonian can be written with ladder operators as follows~\cite{szabo1996modern,helgaker2014electronic}
\begin{equation}
    \hat{\rm H}_{\rm f}^{(2)} = \sum_{p, q} h_{p q} \hat{a}^\dagger_p \hat{a}_q + 
    \frac{1}{2}\sum_{p, q, r, s} h_{p q r s} \hat{a}^\dagger_p \hat{a}^\dagger_q \hat{a}_r \hat{a}_s
\end{equation}
where 
\begin{equation}
\begin{split}
    h_{p q} &= \int {\rm d}\sigma \, \phi^*_p(\sigma) 
    \left(  -\frac{\hbar^2}{2m_e}\nabla^2_r -\sum_{i} \frac{Z_i e^2}{|R_i-r|}
    \right) \phi_q(\sigma)
    \\
    h_{p q r s} &= \int {\rm d}\sigma_1 {\rm d}\sigma_2 \, \frac{\phi^*_p(\sigma_1)\phi^*_q(\sigma_2) \phi_r(\sigma_2)\phi_s(\sigma_1) }{|r_1-r_2|} \,,
    \label{eq:tensor2quant}
\end{split}
\end{equation}
and the latin indexes denote the spin-orbital state, the variable $\sigma$ denotes the position and spin, the ladder operators $(\hat{a},\hat{a}^\dagger)$ are fermionic creation and destruction operators, and the tensor entries $h_{p q}$ and $h_{p q r s}$ are the parameters which identify the system and can be efficiently computed by a classical machine~\cite{psi4}. 

This problem can be translated into the qubit language of the quantum simulator by transforming the previous fermionic Hamiltonian into a spin-like Hamiltonian.
Possible mappings are Jordan-Wigner (JW)~\cite{jordanwigner}, parity~\cite{seeley2012bravyi} and Bravyi-Kitaev~\cite{bravyi2002fermionic,seeley2012bravyi}.
For example, JW transformation maps the previous ladder operators to bosonic operators as follows
\begin{equation}
    \hat{a}_p \to \left( \bigotimes_{k=p+1}^{n} \hat{\mathbf{1}}_k \right) \otimes \frac{\hat{X}_p - {\rm i}\hat{Y}_p}{2} \otimes \left(\bigotimes_{k=1}^{p-1} \hat{Z}_k \right)\,,
    \qquad
    \hat{a}^\dagger_p \to
    \left( \bigotimes_{k=p+1}^{n} \hat{\mathbf{1}}_k \right) \otimes \frac{\hat{X}_p + {\rm i}\hat{Y}_p}{2} \otimes \left( \bigotimes_{k=1}^{p-1} \hat{Z}_k \right)\,,
\end{equation}
where $n$ is the number of qubits and $\{ \mathbf{1}, \hat{X}, \hat{Y}, \hat{Z} \}$ are the identity and Pauli matrices. 
% Note that the right-hand-side of the previous map is given only by strings of Pauli matrices.
Following this recipe, the bosonic Hamiltonian reads
\begin{equation}
    \hat{\rm H}_{\rm b} = \sum_{k} h_k \, \hat{O}_k 
    \quad \,,\qquad \mbox{where}\,\,\,
    \hat{O}_k \equiv \bigotimes_{ \hat{\sigma}_j^{(k)} \in \{ \mathbf{1}, \hat{X}, \hat{Y}, \hat{Z} \}} \hat{\sigma}_j^{(k)} \,,
\end{equation}
where the $\{h_k\}_k$ are given by combination of the tensor entries in Eq.~\eqref{eq:tensor2quant}.
Therefore, we end up with a Hamiltonian, whose expectation value can be used to define the cost function of the Variational Quantum Eigensolver~\cite{Peruzzo_2014,McClean_2016}. The final result is given by the ground energy together with the ground state written in the qubit register.

To test our circuit and algorithm, we decided to solve the Hydrogen molecule.
Using the minimal STO-3G basis set, composed of 1s orbitals of the individual H atoms with two spin values, we have a total number of atomic orbitals (AOs) equal to 4.
In this case there are two commuting groups (CGs): the first is composed of operators made by the identity and $Z$-Pauli matrices and the second with the $X$- and $Y$-Pauli matrices. 
Table~\ref{tab:H2_coeff} shows all the operators divided into the two CGs together with the coefficients $h$s for different atomic distances.

The first CG does not require any rotation, since the computational basis states are their eigenvectors, while the second group is given by operators made of $\hat{Y}\otimes\hat{X}$ and $\hat{X}\otimes\hat{Y}$. The rotation to transform such operators in identity and $Z$-Pauli matrices reads as follows
\begin{equation}
\begin{split}
    & \hat{Y}\otimes \hat{X}= \bar{U}_{\rm R}^{\rm T} \cdot \left( -\mathbb{1}\otimes\hat{Z} \right) \cdot \bar{U}_{\rm R}
    \quad , \quad
    \hat{X}\otimes \hat{Y}= \bar{U}_{\rm R}^{\rm T} \cdot \left( -\hat{Z}\otimes \mathbb{1} \right) \cdot \bar{U}_{\rm R}
    \\
    & \quad \mbox{where} \quad \bar{U}_{\rm R} = 
    \frac{1}{\sqrt{2}} 
    \left(
    \begin{array}{cccc}
     1 & 0 & 0 & {\rm i} \\
     0 & -1 & {\rm i} & 0 \\
     0 & 1 & {\rm i} & 0 \\
     -1 & 0 & 0 & {\rm i} \\
    \end{array}
    \right)\,.
\end{split}
\end{equation}
However, the composition of this rotation with the projectors of the computational basis, $\{ \mathbb{P}_{m}^{(i/s)} \}_{m=1\ldots4}$, cannot be implemented in stages (IV-i) and (IV-s), since such triangular schemes are not universal.
This problem is overcome by noting that we don't need to execute the rotation reported in the previous equation, but we can ask the following less strict requirements
\begin{equation}
\begin{split}
    & \langle 2|\,\hat{Y}\otimes \hat{X}= \langle 2|\,\left(U_{\rm R}^{(2)}\right)^{\rm T} \cdot \left( \mathbb{1}\otimes\hat{Z} \right) \cdot U_{\rm R}
    \quad , \quad
    \langle 2|\,\hat{X}\otimes \hat{Y}\equiv \langle 2|\,\left(U_{\rm R}^{(2)}\right)^{\rm T} \cdot \left(\hat{Z} \otimes \mathbb{1} \right) \cdot U_{\rm R}
    \\
    & \quad \mbox{where} \quad U_{\rm R}^{(2)} = 
    \frac{1}{\sqrt{2}} 
    \left(
    \begin{array}{cccc}
     \sqrt{2} & 0 & 0 & 0 \\
     0 & 1 & {\rm i} & 0 \\
     0 & 1 & -{\rm i} & 0 \\
     0 & 0 & 0 & \sqrt{2} \\
    \end{array}
    \right)\,.
\end{split}
\label{eq:H2rot2CG}
\end{equation}
The previous rotation can be composed with the 16 projectors of the computational basis in order to achieve the manipulation required in stages (IV-i) and (IV-s).
Table~\ref{tab:4rotatedprojection} shows the phase setting of the MZIs belonging to stages (IV-i) and (IV-s) in order to implement the measurement settings corresponding to the two commuting groups.

\begin{table}[!h]
\centering
\scriptsize
\begin{tabular}{|c|ccc|}
\hline
Operation & $\uptheta_1^{i/s}$ & $\uptheta_2^{i/s}$ & $\uptheta_3^{i/s}$ \\ 
\hline
$\mathbb{P}_{1}^{(i/s)}$ & $\frac{\pi}{2}$ & $0$ & $\frac{\pi}{2}$ \\ 
$\mathbb{P}_{2}^{(i/s)}$ & $\frac{\pi}{2}$ & $\frac{\pi}{2}$ & $\frac{\pi}{2}$ \\ 
$\mathbb{P}_{3}^{(i/s)}$ & $0$ & $\frac{\pi}{2}$ & $\frac{\pi}{2}$ \\ 
$\mathbb{P}_{4}^{(i/s)}$ & $0$ & $\frac{\pi}{2}$ & $0$ \\ 
\hline
$\mathbb{P}_{1}^{(i/s)} \cdot U^{(2)}_{\rm R}$ & $\frac{\pi}{2}$ & $0$ & $\frac{\pi}{2}$ \\ 
$\mathbb{P}_{2}^{(i/s)} \cdot U^{(2)}_{\rm R}$ & $\frac{\pi}{4}$ & $\frac{\pi}{2}$ & $\frac{\pi}{2}$ \\ 
$\mathbb{P}_{3}^{(i/s)} \cdot U^{(2)}_{\rm R}$ & $-\frac{\pi}{4}$ & $\frac{\pi}{2}$ & $\frac{\pi}{2}$ \\ 
$\mathbb{P}_{4}^{(i/s)} \cdot U^{(2)}_{\rm R}$ & $0$ & $\frac{\pi}{2}$ & $0$ \\ 
\hline
\end{tabular}
\caption{Phase shifters' setting of the MZIs in stages (IV-i) and (IV-s) in order to execute the measurement settings corresponding to the two commuting groups of operators present in Hydrogen molecules Hamiltonian using the minimal STO-3G basis set.
The first block represents the four projections on the computational basis, while the second one the 'rotated' projection, whose rotation is reported in Eq.~\eqref{eq:H2rot2CG}.
The apexes $(i,s)$ refer to stages (IV-i) and (IV-s), respectively.
The utilized $\{\upphi^{i/s}_k\}_k$ (see Fig.~\ref{fig:MZI_triangle}) are set to zero, even if they enters as global phases.}
\label{tab:4rotatedprojection}
\end{table}

\begin{figure}[t]
    \centering
    \includegraphics[width=\textwidth]{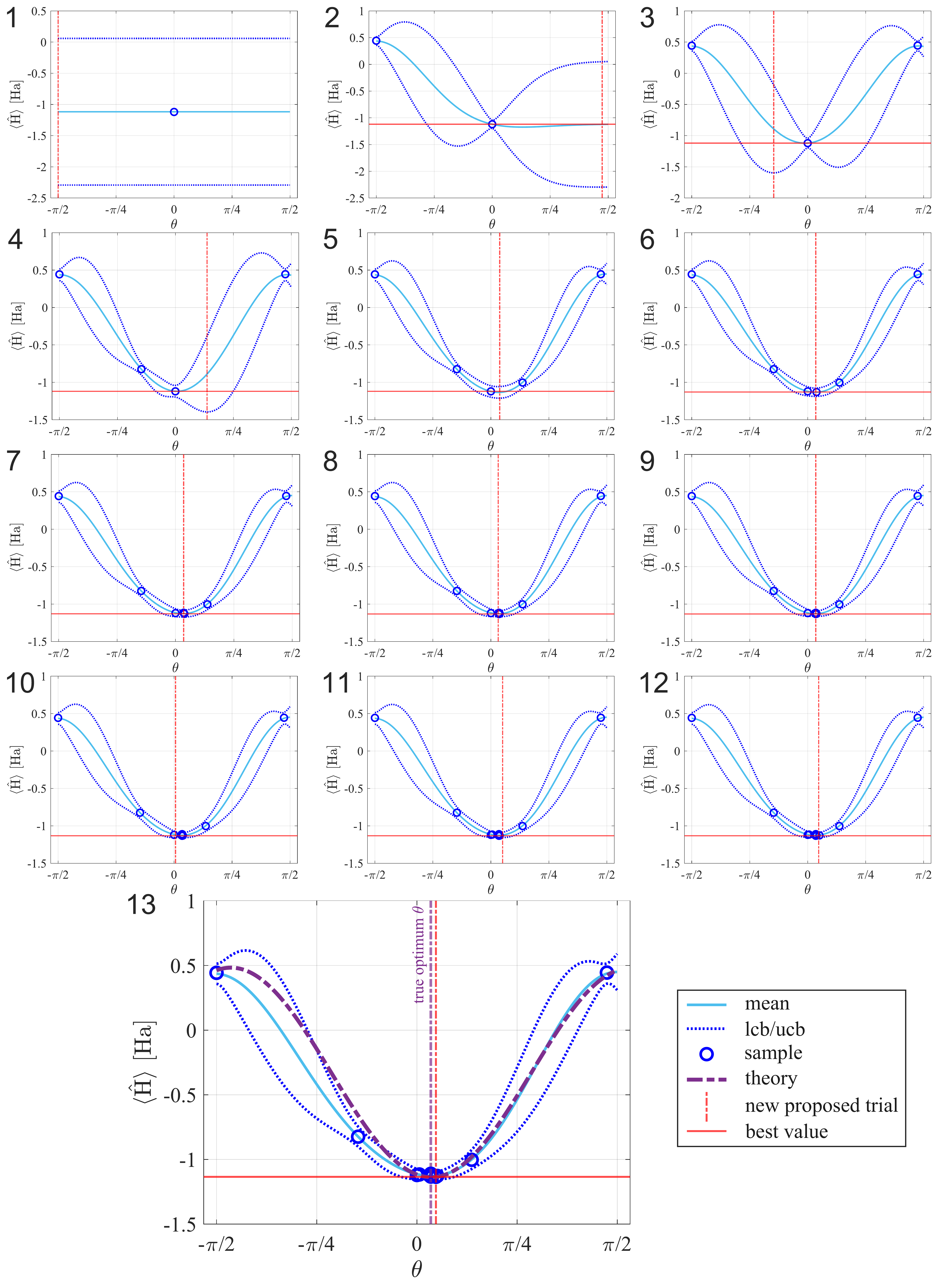}
    \caption{
    Graphs representing the steps of the minimum Bayesian search for the Hydrogen molecule using the minimal STO-3G basis set and the UCC ansatz with an atomic distance equal to 0.736~\AA. At step 1, the experimental datum taken with the Hartree-Fock configuration gives the prior mean and the prior $97.5\%$-confidence bounds of a constant probability distribution.
    In the successive steps, every new sample is used to find the posterior update of the mean and the covariance through the posterior Gaussian Process.
    The new value for the variational parameter $\uptheta$ (vertical dotted red line) is chosen by looking at the lowest value of the lower confidence bound.
    Blue circles are the experimental values estimated through our circuit, light blue solid lines and blue dotted lines are the mean and $2.5\%$-confidence bounds of the Gaussian distribution. The horizontal solid red line indicates the best value found until that specific step. In step 13, the thick purple dotted line is the theoretical curve of the energy as a function of the variational parameter, while the vertical thin dotted purple line the true optimal value for the variational parameter $\uptheta$.
    }
    \label{fig:BO_ite_supp}
\end{figure}

Under the UCC ansatz~\cite{hoffmann1988unitary,bartlett1989alternative,romero2018strategies}, the trial state for ${\rm H}_2$ is given by a linear combination of BMO-BMO and AMO-AMO states. In our qubit register, the UCC trial state reads as follows
\begin{equation}
    | \psi (\uptheta) \rangle
    =
    \cos\uptheta \,|1010\rangle
    - \sin\uptheta \,|0101\rangle \,,
\end{equation}
and the associated expectation value of the ${\rm H}_2$ Hamiltonian has the following dependence as a function of the variational parameter $\uptheta$
\begin{equation}
    \langle \psi (\uptheta) | \,\hat{\rm H}\, | \psi (\uptheta) \rangle
    = g_0({\rm R})
    + g_1({\rm R}) \,\cos\left( 2\uptheta \right) 
    + g_2({\rm R}) \, \sin\left( 2\uptheta \right) \,,
\end{equation}
where R is the atomic distance. The term proportional to $g_1$ comes from the first CG, while the one with $g_2$ from the second CG.
The trial state obtained with $\uptheta=0$ is the Hartree-Fock state, well-known many-body approximation. Note that the ground state is not the Hartree-Fock state if $g_2 \ne 0$. Thus, the ground state is characterized by a non-trivial correlation between the BMO-BMO and AMO-AMO states.

The used optimization routines are gradient-descent and Bayesian optimizations: in both cases, the iteration loop ends when the new variational parameter and the associated cost function change by a value lower than a chosen threshold.
The implemented gradient descent optimization algorithm can be summarized with the following steps:
\begin{enumerate}
    \item[0.] choose the ansatz for the initial value of the variational parameter;
    \item[1.] evaluate the cost function ${\rm C}$ with the set variational parameter $\bar \uptheta$;
    \item[2.] evaluate the cost function ${\rm C}$ with variational parameter $\bar \uptheta +\epsilon$;
    \item[3.] calculate the discrete derivative of the cost function $\dot{\rm C}(\bar \uptheta) = ({\rm C}(\bar \uptheta +\epsilon)-{\rm C}(\bar \uptheta ))/\epsilon$ and update the variational parameter $\bar \uptheta \to \bar \uptheta + \eta \, \dot {\rm C}(\bar \uptheta)$;
    \item[4.] go to step 1;
\end{enumerate}
where $\eta$ is the learning parameter. % and the details on the parameters $(\epsilon,\eta)$ are given in Methods.
We observed that the convergence of this method is strongly limited by the statistical error and the long time needed for steps 2 and 3.
Therefore, we decided to utilize also a Bayesian optimization~\cite{Iannelli_2022rZ}, which has the following steps:
\begin{enumerate}
    \item[0.] choose the prior Gaussian Process and the ansatz for the variational parameter;
    \item[1.] evaluate the cost function ${\rm C}$ with the set variational parameter $\bar \uptheta$;
    \item[2.] calculate the posterior Gaussian Process based on the new datum and the chosen acquisition function and update the variational parameter;
    \item[3.] go to step 1;
\end{enumerate}
where the acquisition function of the Gaussian Process~\cite{mockus2005bayesian,bayopt,Garnett_2023} is given by the lower confidence bound equal to $2.5\%$.
The general idea of this procedure starts by considering the cost function to be evaluated as a multivariate normal joint distribution over functions with a continuous domain of variational parameters.  

% \textcolor{red}{In this case, the procedure starts from a constant prior probability distribution with mean equal to the experimental value achieved for the Hartree-Fock state. Then, the next guess is chosen by looking at the lowest value of the lower confidence bound, which determines the cost function values with 2.5\% probability of being lower, and after every new datum, the posterior probability is calculated.}

We conclude this section by reporting further details on the Bayesian optimization used in the VQE for the Hydrogen molecule, Sec.~\ref{sec:H2}.
We decided to start with a constant prior probability distribution with the mean given by the experimental sample evaluated in the Hartree-Fock configuration ($\uptheta$) and the kernel for the covariant correlation has the following form
\begin{equation}
    K(\uptheta_1, \uptheta_2) = M_K^2 \,\exp \left[ -\frac{|\uptheta_2-\uptheta_1|^2}{2 \sigma_K^2} \right]
\end{equation}
where $M_K = 0.6$ and $\sigma_K = 0.65$ are the values for the hyperparameters. These parameters give good results in the implemented  Bayesian search, but they can be further optimized through maximum likelihood estimation.
In the first step, the experimental sample taken with the Hartree-Fock configuration gives the prior mean and the prior $97.5\%$-confidence bounds of the constant probability distribution.
After the acquisition of new samples, the posterior distribution is updated through the Gaussian Process and the posterior means and the variances are obtained.
At each step, the new value to be sampled is determined by the lowest value of the lower posterior confidence bound.
Fig.~\ref{fig:BO_ite_supp} shows all the steps of the minimum Bayesian search for the Hydrogen molecule using the minimal STO-3G basis set and the UCC ansatz with an atomic distance equal to 0.736~\AA.
The sampled values at each iteration step are reported in Fig.~\ref{fig:BO}.
In the last step, we report the theoretical curve of the energy as a function of the variational parameter together with the true optimal value for the variational parameter $\uptheta$. These values can be compared with the final mean of the posterior probability distribution and the last proposed new sample, respectively.

\section{Variational Quantum Factorizing}
\label{app:35}

After stage (III) the state has the form given in Eq.~\eqref{eq:stageiii_qubit}, and it differs from the desired trial state for the factorization of 35 given in Eq.~\eqref{eq:trialstatefacto}.
Thus the projectors on the computational basis are modified to take into account also the needed linear transformation.
Table~\ref{tab:16projection_facto} shows the phase setting of the MZIs belonging to stages (IV-i) and (IV-s) in order to implement the projectors associated with the computational basis followed by the final transformation to prepare the generic trial state, Eq.~\eqref{eq:trialstatefacto}.
Note that the parameters $\alpha$ in Eq.~\eqref{eq:trialstatefacto} are functions of the phases $\uptheta$ of stage (I) (see Eq.~\eqref{eq:thetatoalpha}), which are the variational parameters.

\begin{table}[!h]
\centering
\scriptsize
\begin{tabular}{|c|ccc|ccc|}
\hline
Operation & $\uptheta_1^i$ & $\uptheta_2^i$ & $\uptheta_3^i$ & $\uptheta_1^s$ & $\uptheta_2^s$ & $\uptheta_3^s$\\ 
\hline
$\mathbb{P}_{1}^{(i)} \otimes \mathbb{P}_{1}^{(s)}$ & $\frac{\pi}{2}$ & $0$ & $\frac{\pi}{2}$ & $0$ & $\frac{\pi}{2}$ & $\frac{\pi}{2}$ \\ 
$\mathbb{P}_{1}^{(i)} \otimes \mathbb{P}_{2}^{(s)}$ & $\frac{\pi}{2}$ & $0$ & $\frac{\pi}{2}$ & $\frac{\pi}{2}$ & $\frac{\pi}{2}$ & $\frac{\pi}{2}$ \\ 
$\mathbb{P}_{1}^{(i)} \otimes \mathbb{P}_{3}^{(s)}$ & $\frac{\pi}{2}$ & $0$ & $\frac{\pi}{2}$ & $0$ & $\frac{\pi}{2}$ & $\frac{\pi}{2}$ \\ 
$\mathbb{P}_{1}^{(i)} \otimes \mathbb{P}_{4}^{(s)}$ & $\frac{\pi}{2}$ & $0$ & $\frac{\pi}{2}$ & $0$ & $\frac{\pi}{2}$ & $0$ \\ 
\hline
$\mathbb{P}_{2}^{(i)} \otimes \mathbb{P}_{1}^{(s)}$ & $\frac{\pi}{2}$ & $\frac{\pi}{2}$ & $\frac{\pi}{2}$ & $\frac{\pi}{2}$ & $0$ & $\frac{\pi}{2}$ \\ 
$\mathbb{P}_{2}^{(i)} \otimes \mathbb{P}_{2}^{(s)}$ & $\frac{\pi}{2}$ & $\frac{\pi}{2}$ & $\frac{\pi}{2}$ & $\frac{\pi}{2}$ & $\frac{\pi}{2}$ & $\frac{\pi}{2}$ \\ 
$\mathbb{P}_{2}^{(i)} \otimes \mathbb{P}_{3}^{(s)}$ & $\frac{\pi}{2}$ & $0$ & $\frac{\pi}{2}$ & $\frac{\pi}{2}$ & $0$ & $\frac{\pi}{2}$ \\ 
$\mathbb{P}_{2}^{(i)} \otimes \mathbb{P}_{4}^{(s)}$ & $\frac{\pi}{2}$ & $\frac{\pi}{2}$ & $\frac{\pi}{2}$ & $0$ & $\frac{\pi}{2}$ & $0$ \\ 
\hline
$\mathbb{P}_{3}^{(i)} \otimes \mathbb{P}_{1}^{(s)}$ & $0$ & $\frac{\pi}{2}$ & $\frac{\pi}{2}$ & $\frac{\pi}{2}$ & $0$ & $\frac{\pi}{2}$ \\ 
$\mathbb{P}_{3}^{(i)} \otimes \mathbb{P}_{2}^{(s)}$ & $0$ & $\frac{\pi}{2}$ & $\frac{\pi}{2}$ & $\frac{\pi}{2}$ & $\frac{\pi}{2}$ & $\frac{\pi}{2}$ \\ 
$\mathbb{P}_{3}^{(i)} \otimes \mathbb{P}_{3}^{(s)}$ & $0$ & $\frac{\pi}{2}$ & $\frac{\pi}{2}$ & $\frac{\pi}{2}$ & $0$ & $\frac{\pi}{2}$ \\ 
$\mathbb{P}_{3}^{(i)} \otimes \mathbb{P}_{4}^{(s)}$ & $0$ & $\frac{\pi}{2}$ & $\frac{\pi}{2}$ & $0$ & $\frac{\pi}{2}$ & $0$ \\ 
\hline
$\mathbb{P}_{4}^{(i)} \otimes \mathbb{P}_{1}^{(s)}$ & $0$ & $\frac{\pi}{2}$ & $0$ & $\frac{\pi}{2}$ & $0$ & $\frac{\pi}{2}$ \\ 
$\mathbb{P}_{4}^{(i)} \otimes \mathbb{P}_{2}^{(s)}$ & $0$ & $\frac{\pi}{2}$ & $\frac{\pi}{2}$ & $0$ & $\frac{\pi}{2}$ & $\frac{\pi}{2}$ \\ 
$\mathbb{P}_{4}^{(i)} \otimes \mathbb{P}_{3}^{(s)}$ & $0$ & $\frac{\pi}{2}$ & $0$ & $0$ & $\frac{\pi}{2}$ & $0$ \\ 
$\mathbb{P}_{4}^{(i)} \otimes \mathbb{P}_{4}^{(s)}$ & $0$ & $\frac{\pi}{2}$ & $0$ & $0$ & $\frac{\pi}{2}$ & $\frac{\pi}{2}$ \\ 
\hline
\end{tabular}
\caption{
Phase shifters' setting of the MZIs in stages (IV-i) and (IV-s) in order to execute the projectors associated with the computational basis followed by the final transformation to prepare the generic trial state, Eq.~\eqref{eq:trialstatefacto}.
The apexes $(i,s)$ refer to stages (IV-i) and (IV-s), respectively.
The utilized $\{\upphi^{i/s}_k\}_k$ (see Fig.~\ref{fig:MZI_triangle}) are set to zero, even if they enters as global phases.
}
\label{tab:16projection_facto}
\end{table}

Table~\ref{tab:facto_coeff} shows all the operators together with the coefficients $\{w_k\}_k$ for three odd semiprime numbers lower 49. The upper bound is given by the chosen ansatz for the prime numbers: $p = 4x_1+2x_2+1$ and $q = 4x_3+2x_4+1$.

The implemented gradient descent optimization algorithm can be summarized with the following steps:
\begin{enumerate}
    \item[0.] choose the ansatz for the initial value of the variational parameters;
    \item[1.] evaluate the cost function ${\rm C}$ with the set variational parameter $(\bar \uptheta_1, \bar \uptheta_2, \bar \uptheta_3)$;
    \item[2a.] evaluate the cost function ${\rm C}$ with variational parameter $(\bar \uptheta_1 +\epsilon, \bar \uptheta_2, \bar \uptheta_3)$;
    \item[2b.] evaluate the cost function ${\rm C}$ with variational parameter $(\bar \uptheta_1 , \bar \uptheta_2+\epsilon, \bar \uptheta_3)$;
    \item[2c.] evaluate the cost function ${\rm C}$ with variational parameter $(\bar \uptheta_1 , \bar \uptheta_2, \bar \uptheta_3+\epsilon)$;
    \item[3.] calculate the discrete gradient of the cost function $\nabla{\rm C}(\bar \uptheta,\epsilon)$ and update the variational parameter $\bar \uptheta \to \bar \uptheta + \eta \, \nabla {\rm C}(\bar \uptheta,\epsilon)$;
    \item[4.] go to step 1;
\end{enumerate}
where $\eta$ is the learning parameter.

\newpage

\begin{landscape}

\begin{table}[!h]
\centering
\scriptsize
\begin{tabular}{|c|ccccccccccc|cccc|}
\hline
 & $\hat{O}_0^{(1)}$ & $\hat{O}_1^{(1)}$ & $\hat{O}_2^{(1)}$ & $\hat{O}_3^{(1)}$ & $\hat{O}_4^{(1)}$ & $\hat{O}_5^{(1)}$ & $\hat{O}_6^{(1)}$ & $\hat{O}_7^{(1)}$ & $\hat{O}_8^{(1)}$ & $\hat{O}_9^{(1)}$ & $\hat{O}_{10}^{(1)}$ & $\hat{O}_{11}^{(2)}$ & $\hat{O}_{12}^{(2)}$ & $\hat{O}_{13}^{(2)}$ & $\hat{O}_{14}^{(2)}$ \\ 
\hline 
 & $\mathbf{1}$ & $\hat{Z}_1$ & $\hat{Z}_3$ & $\hat{Z}_2$ & $\hat{Z}_4$ & $\hat{Z}_1\otimes\hat{Z}_3$ & $\hat{Z}_{12}$ & $\hat{Z}_2\otimes\hat{Z}_3$ & $\hat{Z}_1\otimes\hat{Z}_4$ & $\hat{Z}_{34}$ & $\hat{Z}_2\otimes\hat{Z}_4$ & $\hat{A}_{12}\otimes\hat{A}_{34}$ & $\hat{B}_{12}\otimes\hat{A}_{34}$ & $\hat{A}_{12}\otimes\hat{B}_{34}$ & $\hat{B}_{12}\otimes\hat{B}_{34}$ \\ 
\hline
\hline
R$\,[\mathring{A}]$ & $h_0$ [Ha] & $h_1$ [Ha] & $h_2$ [Ha] & $h_3$ [Ha] & $h_4$ [Ha] & $h_5$ [Ha] & $h_6$ [Ha] & $h_7$ [Ha] & $h_8$ [Ha] & $h_9$ [Ha] & $h_{10}$ [Ha] & $h_{11}$ [Ha] & $h_{12}$ [Ha] & $h_{13}$ [Ha] & $h_{14}$ [Ha] \\ 
\hline 
\hline 
0.1 & 5.0607 & 0.30084 & 0.30084 & -0.72648 & -0.72648 & 0.19297 & 0.15205 & 0.19109 & 0.19109 & 0.15205 & 0.20307 & -0.039043 & 0.039043 & 0.039043 & -0.039043 \\ 
0.2 & 2.3103 & 0.28221 & 0.28221 & -0.64828 & -0.64828 & 0.19101 & 0.14936 & 0.18886 & 0.18886 & 0.14936 & 0.20032 & -0.039493 & 0.039493 & 0.039493 & -0.039493 \\ 
0.3 & 1.3007 & 0.25869 & 0.25869 & -0.54996 & -0.54996 & 0.18801 & 0.14527 & 0.18548 & 0.18548 & 0.14527 & 0.19623 & -0.040205 & 0.040205 & 0.040205 & -0.040205 \\ 
0.4 & 0.74077 & 0.23529 & 0.23529 & -0.45353 & -0.45353 & 0.18422 & 0.1402 & 0.18133 & 0.18133 & 0.1402 & 0.19136 & -0.041129 & 0.041129 & 0.041129 & -0.041129 \\ 
0.5 & 0.37983 & 0.21393 & 0.21393 & -0.36914 & -0.36914 & 0.17993 & 0.13459 & 0.17681 & 0.17681 & 0.13459 & 0.18621 & -0.042218 & 0.042218 & 0.042218 & -0.042218 \\ 
0.6 & 0.13237 & 0.19481 & 0.19481 & -0.2992 & -0.2992 & 0.17533 & 0.12877 & 0.1722 & 0.1722 & 0.12877 & 0.18113 & -0.043433 & 0.043433 & 0.043433 & -0.043433 \\ 
0.7 & -0.042079 & 0.17771 & 0.17771 & -0.24274 & -0.24274 & 0.1706 & 0.12293 & 0.16768 & 0.16768 & 0.12293 & 0.17628 & -0.04475 & 0.04475 & 0.04475 & -0.04475 \\ 
0.736 & -0.097066 & 0.17141 & 0.17141 & -0.22343 & -0.22343 & 0.16869 & 0.12062 & 0.16593 & 0.16593 & 0.12062 & 0.17441 & -0.045303 & 0.045303 & 0.045303 & -0.045303 \\ 
0.8 & -0.16733 & 0.16252 & 0.16252 & -0.19744 & -0.19744 & 0.16583 & 0.1172 & 0.16336 & 0.16336 & 0.1172 & 0.1717 & -0.046157 & 0.046157 & 0.046157 & -0.046157 \\ 
0.9 & -0.25905 & 0.14908 & 0.14908 & -0.16071 & -0.16071 & 0.16114 & 0.11163 & 0.15927 & 0.15927 & 0.11163 & 0.16737 & -0.047643 & 0.047643 & 0.047643 & -0.047643 \\ 
1.0 & -0.32761 & 0.13717 & 0.13717 & -0.13036 & -0.13036 & 0.1566 & 0.10623 & 0.15543 & 0.15543 & 0.10623 & 0.16327 & -0.049198 & 0.049198 & 0.049198 & -0.049198 \\ 
1.1 & -0.37969 & 0.12654 & 0.12654 & -0.10486 & -0.10486 & 0.15229 & 0.10103 & 0.15183 & 0.15183 & 0.10103 & 0.15937 & -0.050806 & 0.050806 & 0.050806 & -0.050806 \\ 
1.2 & -0.4196 & 0.11699 & 0.11699 & -0.083203 & -0.083203 & 0.14827 & 0.096044 & 0.14849 & 0.14849 & 0.096044 & 0.15568 & -0.052448 & 0.052448 & 0.052448 & -0.052448 \\ 
1.3 & -0.45027 & 0.10835 & 0.10835 & -0.064754 & -0.064754 & 0.14457 & 0.091292 & 0.1454 & 0.1454 & 0.091292 & 0.15219 & -0.054104 & 0.054104 & 0.054104 & -0.054104 \\ 
1.4 & -0.4738 & 0.10054 & 0.10054 & -0.049032 & -0.049032 & 0.1412 & 0.086788 & 0.14254 & 0.14254 & 0.086788 & 0.14891 & -0.055755 & 0.055755 & 0.055755 & -0.055755 \\ 
1.5 & -0.49179 & 0.093456 & 0.093456 & -0.035645 & -0.035645 & 0.13818 & 0.082537 & 0.13992 & 0.13992 & 0.082537 & 0.14586 & -0.057384 & 0.057384 & 0.057384 & -0.057384 \\ 
1.6 & -0.50548 & 0.087055 & 0.087055 & -0.024253 & -0.024253 & 0.13547 & 0.078543 & 0.13752 & 0.13752 & 0.078543 & 0.14302 & -0.058975 & 0.058975 & 0.058975 & -0.058975 \\ 
1.7 & -0.51585 & 0.081281 & 0.081281 & -0.014563 & -0.014563 & 0.13306 & 0.074803 & 0.13532 & 0.13532 & 0.074803 & 0.14039 & -0.060518 & 0.060518 & 0.060518 & -0.060518 \\ 
1.8 & -0.52368 & 0.076088 & 0.076088 & -0.0063218 & -0.0063218 & 0.13093 & 0.071308 & 0.13331 & 0.13331 & 0.071308 & 0.13796 & -0.062004 & 0.062004 & 0.062004 & -0.062004 \\ 
1.9 & -0.52955 & 0.071434 & 0.071434 & 0.00068821 & 0.00068821 & 0.12904 & 0.06805 & 0.13148 & 0.13148 & 0.06805 & 0.13573 & -0.063428 & 0.063428 & 0.063428 & -0.063428 \\ 
2.0 & -0.53394 & 0.067279 & 0.067279 & 0.0066513 & 0.0066513 & 0.12737 & 0.065016 & 0.1298 & 0.1298 & 0.065016 & 0.13367 & -0.064785 & 0.064785 & 0.064785 & -0.064785 \\ 
2.1 & -0.53719 & 0.063585 & 0.063585 & 0.011725 & 0.011725 & 0.12588 & 0.062192 & 0.12826 & 0.12826 & 0.062192 & 0.13177 & -0.066073 & 0.066073 & 0.066073 & -0.066073 \\ 
2.2 & -0.5396 & 0.060312 & 0.060312 & 0.016042 & 0.016042 & 0.12457 & 0.059564 & 0.12686 & 0.12686 & 0.059564 & 0.13001 & -0.067294 & 0.067294 & 0.067294 & -0.067294 \\ 
2.3 & -0.54137 & 0.057422 & 0.057422 & 0.019718 & 0.019718 & 0.1234 & 0.05712 & 0.12557 & 0.12557 & 0.05712 & 0.1284 & -0.068446 & 0.068446 & 0.068446 & -0.068446 \\ 
2.4 & -0.54266 & 0.054879 & 0.054879 & 0.022848 & 0.022848 & 0.12236 & 0.054845 & 0.12438 & 0.12438 & 0.054845 & 0.1269 & -0.069531 & 0.069531 & 0.069531 & -0.069531 \\ 
2.5 & -0.5436 & 0.052649 & 0.052649 & 0.025514 & 0.025514 & 0.12142 & 0.052726 & 0.12328 & 0.12328 & 0.052726 & 0.12551 & -0.070553 & 0.070553 & 0.070553 & -0.070553 \\ 
2.6 & -0.54427 & 0.050699 & 0.050699 & 0.027784 & 0.027784 & 0.12057 & 0.050751 & 0.12226 & 0.12226 & 0.050751 & 0.12423 & -0.071513 & 0.071513 & 0.071513 & -0.071513 \\ 
2.7 & -0.54476 & 0.049001 & 0.049001 & 0.029716 & 0.029716 & 0.1198 & 0.048909 & 0.12132 & 0.12132 & 0.048909 & 0.12305 & -0.072414 & 0.072414 & 0.072414 & -0.072414 \\ 
2.8 & -0.5451 & 0.047528 & 0.047528 & 0.031358 & 0.031358 & 0.1191 & 0.047188 & 0.12045 & 0.12045 & 0.047188 & 0.12195 & -0.073261 & 0.073261 & 0.073261 & -0.073261 \\ 
2.9 & -0.54534 & 0.046256 & 0.046256 & 0.03275 & 0.03275 & 0.11845 & 0.045579 & 0.11964 & 0.11964 & 0.045579 & 0.12093 & -0.074056 & 0.074056 & 0.074056 & -0.074056 \\ 
3.0 & -0.54551 & 0.045164 & 0.045164 & 0.033928 & 0.033928 & 0.11784 & 0.044072 & 0.11887 & 0.11887 & 0.044072 & 0.11998 & -0.074803 & 0.074803 & 0.074803 & -0.074803 \\ 
\hline 
\end{tabular}
\caption{List of operators divided into the two commuting groups together with the coefficients $h$s for different atomic distances in the case of the Hydrogen molecule, using the minimal STO-3G basis set~\cite{psi4}. To simplify the writing, we define the operators $\hat{Z}_{mn} \equiv \hat{Z}_m\otimes\hat{Z}_n$, $\hat{A}_{mn} \equiv \hat{Y}_m\otimes\hat{X}_n$ and $\hat{B}_{mn} \equiv \hat{X}_m\otimes\hat{Y}_n$, where the indices refer to the qubit numbers, and identity operators are omitted.
}
\label{tab:H2_coeff}
\end{table}

\end{landscape}

\begin{landscape}

\begin{table}[!h]
\centering
\scriptsize
\begin{tabular}{|c|cccccccccccccccc|}
\hline
 & $\hat{O}_0$ & $\hat{O}_1$ & $\hat{O}_2$ & $\hat{O}_3$ & $\hat{O}_4$ & $\hat{O}_5$ & $\hat{O}_6$ & $\hat{O}_7$ & $\hat{O}_8$ & $\hat{O}_9$ & $\hat{O}_{10}$ & $\hat{O}_{11}$ & $\hat{O}_{12}$ & $\hat{O}_{13}$ & $\hat{O}_{14}$ & $\hat{O}_{15}$ \\ 
\hline 
 & $\mathbf{1}$ & $\hat{Z}_1$ & $\hat{Z}_2$ & $\hat{Z}_3$ & $\hat{Z}_4$ & $\hat{Z}_{12}$ & $\hat{Z}_1\otimes\hat{Z}_3$ & $\hat{Z}_1\otimes\hat{Z}_4$ & $\hat{Z}_2\otimes\hat{Z}_3$ & $\hat{Z}_2\otimes\hat{Z}_4$ & $\hat{Z}_{34}$ & $\hat{Z}_{12}\otimes\hat{Z}_{3}$ & $\hat{Z}_{12}\otimes\hat{Z}_{4}$ & $\hat{Z}_{1}\otimes\hat{Z}_{34}$ & $\hat{Z}_{2}\otimes\hat{Z}_{34}$ & $\hat{Z}_{12}\otimes\hat{Z}_{34}$ \\ 
\hline
\hline
$N$ & $w_0$ & $w_1$ & $w_2$ & $w_3$ & $w_4$ & $w_5$ & $w_6$ & $w_7$ & $w_8$ & $w_9$ & $w_{10}$ & $w_{11}$ & $w_{12}$ & $w_{13}$ & $w_{14}$ & $w_{15}$ \\ 
\hline 
\hline 
15 & 186 & -96 & -48 & -96 & -48 & 84 & 136 & 68 & 68 & 34 & 84 & -64 & -32 & -64 & -32 & 16\\
21 & 210 & 0 & 0 & 0 & 0 & 84 & 88 & 44 & 44 & 22 & 84 & -64 & -32 & -64 & -32 & 16 \\
35 & 546 & 224 & 112 & 224 & 112 & 84 & -24 & -12 & -12 & -6 & 84 & -64 & -32 & -64 & -32 & 16 \\
\hline 
\end{tabular}
\caption{
List of operators in the case of three semiprime numbers lower than 49. To simplify the writing, we define the operators $\hat{Z}_{mn} \equiv \hat{Z}_m\otimes\hat{Z}_n$, where the indices refer to the qubit numbers, and identity operators are omitted. }
\label{tab:facto_coeff}
\end{table}

\end{landscape}

\end{appendices}

% \end{comment}

\begin{backmatter}

\bmsection{Funding}
Horizon 2020 Framework Programme (899368);
Horizon Widera 2023 (101160101);
Provincia Autonoma di Trento through the Q@TN joint laboratory.

\bmsection{Acknowlegments}
The work was supported by the Horizon 2020 Framework Programme (899368), Horizon Widera 2023 (101160101) and by the Provincia Autonoma di Trento through the Q@TN joint laboratory.
We acknowledge the work of Nicola Furlan (University of Trento) for the PCB design, Sara Ferrari (Fondazione Bruno Kessler) for the wire-bonding service, and the master students Nicolò Broseghini and Leonardo Cattarin for the self-alignment code and for the initial work on the setup, respectively.

\bmsection{Disclosures}
The authors declare no conflicts of interest.

\bmsection{Data availability}
No data were generated or analyzed in the presented research.

\end{backmatter}

%%%%%%%%%%%%%%%%%%%%%%% References %%%%%%%%%%%%%%%%%%%%%%%%%
\bibliography{vqa_biblio}

\end{document}